\documentclass[aps,prx,letterpaper,twocolumn,floats,superscriptaddress,tighten,eqsecnum]{revtex4}

\UseRawInputEncoding

\usepackage{amssymb}
\usepackage{amsbsy}
\usepackage{amsmath}
\usepackage{graphicx}
\usepackage{graphics}
\usepackage{setspace}
\usepackage{array}
\usepackage{color}
\usepackage{xcolor}
\usepackage{fontenc}
\usepackage{textcomp}
\usepackage{rotating}
\usepackage{bm}
\usepackage{hyperref}
\usepackage{placeins}
\hypersetup{pdfpagemode=UseNone}
\usepackage{subfigure}
\usepackage{chngcntr}

\newcommand{\e}{\varepsilon}

\newcommand{\vex}[1]{\bm{\mathrm{#1}}}

\newcommand{\bsub}{\begin{subequations}}
\newcommand{\esub}{\end{subequations}}

\newcommand{\be}{\begin{equation}}
\newcommand{\ee}{\end{equation}}
\newcommand{\bea}{\begin{eqnarray}}
\newcommand{\eea}{\end{eqnarray}}

\begin{document}
\title{Phonon-limited resistivity of multilayer graphene systems}
\author{Seth M.\ Davis}
\email{smdavis1@umd.edu}
\affiliation{Condensed Matter Theory Center and Joint Quantum Institute, Department of Physics, University of Maryland, College Park, MD 20742, USA}
\author{Yang-Zhi\ Chou}
\affiliation{Condensed Matter Theory Center and Joint Quantum Institute, Department of Physics, University of Maryland, College Park, MD 20742, USA}
\author{Fengcheng\ Wu}
\affiliation{School of Physics and Technology, Wuhan University, Wuhan 430072, China}
\affiliation{Wuhan Institute of Quantum Technology, Wuhan 430206}
\author{Sankar\ Das\ Sarma}
\affiliation{Condensed Matter Theory Center and Joint Quantum Institute, Department of Physics, University of Maryland, College Park, MD 20742, USA}
\date{\today}

\begin{abstract}
We calculate the theoretical contribution to the doping and temperature ($T$) dependence of electrical resistivity due to scattering by acoustic phonons in Bernal bilayer graphene (BBG) and rhombohedral trilayer graphene (RTG). We focus on the role of nontrivial geometric features of the detailed, anisotropic $\vex{k}\cdot\vex{p}$ band structures of these systems - e.g. Van Hove singularities, Lifshitz transitions, Fermi surface anisotropy, and band curvature near the gap - whose effects on transport have not yet been systematically studied. We find that these geometric features strongly influence the temperature and doping dependencies of the resistivity. In particular, the band geometry leads to a nonlinear $T$-dependence in the high-$T$ equipartition regime, complicating the usual $T^4$ to $T$ Bloch-Gr\"{u}neisen crossover. Our focus on BBG and RTG is motivated by recent experiments in these systems that have discovered several exotic low-$T$ superconductivity proximate to complicated hierarchies of isospin-polarized phases. These interaction-driven phases are intimately related to the geometric features of the band structures, highlighting the importance of understanding the influence of band geometry on transport. While resolving the effects of the anisotropic band geometry on the scattering times requires nontrivial numerical solution, our approach is rooted in intuitive Boltzmann theory. We compare our results with recent experiment and discuss how our predictions can be used to elucidate the relative importance of various scattering mechanisms in these systems.
\end{abstract}
\maketitle





\section{Introduction}
\label{Section-Introduction}

Rapid progress in the ability to produce clean, stable, 2D layered van der Walls heterostructures made up of graphene and/or transition metal dichalcogenides (TMDs) has opened a new subfield of condensed matter physics \cite{Geim_2013, Novoselov_2006, Bistritzer_2011, Morell_2010, Li_2019, Kim_2017, Cao_2018a, Cao_2018b, Cao2020PRL, Cao2021, Yankowitz_2019,Kerelsky_2019, Lu_2019, Stepanov2020untying, Sharpe_2019, Chen_2020, Rozen2021entropic, AndreaYoungBernal, AndreaYoungRhombo, AndreaYoungRhombo2, Serlin_2020, Wu_2018, Wu_2019_TIPRL, KinFaiMak_TopologyTMD, Polshyn_2020, TBLGStrangeMetalExperiment1, TBLGStrangeMetalExperiment2, TBLGStrangeMetalExperiment3, SankarFengchengStrangeMetal, CalTechBernal, CalTechSymmetryBreakingTBLG, CalTechTwisted, Xie_2020, MacDonaldTLBGReview, Li_2021, Ghiotto_2021, Pan_2020, Pan_2021, Morales_Dur_n_2021, Ahn_2022, Kerelsky2021moireless, Khalaf2019}. The sensitivity of the band structures of these systems to external control parameters, especially twist angle and displacement field, gives an unprecedented experimental ability to engineer flat bands and control the location of geometric band features (e.g. Van Hove singularities and Lifshits transitions), and thus to tune the relative strength of interaction-driven physics.

This family of systems has already shown various correlated insulating states \cite{Cao_2018a, Lu_2019}, ferromagnetism \cite{Sharpe_2019, Chen_2020}, correlation-driven valley and iso-spin polarization \cite{AndreaYoungBernal, AndreaYoungRhombo}, anomalous quantum hall physics \cite{Serlin_2020}, topological insulator physics \cite{Wu_2019_TIPRL, KinFaiMak_TopologyTMD, Polshyn_2020}, metal-insulator transitions \cite{Li_2021, Ghiotto_2021, Pan_2020, Pan_2021, Morales_Dur_n_2021, Ahn_2022}, possible ``strange metal" resistance scaling at very low temperature \cite{TBLGStrangeMetalExperiment1, TBLGStrangeMetalExperiment2, TBLGStrangeMetalExperiment3, SankarFengchengStrangeMetal}, and most conspicuously, possibly-exotic superconductivity \cite{Cao_2018b, Wu_2018, AndreaYoungRhombo2, AndreaYoungRhombo, AndreaYoungBernal, CalTechBernal, CalTechSymmetryBreakingTBLG, CalTechTwisted}, including phases with verified non-spin-singlet pairing \cite{AndreaYoungBernal, AndreaYoungRhombo2}. The rich phase diagrams and high experimental control that characterize these systems has quickly made them into one of the most studied platforms in condensed matter physics. The above-listed discoveries demonstrate that geometric band features can have a profound influence on the effects of interactions on transport properties. In turn, this highlights the need for a refinement of the basic theories of phonon-limited resistivity as applied to these materials, accurately taking complex band geometry into account.

\begin{figure}[t!]
\includegraphics[angle=0,width=.45\textwidth]{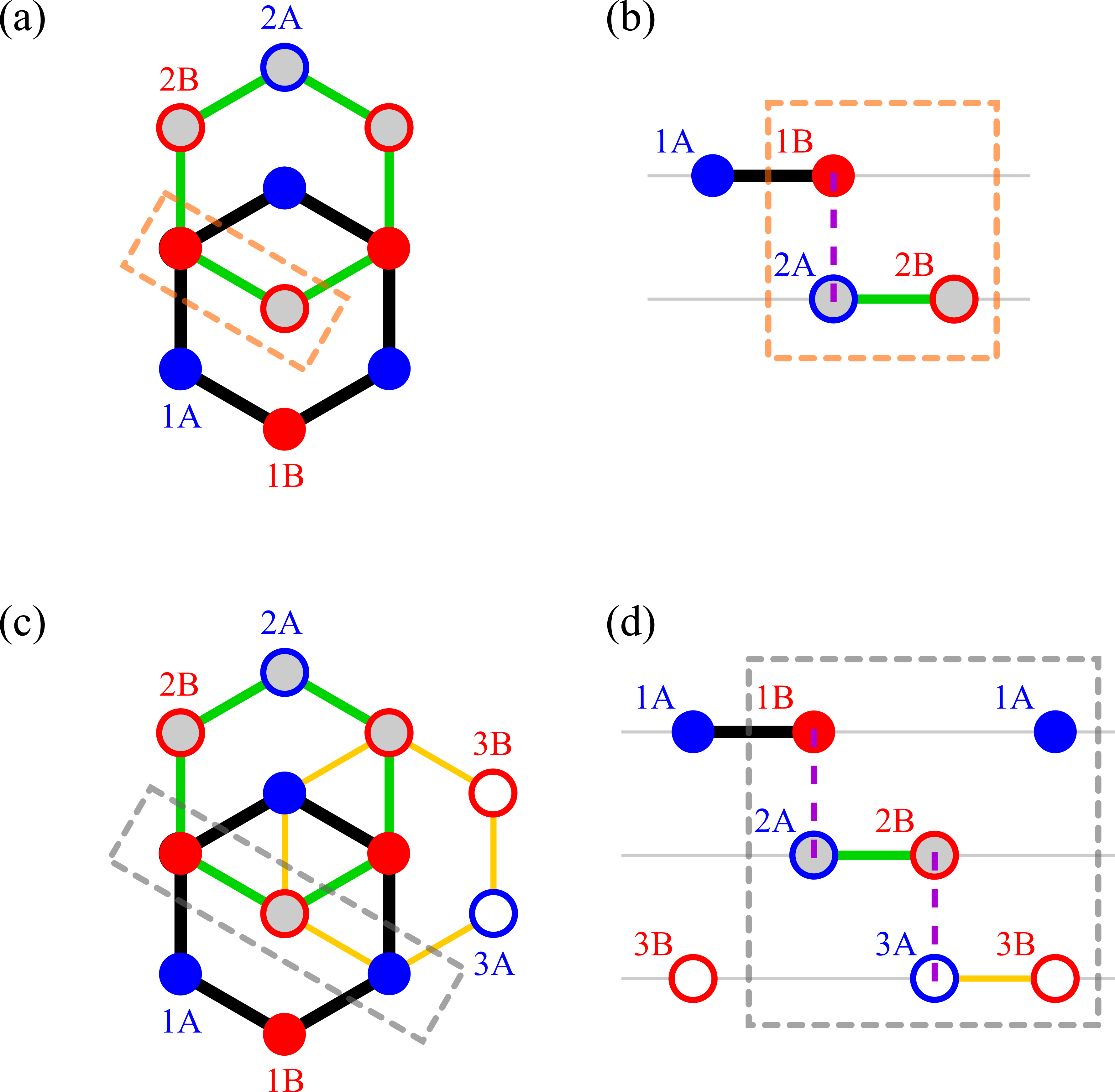}
\caption{We depict the lattice structure of Bernal bilayer graphene (top) and rhombohedral trilayer graphene (bottom). The left side of the image shows top-down views of the xy-plane, labelling atoms with their layer number ($\{1,2\}$) and sublattice index ($\{A,B\}$). The right side of the figure shows the stacking from a cross-section view.}
\label{GrapheneStackGeometries}
\end{figure}

\begin{figure*}[t]
\includegraphics*[angle=0,width=.9\textwidth]{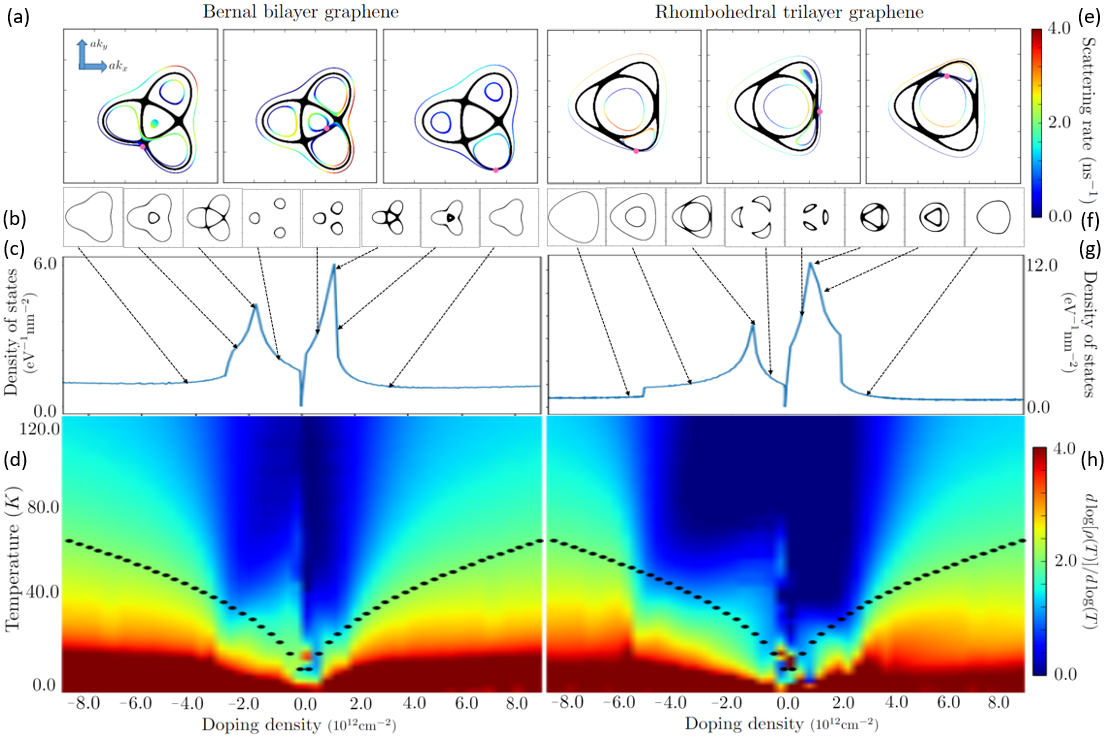}
\caption{
Overview of phonon scattering in Bernal bilayer and rhombohedral trilayer layered graphene systems. The top row (a,e) depicts qualitatively distinct kinematically-allowed scattering manifolds for different Bloch states at the same energy. In these figures, the black curve depicts a Fermi surfaced (near the hole-doped VHS), the pink dot denotes a reference Bloch state, and the colored points denote the points in $\vex{k}$-space that the reference state can scatter too while conserving conservation of energy and momentum. These are the ``scattering manifolds" [Sec.~\ref{Subsection-BasicTheory}], which depend on the geometry of the system. The coloring of the scattering manifolds encodes the transitions rates from the reference state. These plots demonstrate the nontrivial kinematics and geometry at play in scattering in these systems.
The middle layers (b,c,f,g) shows the density of states of the two systems, with labels showing how the Fermi surface geometry changes as the sample is doped.
The bottom layer (d,h) gives the central results of this work, the scaling of the resistivity with temperature in various regions of $n-T$ space. This is captured by a heat map of $d\log[\rho(n,T)]/d\log T.$ We emphasize the clear connection between the scaling behavior and the geometric features in the density of states. The doted line gives the naive $T^*_{BG}$ calculated with Eq.~(\ref{BGTransitionTemperature}) for an isotropic system. We see that at sufficiently large dopings, the color contours begin to follow the $\sqrt{|n|}$ profile traced out by the black dotted $T^*_{BG}$ line, reflecting the fact that at large dopings, the Fermi surface becomes roughly circular. However, our system exhibits a surprising suppression of $T^*_{BG}$ as doping is decreased and the Fermi surface qualitatively changes. These results are calculated with an inter-layer potential of $\Delta = 0.07\ eV$, and should be compared with Fig.~\ref{ResultsSummary-GaplessBernalAndRhomboPhaseDiagram}, which treats the zero-field case. The analogous results for a simple Dirac cone (gapped and ungapped) are given in Fig.~\ref{ResultsSummary-GappedAndGaplessDiracConePhaseDiagram} for further comparison.
} 
\label{IntroductionSuperfigure-GappedBernalAndRhomboPhaseDiagram}
\end{figure*}

\begin{figure*}[t!]
\includegraphics[angle=0,width=.9\textwidth]{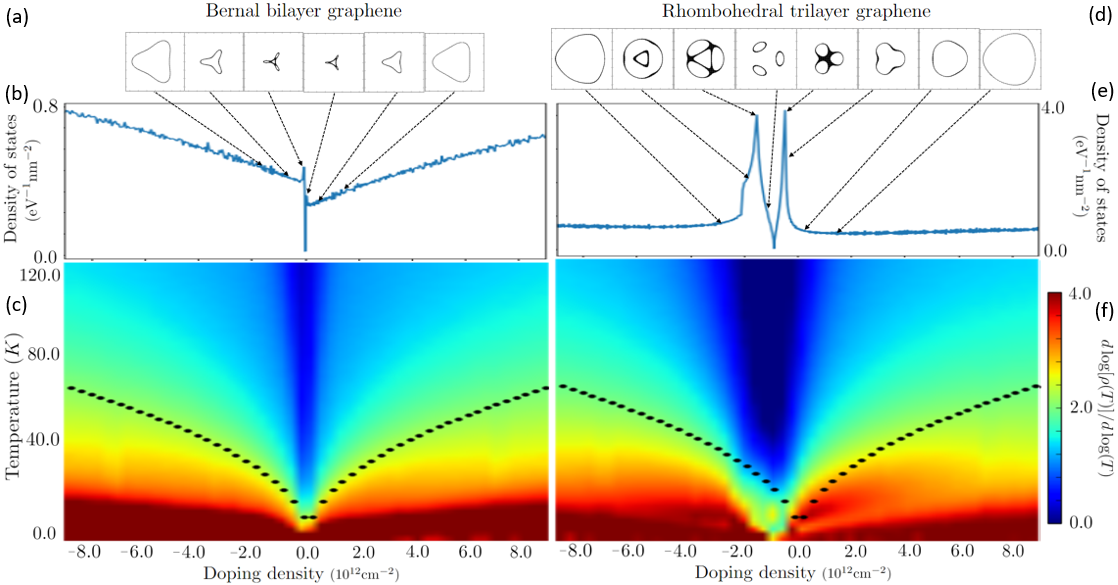}
\caption{
Resistivity scaling with $T$ due to phonon scattering in ungapped BBG and RTG, as in the absence of the inter-layer potential, to be compared with Fig.~\ref{IntroductionSuperfigure-GappedBernalAndRhomboPhaseDiagram}. As in Fig.~\ref{IntroductionSuperfigure-GappedBernalAndRhomboPhaseDiagram}, the top layers (a,b,d,e) plot the density of states of the system for various doping levels, labelled with Fermi surface geometries. The bottom row (c,f) provides a heat map of $d \log[\rho]/d \log T$ over $n-T$ space, mapping out the various regimes of resistivity scaling. As in the case of the gapped systems (Fig.~\ref{IntroductionSuperfigure-GappedBernalAndRhomboPhaseDiagram}), we see that the BG transition is strongly affected by the band geometry. Though this is more subtle without the applied field, we still see the effects clearly in the RTG case, which still exhibits an annular Fermi surface over a small doping window. As before, the dotted line gives the expected BG crossover for an isotropic system; for the ungapped systems plotted here, this estimate is quite accurate as long as the sample is sufficiently doped.
}
\label{ResultsSummary-GaplessBernalAndRhomboPhaseDiagram}
\end{figure*}

\begin{figure}[b!]
\includegraphics[angle=0,width=.45\textwidth]{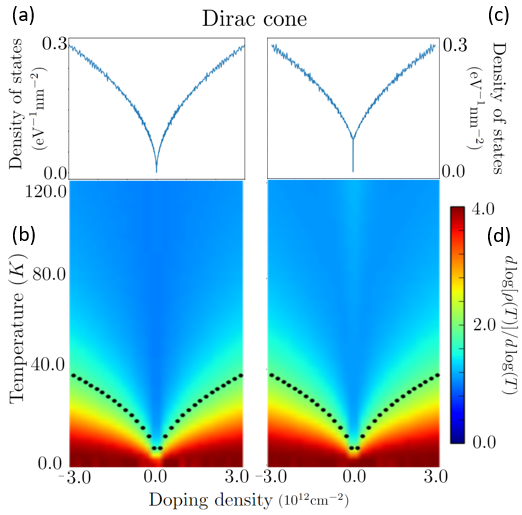}
\caption{
Resistivity scaling with $T$ due to phonon scattering in simple gapped (right) and ungapped (left) Dirac cones, to be compared with Figs.~\ref{IntroductionSuperfigure-GappedBernalAndRhomboPhaseDiagram} and \ref{ResultsSummary-GaplessBernalAndRhomboPhaseDiagram}. As in those figures, the bottom row provides a heat map of $d \log[\rho]/d \log T$ over $n-T$ space. The top row plots the density of states of the system for various doping levels. The dotted line gives $T^*_{BG}$ [Eq.~(\ref{BGTransitionTemperature})].
}
\label{ResultsSummary-GappedAndGaplessDiracConePhaseDiagram}
\end{figure}

In particular, recent experiments in ABC-stacked rhombohedral trilayer graphene (RTG) and AB-stacked Bernal bilayer graphene (BBG) (Fig.~\ref{GrapheneStackGeometries}) have discovered superconductivity (SC) proximate to several correlated, iso-spin polarized phases \cite{AndreaYoungBernal, AndreaYoungRhombo, AndreaYoungRhombo2, CalTechBernal} in the vicinity of Van Hove singularities and Lifshitz transitions in the band structures. Additionally, there is evidence that some of the superconducting phases host unconventional, non-spin-singlet pairing. 
Theories of SC in RTG and BBG based on Cooper pairing mediated by interaction with acoustic phonons have been put forth that propose likely explanations for the SC, explaining the presence of both spin-triplet and spin-singlet phases and providing roughly accurate transition temperatures \cite{YZBernal, YZRhombo, YZLongPaper}. Additionally, the proximity of SC phases to various interaction-driven phases has spurred comparison to strong correlation physics, and several other explanations centering $e-e$ interactions have been proposed \cite{EEAshvin,EEBerg,EEBitan,EEGuinea,EELeonid,EEZalatel,Szabo2021BBG, Qin2022, Dai2022}. 




A time-tested method for ascertaining the relative importance of various scattering mechanisms in a material is to look for clues in the temperature dependence of the resistivity. This is because different mechanisms generally produce various characteristic contributions and finite-$T$ crossovers between these. For example, this debate is currently unfolding for twisted bilayer graphene, where it is still unclear whether observed linear-in-$T$ resistivity dependence is caused by phonons or an interaction-driven strange metal state, analogous to that famously seen in several highly-correlated systems \cite{SankarFengchengStrangeMetal, TBLGStrangeMetalExperiment1, TBLGStrangeMetalExperiment2, TBLGStrangeMetalExperiment3}.

The recent experiments in BBG and RTG show that moir\'{e}-induced correlation effects are not a necessary ingredient for SC in layered graphene systems, leaving phonon-induced pairing as the de-facto leading candidate for a universal SC mechanism in these systems. Especially since acoustic phonons give a consistent theory of SC in both BBG and RTG, it is important to understand and isolate the contribution to the resistivity that should be expected due to acoustic phonons in the absence of $e-e$ effects. In conventional superconductors, electron-phonon couplings extracted from SC tend to agree well with those extracted from transport measurements. Thus, an extensive quantitative comparision of the SC data and the transport data is an important step in elucidating the nature of the SC pairing. Further, since these systems demonstrate that superconductivity in 2D layered systems can be intertwined with the nontrivial Fermi surface geometry, they offer an arena to understand the extent to which these geometric features effect transport generally.

The general paradigm of acoustic-phonon-limited resistivity in isotropic (semi)metals is as follows \cite{SankarGrapheneKT1, SankarGrapheneKT2, SankarGrapheneKT3, SankarGrapheneKT4, SankarGrapheneKT5, Ziman, AshcroftAndMermin}. In the low-T regime, where the quantum statistics of the phonon are important, we expect $\rho \approx T^{d+2}$, where $d$ is the dimension of the sample. This characterizes the ``Bloch-Gr\"{u}neisen" (BG) regime, which corresponds to
\begin{align}
\label{BGTransitionTemperature}
    k_BT \ll k_BT^*_{BG} = \mathcal{C}_{BG}\cdot(2\hbar v_p k_F),
\end{align}

\FloatBarrier
\noindent
where $v_p$ is the phonon velocity, $k_F$ is the Fermi momentum, and $\mathcal{C}_{BG} \approx \mathcal{O}(1)$ is a material-specific constant. (Further, $k_B T_{BG} \equiv 2\hbar v_p k_F$ is usually defined.) In the high-$T$ ``equipartition" (EP) regime, $T >> T^*_{BG}$, we instead expect linear-in-T resistivity. We note that single-layer graphene displays these properties elegantly, with $\mathcal{C}_{BG} \approx 1/6$ \cite{SankarGrapheneKT1, Efetov_2010}.

The goal of this paper is to give a precise theoretical calculation of the resistance due to acoustic phonon scattering in BBG and RTG systems in the presence of an inter-layer potential ($\Delta$). We give concrete predictions for the doping ($n$) and temperature ($T$) dependence of the resistivity of these systems in the limit of phonon-dominated transport. The inter-layer potential (produced by a displacement field) is required to induce SC in BBG, and tuning this potential can significantly alter the band structure and control the location of the Van Hove singularities, affecting both the SC and the interaction-driven phases \cite{AndreaYoungBernal, AndreaYoungRhombo}.

\begin{figure}[t!]
\includegraphics[angle=0,width=.47\textwidth]{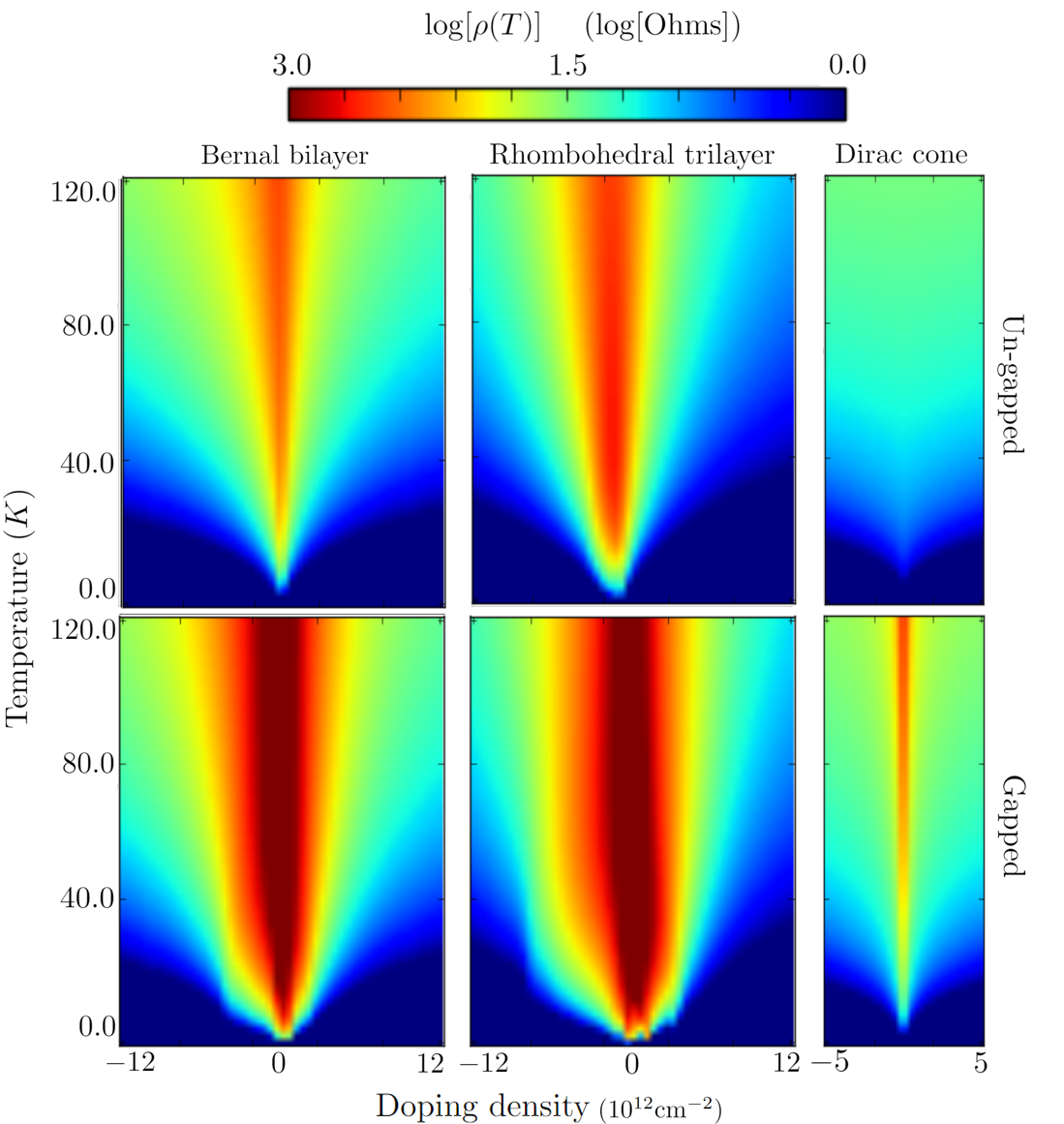}
\caption{
We plot a heat map of $\log[\rho]$ over doping density and temperature for the BBG (left), RTG (middle), and Dirac (right) systems for both the gapless (top) and gapped (bottom) cases. For BBG and RTG, the gapped systems correspond to a displacement field $\Delta = 0.07\ eV$, and the gapped Dirac system has $M = 0.05\ eV$. Several features are prominent. The dark blue in the lower corners shows the universal features of the BG-transition. All systems but the ungapped Dirac cone display density dependence throughout the high-$T$ regime, with a significant increase in resistivity near charge neutrality. This is especially prominent for the gapped RTG and BBG systems, where the resistivity spikes to $\rho\approx 3000 \Omega$. 
}
\label{ResultsSummary-LogRhoLowT}
\end{figure}
\begin{figure}[t!]
\includegraphics[angle=0,width=.46\textwidth]{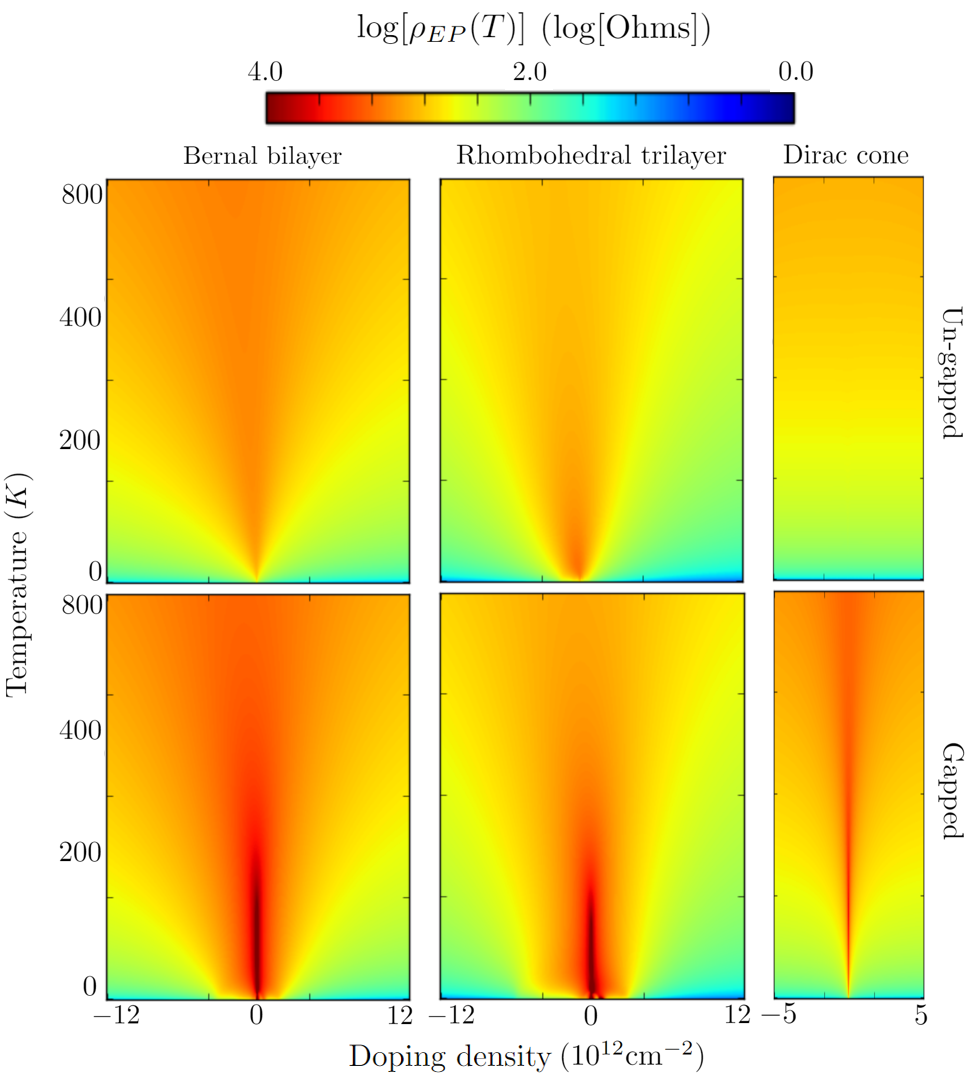}
\caption{
Accurate approximate calculations of resistivity in the equipartition regime can be made efficiently with the protocol discussed in Sec.~\ref{Subsection-BGAndEPRegimes}. Here we plot these approximate results over a large range of $T$, extending the scope of Fig.~\ref{ResultsSummary-LogRhoLowT}. We see that the large spike in resistivity near charge neutrality is a relatively low-$T$ behavior and that the resistivity decreases at higher temperatures, as $T$ get high enough to excite carriers in the conduction band. We emphasize that even at high-$T$, the resistivity does \textit{not} return to simple linear scaling, but instead asymptotes to a constant value. [See also Figs.~\ref{ResultsSummary-Slices} and \ref{AdditionalData-ParticleDopedResistivitySlices}.]
}
\label{ResultsSummary-LogRhoHighT}
\end{figure}

We carry out our calculation in the framework of Boltzmann kinetic theory, treating the acoustic phonons via the Debye approximation but retaining the full electronic band structure obtained by the diagonalization of $\vex{k}\cdot\vex{p}$ Hamiltonians \cite{MacDonaldBandstructureBernal, MacDonaldBandstructureRhombo}. We are able to numerically solve the linearized Boltzmann equation in the anisotropic band geometry and give quantitative predictions for the resistance and thus for the BG crossover temperature, $T^*_{BG}.$ We emphasize that accurately treating the non-isotropic band structure is a significant technical complication, beyond the techniques of prominent earlier treatments of resistivity in 2D layered graphene structures \cite{SankarGrapheneKT1, SankarGrapheneKT2, SankarGrapheneKT3, SankarGrapheneKT4, SankarGrapheneKT5}. Further, these earlier treatments of multi-layer graphene do not include the effects of the inter-layer potential.

We find that the electronic structure of the layered systems significantly distorts the BG paradigm explained above. In particular, while the high-$T$ behavior of the scattering rate of an individual Bloch state is linear, $1/\tau_{\vex{k}} \propto k_B T$, band curvature effects can lead to a complicated non-linear T-dependence of the resistivity curves. In the cases of gapped systems, (the displacement field generating $\Delta > 0$ opens a gap), there is a large spike in resistivity near charge neutrality.  These band-curvature effects interfere with the BG crossover, and we find that 

\begin{figure*}[p!]
\includegraphics*[angle=0,width=.9\textwidth]{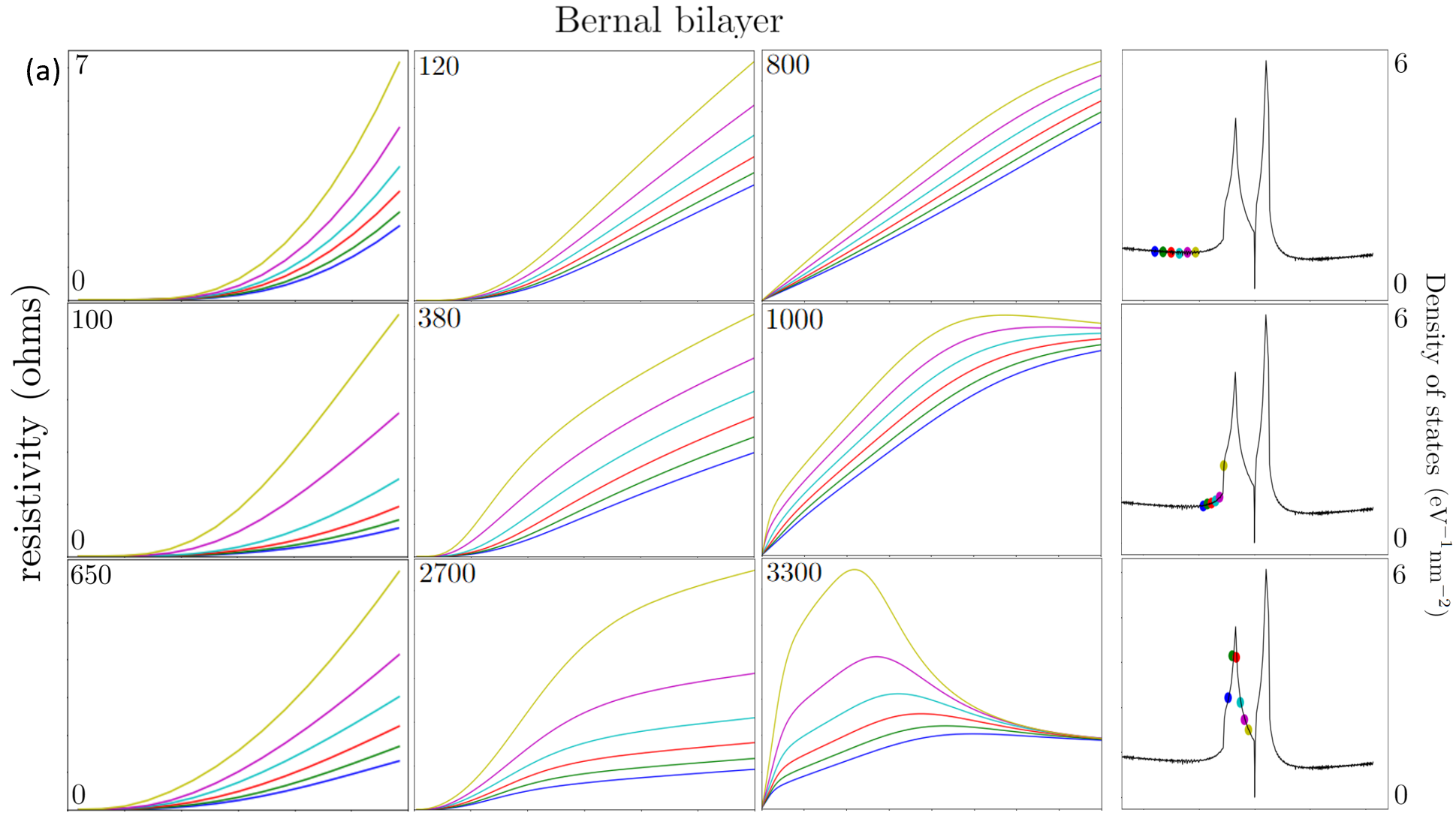}
%
\includegraphics*[angle=0,width=.9\textwidth]{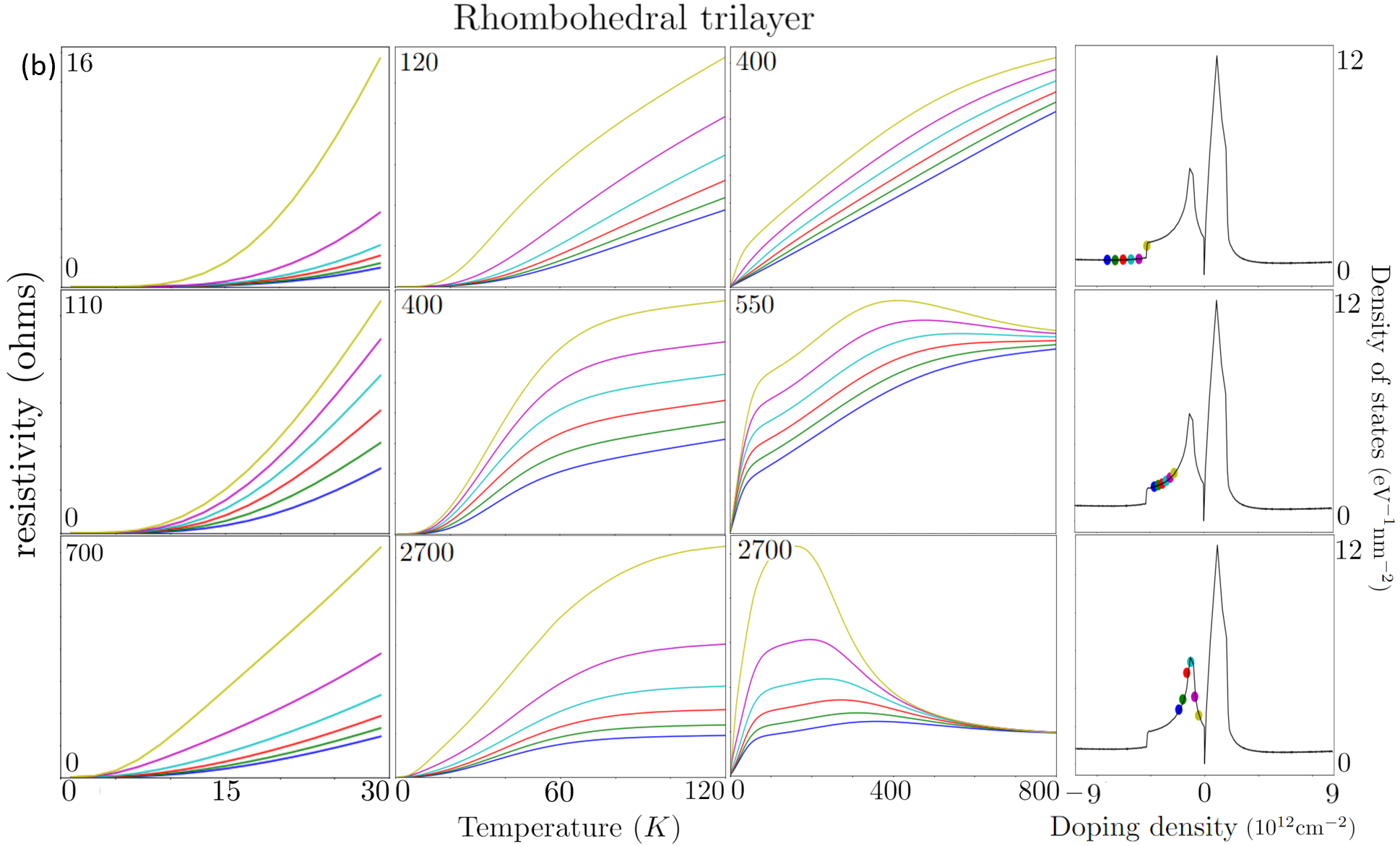}
\caption{
Resistivity data for hole-doped Bernal bilayer (top) and rhombohedral trilayer (bottom) graphene stacks, evaluated with the displacement field at $\Delta = 0.07\ eV$. Resistivity is given in ohms on a linear scale. The leftmost two columns give the results of our full numerical calculation for the resistivity of the two systems at various doping levels up to $30K$ and $120K$, respectively. The third column gives the same resistivity curves extended to $800K$, making use of the equipartition assumption discussed in Sec.~\ref{Subsection-NumericalComputation}. The far right columns indicate the doping levels of the curves shown in each row. From the full low-$T$ results (left) we may extract the effective BG crossover. We see that for large dopings, $\rho(T)$ exhibits a BG-EP crossover temperature $T^*_{BG}$ as high as $40-60 K$; however, there is a sharp drop in $T^*_{BG}$ to around $20 K$ upon the Lifshitz transition to the annular Fermi surface. The $T^*_{BG}$ continues to drop as we approach charge neutrality, dropping as low as $5-10 K$. We note that there is no sharp or discontinuous behavior at the Van Hove singularities. The high-$T$ equipartition results (center) show how band curvature effects lead to nonlinearity in $\rho(n,T)$ and how the effects can complicate the BG crossover. We see also that for dopings close to charge neutrality, we should expect a large spike in resistivity at moderate temperatures ($100-300 K$).
} 
\label{ResultsSummary-Slices}
\end{figure*}
\FloatBarrier

\noindent
the approximate power law for resistivity scaling is strongly influenced by the band structure geometry. This is demonstrated in Figs.~\ref{IntroductionSuperfigure-GappedBernalAndRhomboPhaseDiagram} and \ref{ResultsSummary-GaplessBernalAndRhomboPhaseDiagram} (and most others in this paper). Further, we note that the anisotropy (i.e., trigonal warping in graphene systems) in the band structure alters the low-$T$ BG relaxation rate $T^4$ power law to a non-universal, $\vex{k}$-dependent $T$-dependence. While this nonlinear-in-$T$ equipartition-regime phonon-limited resistivity is unexpected in the context of Boltzmann theory, we note that it has been detected experimentally in both bilayer and trilayer twisted graphene systems \cite{Polshyn2019, Siriviboon_2021}.

Our paper is organized as follows. In Sec.~\ref{Section-SummaryOfMainResults}, we present an overview of the main results of the work, emphasizing the most important quantitative aspects for comparison with experiment and qualitative results that run counter to common expectations. We then provide a concise review of acoustic phonon scattering in kinetic theory and present an overview of the calculation of relaxation times in the BBG and RTG systems in Sec.~\ref{Section-ResistivityViaBoltzmannKineticTheory}. We emphasize the roles of anisotropy and band curvature, which requires more care than the case of an isotropic band. The non-linear $T$-dependence we report in the equipartition regime is unexpected ; Section ~\ref{Section-NonlinearHighTResistivity} provides more intuition for these effects. Our concluding discussion is presented in Sec.~\ref{Section-DiscussionAndConclusions}.

Some supporting details are relegated to appendices. Appendix~\ref{Appendix-HamiltoniansForStackedGrapheneSystems} presents the $\vex{k}\cdot\vex{p}$ Hamiltonians used to calculate the band structure of BBG and RTG. Appendix~\ref{Appendix-NumericalImplementationOfResistivityCalculation} discusses the numerical solution for the relaxation rates in the solution of the linearized Boltzmann equation. Appendix~\ref{Appendix-RelaxationTimeApproximationInNonIsotropicSystems} discusses the role of the relaxation time approximation in resistivity calculations. Finally, Appendix~\ref{Appendix-AdditonalData} presents additional $\rho(n,T)$ data for the systems of interest, supplementing the results presented in Sec.~\ref{Section-SummaryOfMainResults}.




\section{Summary of main results}
\label{Section-SummaryOfMainResults}

Our central results are the calculations of the doping ($n$) and temperature ($T$) dependence of the resistivity [$\rho(n,T)$] for Bernal bilayer and rhombohedral trilayer graphene in the presence of a displacement field, under the assumption that scattering is limited to acoustic phonons (which we treat in the Debye approximation.) In particular, we give quantitative predictions for the crossover from the Bloch-Gr\"{u}neisen regime to the equipartition regime. 

We plot $log[\rho(n,T)]$ for low $T$ ($0-120K$) in Fig.~\ref{ResultsSummary-LogRhoLowT} and an approximate extension of these results to higher-$T$ ($0-800K$) in Fig.~\ref{ResultsSummary-LogRhoHighT}. Individual curves of $\rho(n,T)$ for fixed $n$ are given in Figs.~\ref{ResultsSummary-Slices}. The most obvious feature in this data is a strong spike in resistivity ($\rho \approx 3000 \Omega$) near to charge neutrality at low $T$. From Figs.~\ref{ResultsSummary-LogRhoHighT} and \ref{ResultsSummary-Slices}, we see that this is a low-$T$ phenomenon and that resistivity drops and levels out at higher $T$. However, we note that the high-$T$ resistivity is definitely \textit{not} given by a simple $T$-linear power law above the BG regime. In Figs.~\ref{IntroductionSuperfigure-GappedBernalAndRhomboPhaseDiagram}, \ref{ResultsSummary-GaplessBernalAndRhomboPhaseDiagram}, and \ref{ResultsSummary-GappedAndGaplessDiracConePhaseDiagram}, we plot $d\log[\rho(n,T)]/d\log(T)$ as an approximate scaling exponent for the resistivity. These plots act as a sort of ``phase diagram" for the various regimes of $T$-dependence in the resistivity profile. In particular, we find there is a region where the resistivity curve flattens out to be essentially constant with $T$, sometimes after a downturn. While this is counter to high-$T$ phonon expectations, this behavior has been measured in twisted bilayer \cite{Polshyn2019} and trilayer \cite{Siriviboon_2021} graphene systems. We stress that this is an effect entirely due to band curvature, which we discuss further in Sec.~\ref{Section-NonlinearHighTResistivity}.


Figures \ref{IntroductionSuperfigure-GappedBernalAndRhomboPhaseDiagram}, \ref{ResultsSummary-Slices}, \ref{ResultsSummary-LogRhoLowT},  \ref{ResultsSummary-GaplessBernalAndRhomboPhaseDiagram}, and \ref{ResultsSummary-GappedAndGaplessDiracConePhaseDiagram} all demonstrate the BG crossover mentioned in the introduction. At high dopings, where the Fermi surface is roughly circular, we find a $T^*_{BG} \approx 40-60K$, in line with expectations for a circular Fermi surface \cite{SankarGrapheneKT1, SankarGrapheneKT2, SankarGrapheneKT3, SankarGrapheneKT4, SankarGrapheneKT5}. However, we see a sharp drop to around $20 K$ at the Lifshitz transition to an annular Fermi surface, and $T^*_{BG}$ continues to drop as we approach charge neutrality. From Figs.~\ref{IntroductionSuperfigure-GappedBernalAndRhomboPhaseDiagram}, \ref{ResultsSummary-GaplessBernalAndRhomboPhaseDiagram}, and \ref{ResultsSummary-GappedAndGaplessDiracConePhaseDiagram}, it is clear that the band geometry created by applying a displacement field ($\Delta$) to the graphene layers causes significant alterations to the standard BG transition profile. Additionally, the curve-flattening discussed in the last paragraph can come into effect at $T$ comparable to the crossover temperature $T^*_{BG}$, making the $T^4 \rightarrow T$ transition difficult to observe.

Nevertheless, we predict that the phonon contribution to resistivity should become important at temperatures that vary between $10 K$ and $60 K$, depending on the doping, as shown in Fig.~\ref{ResultsSummary-Slices}. This should be compared with what is currently known from experiment: linear-in-$T$ resistivity dependence has not been observed under 20$K$ in RTG or under 1.5$K$ in BBG. It is important to note that the zero-$T$ contribution to resistivity from disorder ranges from about 30$\Omega$ to 70$\Omega$ in these systems \cite{AndreaYoungBernal, AndreaYoungRhombo, AndreaYoungRhombo2, CalTechBernal}.

We also report results for phonon scattering in BBG and RTG in the absence of the displacement field. The data is all given in Fig.~\ref{ResultsSummary-LogRhoHighT}, and the effective resistivity power law is extracted in Fig.~\ref{ResultsSummary-GaplessBernalAndRhomboPhaseDiagram}, which should be compared with Fig.~\ref{IntroductionSuperfigure-GappedBernalAndRhomboPhaseDiagram}. We note that in the absence of the applied field, the high-resistivity spike near charge neutrality is significantly diminished. However, we still find high-$T$ nonlinearity in the resistivity curves. Resistivity curves for the ungapped cases analogous to Fig.~\ref{ResultsSummary-Slices} can be found in Appendix~\ref{Appendix-AdditonalData}.




\section{Resistivity via Boltzmann kinetic theory}
\label{Section-ResistivityViaBoltzmannKineticTheory}

The use of Boltzmann kinetic theory to calculate linear response resistivities due to phonon collisions with Bloch state electrons is well-established \cite{AshcroftAndMermin, Ziman, SankarGrapheneKT1, SankarGrapheneKT2}. In this section, we outline the structure of the theory and explain our calculation, appealing to the Dirac cone of single-layer graphene to display concepts and highlight departures of our theory from previous work. We first introduce the model in Sec.~\ref{Subsection-Model}, then we state the main results of the kinetic theory in Sec.~\ref{Subsection-BasicTheory} and use these results to give intuition into the Bloch-Gr\"{uneisen} crossover in Sec.~\ref{Subsection-BGAndEPRegimes}. Finally, Sec.~\ref{Subsection-NumericalComputation}
discusses the actual computation of the resistivity.

\subsection{Model}
\label{Subsection-Model}
We use the electronic single-particle Hamiltonian
\begin{align}
\label{SingleParticleHamiltonian}
    H^e &= 
    \frac{1}{L^2}
    \sum_{\vex{k}}
    c^\dagger_{\vex{k}}
    H^e_{\vex{k}}
    c_{\vex{k}},
\end{align}
where $c^\dagger \equiv c^\dagger_{s,\xi,\sigma,l,\vex{k}}$ creates an electron with crystal momentum $\vex{k}$ (relative to Dirac point), spin $s$, valley $\xi$, sublattice $\sigma$, and layer $l$. In our models, $H^e_{\vex{k}} \equiv \delta_{s,s'}\delta_{\xi,\xi'}H^e_{\sigma,l,\sigma',l',\vex{k}}$ is a $\vex{k}$-dependent matrix coupling together layer and sublattice degrees of freedom, which are given in Appendix~\ref{Appendix-HamiltoniansForStackedGrapheneSystems}. This is a $\vex{k}\cdot\vex{p}$ continuum Hamiltonian from \cite{MacDonaldBandstructureBernal, MacDonaldBandstructureRhombo}, which is very accurate within $1 eV$ of the charge neutrality point. The four degenerate spin-valley flavors remain decoupled in our calculation and contribute equally to the conductivity (inverse resistivity).

We are interested in the effects of the electron bands, so we restrict our model to in-plane longitudinal acoustic phonons and adopt a simple Debye description. We thus take the phonon Hamiltonian to be
\begin{align}
\label{PhononHamiltonian}
H^p = \sum_{l,\vex{q}}\hbar\omega_{\vex{q}}a_{l,\vex{q}}^\dagger a_{l,\vex{q}},
\end{align}
where $\omega_{\vex{q}}$ is the phonon dispersion and we use the Debye approximation $\omega_{\vex{q}} \approx v_{p}|\vex{q}|$, where $v_p$ is the phonon velocity. This treatment neglects optical phonons, which should give a quantitative correction above some temperature. Since optical phonons have a large excitation gap in graphene, ranging from about $0.15$ to $0.20 eV$ \cite{Sohier2014}, they will become relevant at higher temperatures than we are concerned about here (approximately $1500K$) \cite{Sohier2014, Xie2016, Ghahari2016}. Our neglect of optical phonons is further justified by the fact that the electron-optical-phonon couplings are weak in graphene multilayers due to sublattice polarization \cite{Wu_2018}.

We couple the electrons to phonons via the well-known deformation potential coupling Hamiltonian \cite{Ziman, SankarGrapheneKT1, Coleman2015introduction}:
\begin{align}
\label{CouplingHamiltonian}
    H^{epc} &= 
    \sqrt{\frac{D^2 \hbar}{2\rho_M L^2}}
    \sum_{l,\vex{q}}
    \frac{\hat{n}_{\vex{q},l}}{\sqrt{\omega_{\vex{q}}}}(-i\vex{q}\cdot\hat{e}_{\vex{q}})(a_{\vex{q},l} + a_{-\vex{q},l}^\dagger).
\end{align}
Above, $D$ is the deformation potential, $\rho_M$ is the mass density of monolayer graphene, and $\hat{e}_{\vex{q}}$ is the desplacement unit vector of the phonon. Throughout this work, we set $D = 25$ eV, $\rho_M = 7.6 \cdot 10^{-8} g/cm^2$, and $v_p = 2.6 \cdot 10^6 cm/s$ \cite{SankarGrapheneKT1, SankarGrapheneKT2, SankarGrapheneKT3, Efetov_2010}. Finally, the electron density operator is 
\begin{align}
\label{DensityOperator}
    \hat{n}_{\vex{q},l} \equiv \sum_{\vex{k}}c^{\dagger}_{(\vex{k}+\vex{q}),l}c_{\vex{k},l}.
\end{align}
In Eqs.~(\ref{SingleParticleHamiltonian}) and (\ref{DensityOperator}), sums over unwritten $s,\sigma,\xi,l$ indices are implicit.

\begin{figure}[t!]
\includegraphics[angle=0,width=.48\textwidth]{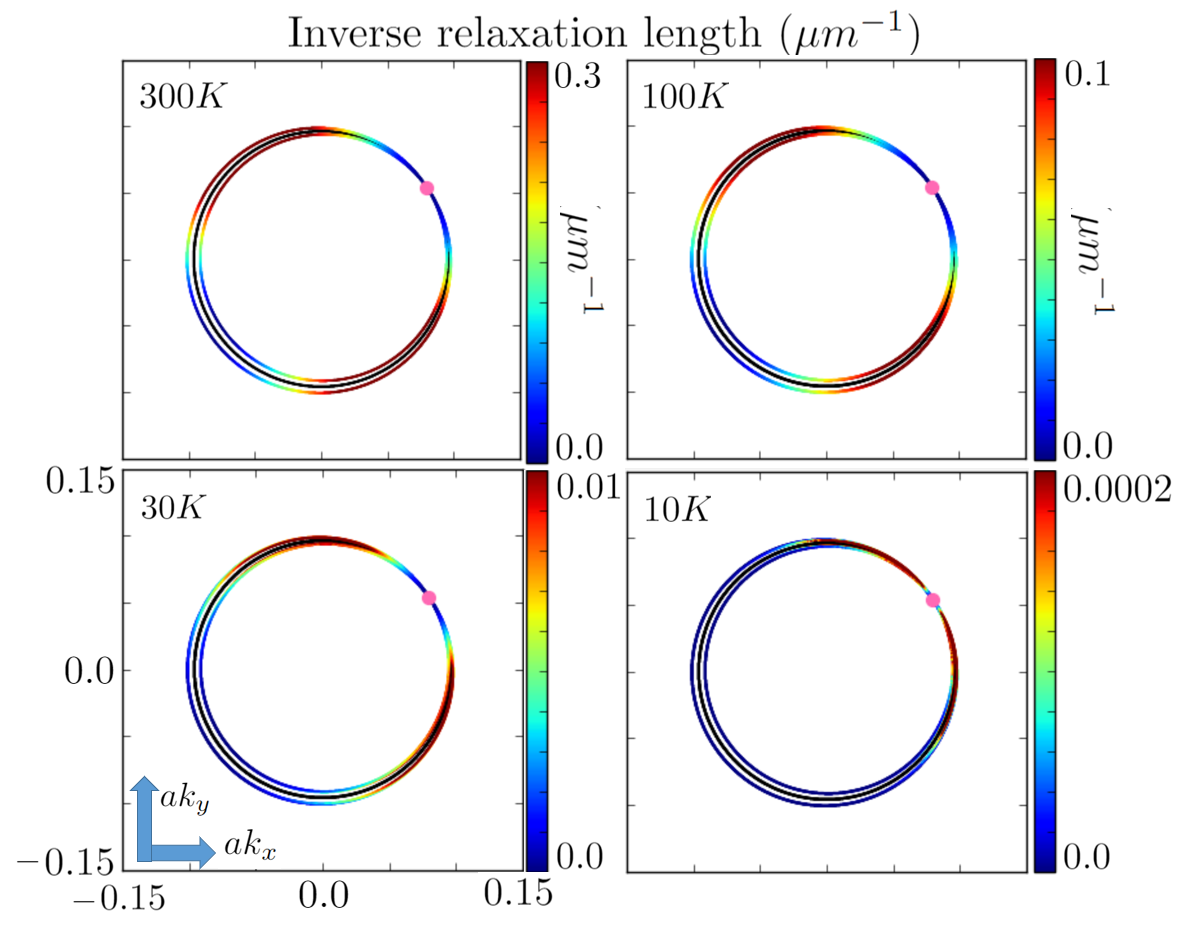}
\caption{
We plot the kinematically-allowed scattering manifolds for phonon-scattering on a Dirac cone, given by Eq.~(\ref{EnergyConservation}). The Fermi surface at $\mu = -0.25 eV$ is plotted in black and a reference point on the Fermi surface is identified with a pink dot. The colored points mark the set of k-space points that the reference point can scatter too while conserving energy and momentum. The color coding on the scattering manifold are proportional to the scattering rate between the two points. The figure is shown for descending temperatures: $300K$, $100K$, $30K$, $10K$. The Bloch-Gr\"{u}neisen transition is demonstrated by the fact that the $300K$ and $100K$ figures (top) only differ quantitatively by the scale of the color bar, while they are qualitatively distinct from the lower-temperature versions (bottom).
}
\label{KineticTheory-DiracScatteringManifold}
\end{figure}

\subsection{Kinetic theory}
\label{Subsection-BasicTheory}

In the so-called ``relaxation time approximation" \cite{AshcroftAndMermin} [also see Appendix~\ref{Appendix-RelaxationTimeApproximationInNonIsotropicSystems}] to Boltzmann kinetic theory, the resistivity tensor ($\rho$) is given by
\begin{align}
    \label{ResistivityDefinition}
    [\rho^{ij}(n,T)]^{-1} 
    &= 
    \frac{4e^2}{TL^2}
    \sum_{\vex{k} \in BZ}
    \tau_{\vex{k}}
    v^i_{\vex{k}}v^j_{\vex{k}}
    f(\e_{\vex{k}})[1-f(\e_{\vex{k}})],
\end{align}
where $T$ is temperature, $L$ is system length, $e$ is the electron charge, $v_{\vex{k}}^j$ are components of the velocity of the Bloch state $\vex{k}$, $f(\e)$ is the Fermi distribution function, and the $\tau_{\vex{k}}$ are the \textit{relaxation times} of the various Bloch states. If the band structure and Bloch states are known, the main challenge in the computation of the resistivity is the computation of the relaxation times. The leading factor of $4$ follows from the spin and valley degeneracies of the problem.

In Eq.~(\ref{ResistivityDefinition}), we have suppressed the band index ($b$) and taken the sum over $\vex{k}$ to mean a sum over all Bloch states: $\vex{k} \rightarrow (\vex{k},b)$. We will continue to use this notation and will explicitly mention when interband excitations or transitions are important.

Enforcing self-consistency of the relaxation time approximation on the Boltzmann equation [Appendix~(\ref{Appendix-RelaxationTimeApproximationInNonIsotropicSystems})], we find that
\begin{align}
    \label{RelaxationLengthSelfConsistency}
    \frac{1}{|\vex{v}_{\vex{k}}| L^2}
    \sum_{\vex{k'}\in BZ}
    \mathcal{W}_{\vex{k}\rightarrow\vex{k'}}
    \frac{1-f(\e_{\vex{k'}})}{1-f(\e_{\vex{k}})}
    \left[
    l_{\vex{k}} - l_{\vex{k'}}\cos\theta_{\vex{v}}
    \right]
    &= 1,
\end{align}
where $l_{\vex{k}} \equiv |\vex{v}_{\vex{k}}|\tau_{\vex{k}}$ are the ``relaxation lengths" (mean free paths), $\theta_{\vex{v}}$ is the angle between the Bloch velocities $\vex{v}_{\vex{k}}$ and $\vex{v}_{\vex{k'}}$, and $\mathcal{W}_{\vex{k}\rightarrow\vex{k'}}$ is the transition rate from state $\vex{k}$ to $\vex{k'}$. In the thermodynamic limit, Eq.~(\ref{RelaxationLengthSelfConsistency}) becomes an integral equation. For a finite-size system, it is a matrix equation that can be inverted to find the relaxation lengths. Again, band indices have been suppressed, but $\vex{k}$ and $\vex{k'}$ should be taken to stand for the Bloch states $(\vex{k},b)$ and $(\vex{k'},b')$.

In the case of the standard deformation potential phonon coupling Hamiltonian, a standard Fermi's golden rule calculation gives the transition rates
\begin{align}
    \nonumber
    \mathcal{W}_{\vex{k}\rightarrow\vex{k'}}
    &=
    \frac{\pi D^2}{\rho_M v_p}
    |\vex{k'}-\vex{k}|
    \Delta(\e_{\vex{k}},\e_{\vex{k'}})
    \sum_{l}
    \bigg|\langle \psi_{\vex{k'},l}|\psi_{\vex{k},l}\rangle\bigg|^2
    \\
    \label{TransitionRates}
    &\equiv
    \hbar v_p|\vex{q}|
    \Delta(\e_{\vex{k}},\e_{\vex{k'}})
    \mathcal{C}_{\vex{k},\vex{k'}}.
\end{align}
with 
\begin{align}
    \vex{q} &\equiv \vex{k'}-\vex{k},
    \\
\label{EnergyConservation}    
    \Delta(\e_{\vex{k}},\e_{\vex{k'}}) 
    &\equiv
    \begin{aligned}
    N_{\vex{q}} &\delta(\e'-\e - \hbar v_p |\vex{q}|) 
    \\
    + 
    (N_{\vex{q}} + 1 ) &\delta(\e'-\e + \hbar v_p |\vex{q}|)
    \end{aligned}
    ,
    \\
\label{OccupationNumber}
    N_{\vex{q}} &\equiv \frac{1}{\exp(\hbar v_p |\vex{q}|/k_BT) - 1}.
\end{align}

The Dirac $\delta$-functions in Eq.~(\ref{EnergyConservation}) enforce conservation of energy and momentum and $N_{\vex{q}}$ gives the occupation numbers of phonons available for scattering. The first line in Eq.~(\ref{EnergyConservation}) refers to phonon absorption processes while the second refers to phonon emission. 

For a given energy band geometry, the conservation laws in Eq.~(\ref{EnergyConservation}) determine a set of \textit{scattering manifolds} for each Bloch state, corresponding to absorption and emission of phonons. The summand in Eq.~(\ref{RelaxationLengthSelfConsistency}) then determines the rate of transition to each point on the scattering manifold. Written as a sum over the scattering manifold  ($SM$), Eq.~(\ref{RelaxationLengthSelfConsistency}) takes the form
\begin{align}
    \label{RelaxationLengthSelfConsistency2}
    \frac{\hbar v_p}{|\vex{v}_{\vex{k}}| L^2}
    \sum_{\vex{k'}\in SM}
    |\vex{q}|
    \mathcal{C}_{\vex{k},\vex{k'}}
    \mathcal{F}_{\vex{k},\vex{k'}}^{\mu,T} 
    \left[
    l_{\vex{k}} - l_{\vex{k'}}\cos\theta_{\vex{v}}
    \right]
    &= 1,
\end{align}
with
\begin{align}
    \mathcal{F}_{\vex{k},\vex{k'}}^{\mu,T}
    \equiv 
    \frac{1-f(\e_{\vex{k'}})}{1-f(\e_{\vex{k}})}
    \times
    \left\{
    \begin{aligned}
    N_{\vex{q}} \ \ \ \ \e_{\vex{k'}} > \e_{\vex{k}}
    \\
    N_{\vex{q}} + 1 \ \ \e_{\vex{k'}} < \e_{\vex{k}}
    \end{aligned}
    \right\},
\end{align}
where $\vex{k} \in SM$ indicates a summation over the scattering manifold of states picked out by the delta functions in Eq.~(\ref{EnergyConservation}). We emphasize that all implicit dependence of the relaxation lengths on the temperature or chemical potential are due to $\mathcal{F}_{\vex{k},\vex{k'}}^{\mu,T}$.

From Eqs.~(\ref{ResistivityDefinition}-\ref{TransitionRates}) we see that resistance scales linearly with $D^2/\rho_M$, so our results are easy to adjust for different values of these parameters. The dependence on $v_p$ is more involved, since it also affects the geometry of the scattering manifolds.

Fig.~\ref{KineticTheory-DiracScatteringManifold} shows how scattering rates can vary across the scattering manifold, using a Dirac cone as a simple example. It also visually demonstrates the transition between the BG and EP regimes, which we discuss next. 

\begin{figure}[b]
\includegraphics[angle=0,width=.48\textwidth]{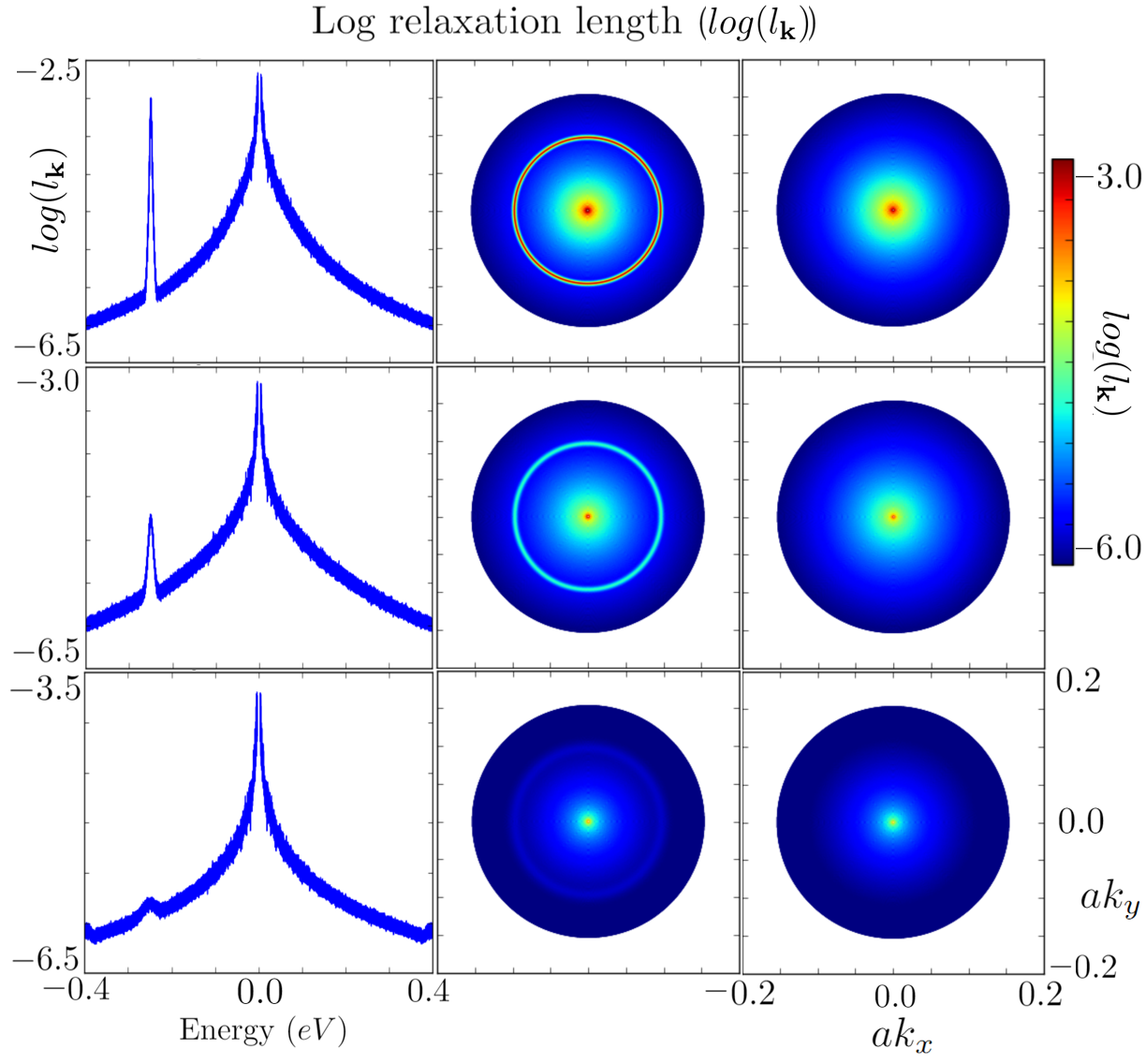}
\caption{
We plot the log of the relaxation lengths $\{l_{\vex{k}}\}$ for a Dirac cone band structure for various temperatures. We fix the chemical potential at $\mu = -0.25\ eV$ and perform the calculation at $T = 10\ K$ (top), $T = 30\ K$ (middle), and $T = 100\ K$ (bottom). The far left column shows the relaxation lengths plotted over energy, while the central and rightmost columns give a heat map of the relaxation lengths in momentum space for the bottom (center) and top (right) bands, respectively. We emphasize that at low T, states near the Fermi level become long-lived. Further, states near the Dirac point are always long-lived due to a vanishing scattering manifold.
}
\label{KineticTheory-DiracScatteringLengths}
\end{figure}


\subsection{Bloch-Gr\"{u}neisen and Equipartition regimes}
\label{Subsection-BGAndEPRegimes}

The low-$T$ BG regime is best understood in the case of an isotropic ($l_{\vex{k}} \rightarrow l_{\e_{\vex{k}}}$ and $\vex{v}_{\vex{k}} \parallel \vex{k}$) and quasi-elastic ($\e' \approx \e$) system, such as graphene \cite{SankarGrapheneKT1}. In this case, we can replace the velocity angle with the momentum angle ($\theta_{\vex{v}} = \theta_{\vex{k}}$) and
Eq.~(\ref{RelaxationLengthSelfConsistency}) simplifies to a direct formula for the relaxation time: 
\begin{align}
    \label{IsotropicRelaxationTime}
    \frac{1}{\tau_{\vex{k}}}
    &=
    \frac{\hbar v_p}{L^2}
    \sum_{\vex{k'}\in FS}
    |\vex{q}|
    \mathcal{C}_{\vex{k},\vex{k'}}
    \mathcal{F}_{\vex{k},\vex{k'}}^{\mu,T}
    \left[
    1 - \cos\theta_{\vex{k}}
    \right],
\end{align}
where $\vex{k} \in FS$ indicates a summation over the Fermi surface, which is taken to be indistinguishable from the scattering manifold in the quasi-elastic approximation.

For small $\vex{q}$, $1-\cos\theta_{\vex{k}} \approx |\vex{q}|^2$ and $\mathcal{C}_{\vex{k},\vex{k'}} \approx 1$, and the summand of Eq.~(\ref{IsotropicRelaxationTime}) scales with $\vex{q}$ roughly as $|\vex{q}|^3$. For low $T$, the Fermi functions $1-f(\e_{\vex{k'}})$ and the phonon occupation function $N_{\vex{q}}$ effectively restrict the sum in Eq.~(\ref{IsotropicRelaxationTime}) to $\vex{k'}$ with $\hbar v_p |\vex{q}| \leq k_B T$. Summing $|\vex{q}|^3$ over the portion of the [$(d-1)$-dimensional] scattering manifold within a radius proportional to $T$ gives the famous power-law defining the BG regime:
\begin{align}
    \frac{1}{\tau_{\vex{k}}} \propto T^{d+2}.
\end{align}
However, if we do \textit{not} assume isotropy, then we must restore
\begin{align}
\label{Anisotropy}
    1 - \cos\theta_{\vex{k}} \rightarrow 1 - \frac{l_{\vex{k'}}}{l_{\vex{k}}}\cos\theta_{\vex{v}}
\end{align}
in Eq.~(\ref{IsotropicRelaxationTime}). The small-$|\vex{q}|$ limit of the right hand side of Eq.~(\ref{Anisotropy}) is not necessarily proportional to $|\vex{q}|^2$, since it depends on the way $l_{\vex{k'}} \rightarrow l_{\vex{k}}$ and $\vex{v}_{\vex{k'}} \rightarrow \vex{v}_{\vex{k}}$ as $\vex{k'} \rightarrow \vex{k}.$ We therefore expect anisotropy to introduce non-universal, $\vex{k}$-dependent modifications of the BG power law in the $T$-dependence of each relaxation time $\tau_{\vex{k}}.$

In the high-$T$ limit, expanding in small $\Delta\e/T$, we find
\begin{align}
\label{FExpansion}
    \mathcal{F}_{\vex{k},\vex{k'}}^{\mu,T} = \frac{k_B T}{\hbar v_p |\vex{q}|} + \mathcal{O}(\Delta\e/T),
\end{align}
and inserting into Eq.~(\ref{RelaxationLengthSelfConsistency}) gives
\begin{align}
    \label{RelaxationLengthHighT}
    \frac{k_B T}{|\vex{v}_{\vex{k}}| L^2}
    \sum_{\vex{k'} \in SM}
    \mathcal{C}_{\vex{k},\vex{k'}}
    \left[
    l_{\vex{k}} - l_{\vex{k'}}\cos\theta_{\vex{v}}
    \right]
    &= 
    1 + \mathcal{O}(\Delta\e/T)^2.
\end{align}
Solving Eq.~(\ref{RelaxationLengthHighT}) order-by-order in $1/T$, we see that the high-$T$ form of the relaxation length is 
\begin{align}
\label{EquipartitionLengths}
    l_{\vex{k}} &= \frac{c_{\vex{k}}}{k_B T} + \mathcal{O}(\Delta\e/T)^3.
\end{align}
We note that the $\mathcal{O}(1)$ term in the $\Delta\e/T$ expansion of $\mathcal{F}_{\vex{k},\vex{k'}}^{\mu,T}$ in Eq.~(\ref{FExpansion}) rather remarkably vanishes, preventing a  $\mathcal{O}(\Delta\e/T)^2$ term in Eq.~(\ref{EquipartitionLengths}). This implies that the high-$T$ scattering rate (due to phonons) of a given Bloch state should be purely linear, going to zero in the $T \rightarrow 0$ extrapolation.

The equipartition regime is the range of temperature for which Eq.~(\ref{EquipartitionLengths}) holds for all Block states $\vex{k}$. Unlike the case in the BG regime, the linear-in-$T$ power law for the relaxation rate of the EP regime is not affected by anisotropy - all band structure information is encoded in the ``length constants" $c_{\vex{k}}$.

\subsection{Resistivity computation}
\label{Subsection-NumericalComputation}

Equations (\ref{ResistivityDefinition}-\ref{OccupationNumber}) combined with knowledge of the Bloch states give all the tools necessary to make a resistivity prediction. We solve Eqs.~(\ref{RelaxationLengthSelfConsistency}) for scattering lengths for each Bloch state [see Appendix~\ref{Appendix-NumericalImplementationOfResistivityCalculation} for discussion.] We emphasize that in general, the relaxation lengths $\{l_{\vex{k}}\}$ implicitly depend on temperature and chemical potential through the Fermi functions and phonon occupation number ($N_{\vex{q}}$) in Eq.~(\ref{RelaxationLengthSelfConsistency}). Once the $\{l_{\vex{k}}\}$ are known for a given pair $(n,T)$, the resistivity can be computed through Eq.~(\ref{ResistivityDefinition}). We plot the relaxation lengths for a Dirac cone band structure in Fig.~\ref{KineticTheory-DiracScatteringLengths}, keeping $\mu$ fixed as we vary $T$. These results illustrate that states near the Fermi surface become long-lived at low $T$.

In Secs.~\ref{Subsection-BasicTheory}-\ref{Subsection-BGAndEPRegimes}, we have suppressed the band index in summations over Bloch states. The Bernal bilayer and rhombohedral trilayer $\vex{k}\cdot\vex{p}$ Hamiltonians have four and six bands, respectively, while the Dirac cone model has two. In each case, we have two ``low energy" bands near charge neutrality: a ``valence" (hole) band and a ``conduction" (particle) band. The higher energy bands, when present, are over $3.5eV$ from charge neutrality. We note that the Fermi distributions in Eq.~(\ref{ResistivityDefinition}) suppress excitations in these higher energy bands for the temperatures and dopings we are interested in. However, it is important to keep both the conduction and valence bands as charge carriers may be excited in both bands, especially in the gapless systems. In all the models we study, interband transitions between the conduction and valence bands are forbidden by kinematics (i.e. the phonon velocity is too low). Interband transitions into higher energy bands are kinematically allowed, but thermally irrelevant.

It is important to note that as we scan $T$ for fixed $n$, $\mu(n,T)$ can change, and this can be quite drastic near a gap. We must therefore calculate $\mu(n,T)$ self-consistently via 
\begin{align}
    \label{SelfConsistentChemicalPotential}
    n = \frac{4}{L^2}\sum_{\vex{k} \in BZ}f(\e_{\vex{k}}).
\end{align}
The prefactor 4 above follows from the spin and valley degeneracies. We stress that accurately computing the $T$-dependence of $\mu(n,T)$ near the band edge requires keeping both the valence and conduction bands, even if $T$ is far too low to excite carriers across the gap.

\begin{figure}[t]
\includegraphics[angle=0,width=.48\textwidth]{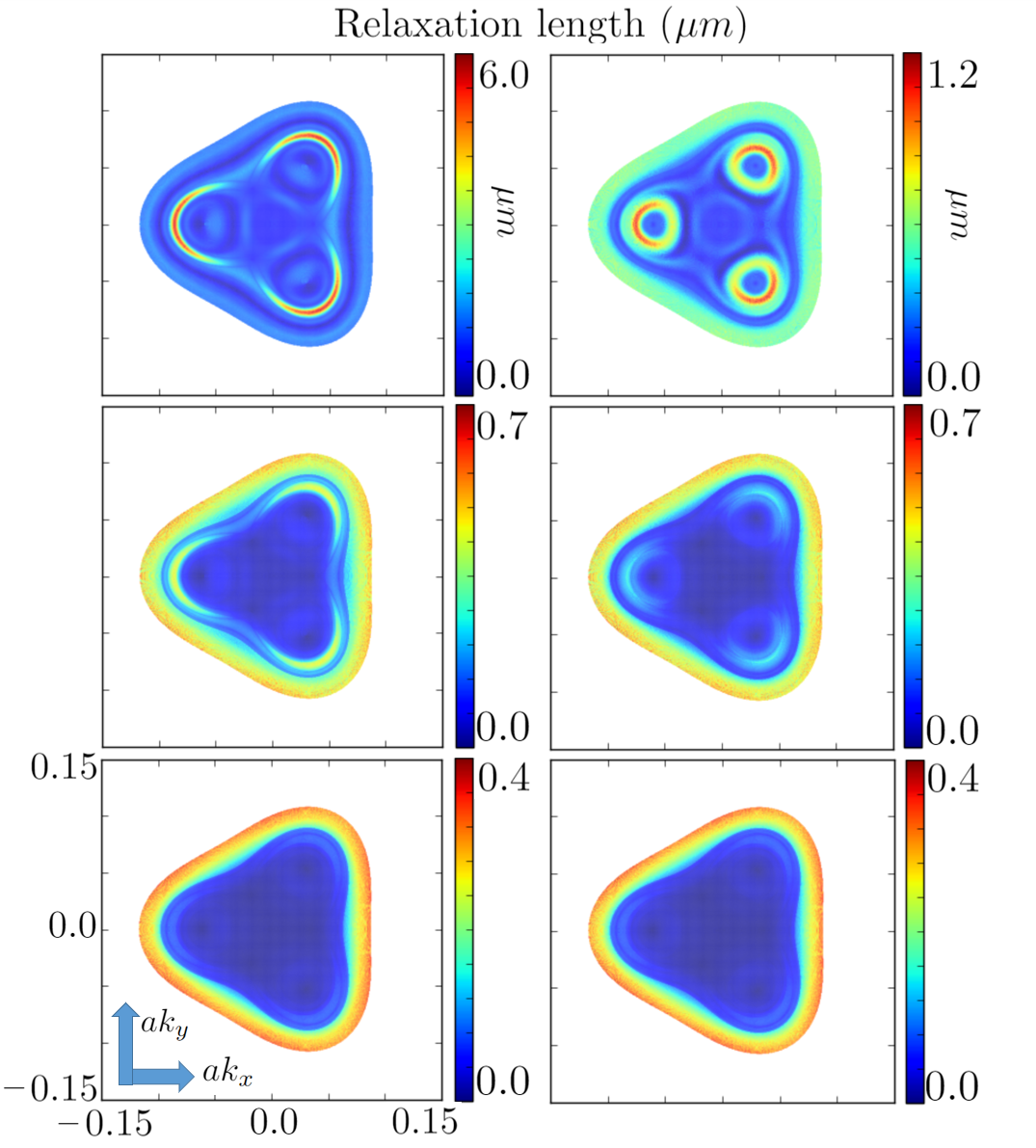}
\caption{
We plot heat maps of the relaxation length in k-space for the Bernal bilayer band structure. We set $\mu = -0.066\ eV$ (left) and $\mu = -0.058\ eV$ (right). For each $\mu$, we do the calculation for $T = 10\ K$ (top), $T = 30\ K$ (middle), and $T = 100\ K$ (bottom). and We see that at low $T$, the specific geometry of the Fermi surface is very important to relaxation, but that this information tends to get washed out at higher $T$. 
}
\label{KineticTheory-BernalScatteringLengths}
\end{figure}

\begin{figure}[t]
\includegraphics[angle=0,width=.48\textwidth]{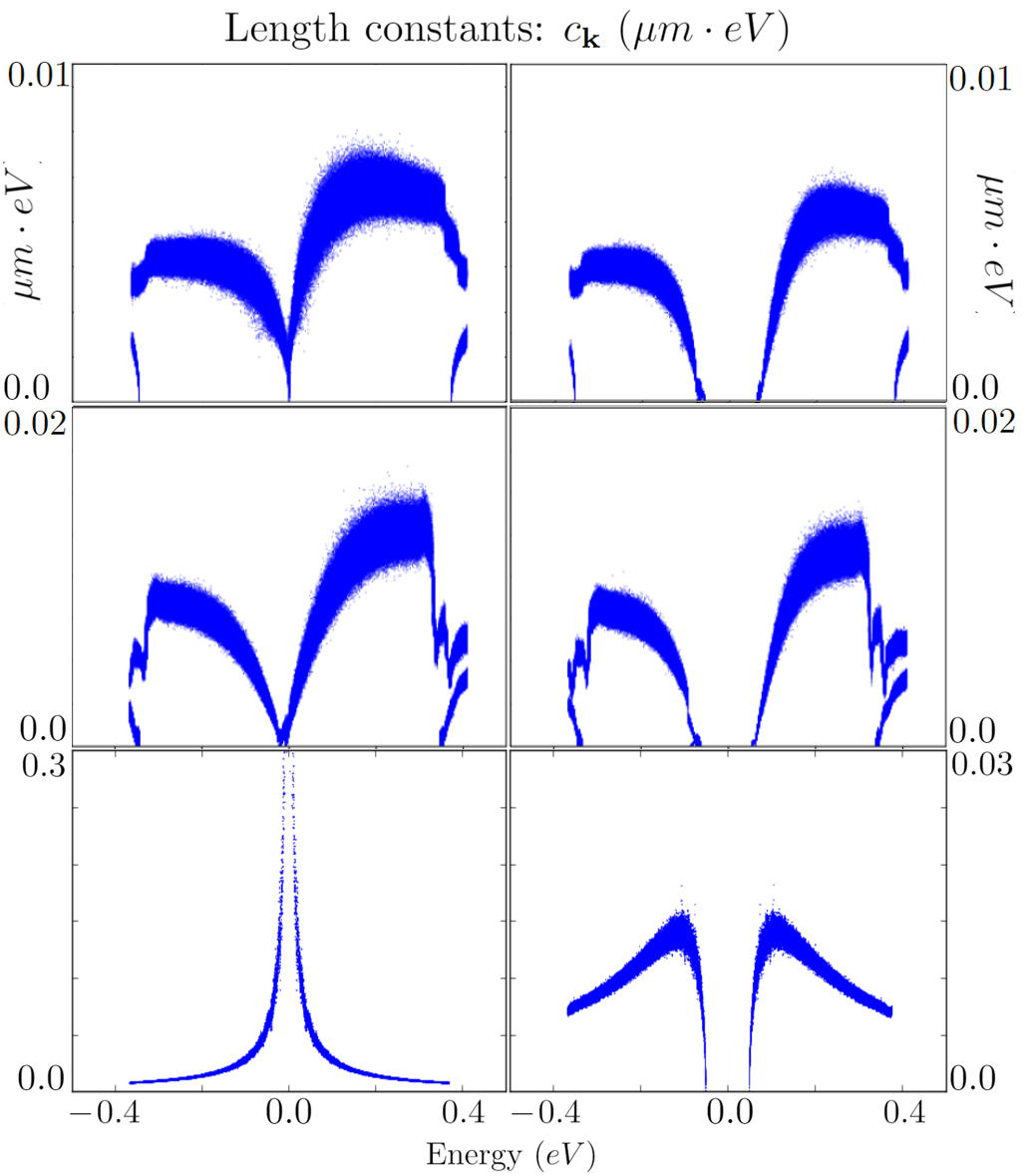}
\caption{
We plot the equipartition ``length constants" over a wide range of energy for the various systems under study. The top row gives Bernal bilayer graphene, the middle row gives rhombohedral trilayer graphene, and the bottom row gives standard Dirac cone graphene. Gapless systems are on the left and gapped systems are on the right. With these values stored we can efficiently compute the resistivity of these systems up to very high temperatures using Eq.~\ref{EquipartitionLengths}, though our results will miss the low-$T$ BG physics, as discussed in Sec.~\ref{Subsection-BGAndEPRegimes}. This is how we generate the high-T results in Figs.~\ref{ResultsSummary-Slices}, \ref{ResultsSummary-LogRhoHighT}. We emphasize that the ungapped Dirac cone graphene, which has diverging $c_{\vex{k}}$ near charge neutrality due to a vanishing scattering manifold, is the outlier here. All other systems we consider have band curvature effects near charge neutrality that more than compensate for the vanishing scattering manifolds and suppress the divergence of $c_{\vex{k}}.$
}
\label{KineticTheory-EPLengthConstantPlots}
\end{figure}

\begin{figure}[t]
\includegraphics[angle=0,width=.46\textwidth]{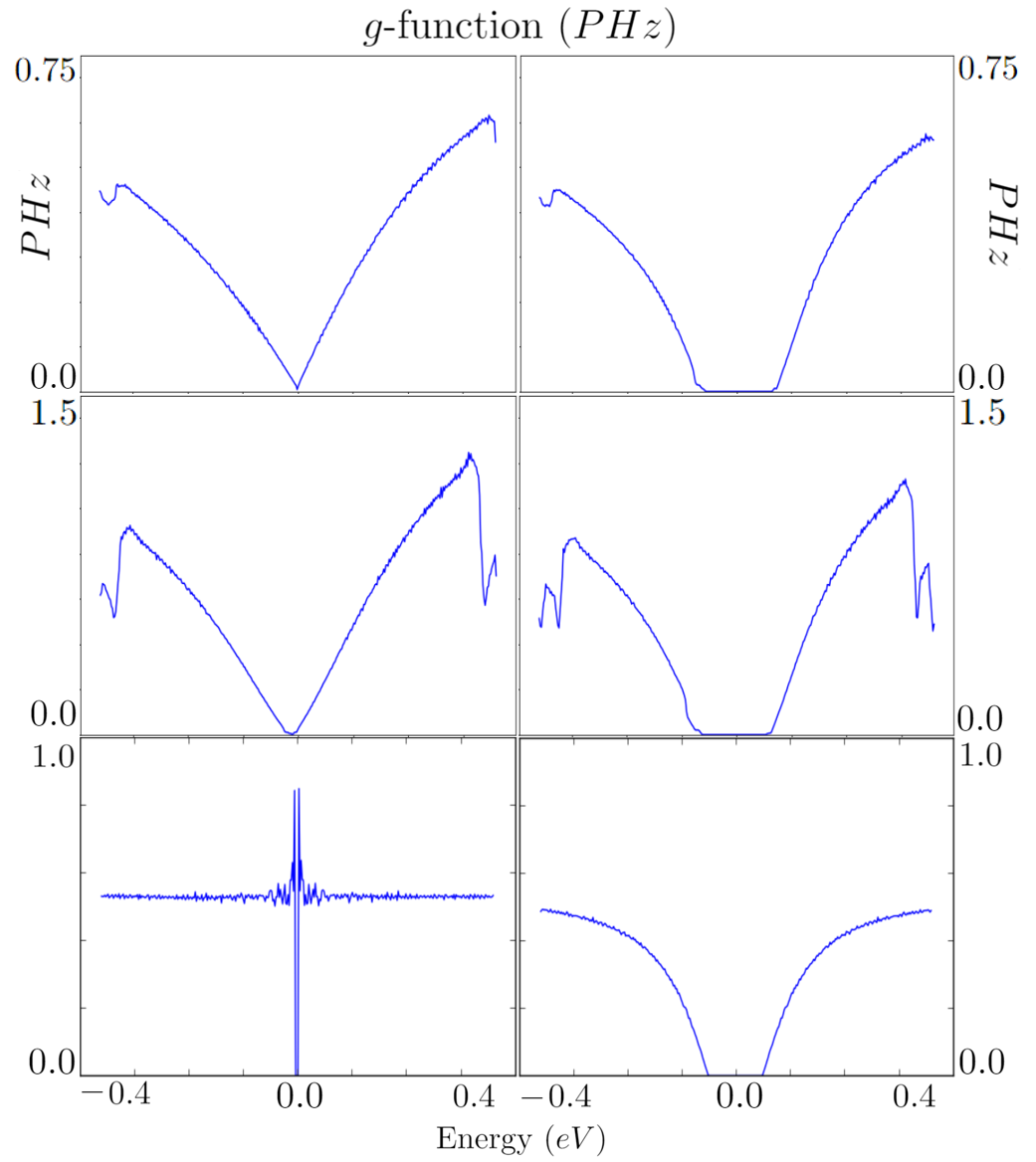}
\caption{
We plot the ``$g$-function", [$g(\e)$], defined in Eqs.~(\ref{GFunctionDefinition}-\ref{ConductivityFromG}) for Bernal bilayer (top), rhombohedral trilayer (center), and Dirac cone graphene (bottom). We plot $g(\e)$ for both ungapped (left) and gapped (right) cases. These graphs demonstrate clearly why single ungapped Dirac cone graphene has such a robust high-$T$ linear resistivity and why all the other systems display nonlinear resistivity effects at high-$T$.
}
\label{KineticTheory-GFunctionPlots}
\end{figure}

\begin{figure}[t]
\includegraphics[angle=0,width=.48\textwidth]{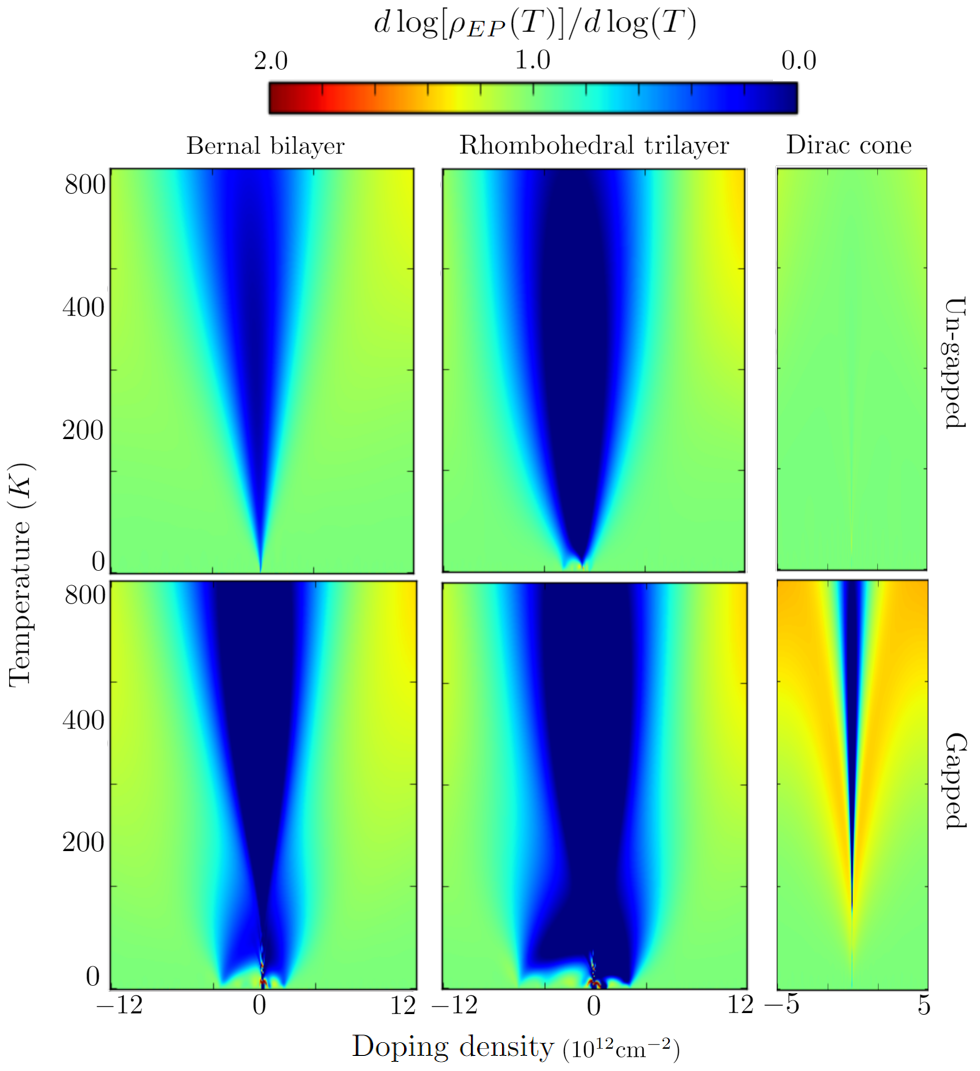}
\caption{
We plot heat maps of $d\log[\rho(n,T)]/d\log T$ for the various systems using the resistivities calculated in the high-$T$ equipartition calculation (the data from Fig.~\ref{ResultsSummary-LogRhoHighT}). This should be compared with Figs.~\ref{IntroductionSuperfigure-GappedBernalAndRhomboPhaseDiagram},\ref{ResultsSummary-GaplessBernalAndRhomboPhaseDiagram}, and \ref{ResultsSummary-GappedAndGaplessDiracConePhaseDiagram}. Since this is based on resistivity data from the equipartition regime, it does not contain any BG physics, and any deviations from linear scaling are due to band curvature effects. We emphasize that these figures show linear-in-$T$ scaling for dopings away from charge neutrality, but then demonstrate a flattening of the $\rho(T)$. We further emphasize that the flattening seen near the van-Hove singularities is present here as well, indicating it is an effect of band geometry in the thermal averaging and not a transition in the nature of the scattering. 
}
\label{KineticTheory-EquipartitionPhaseDiagram}
\end{figure}





The main result of this work is the application of the above analysis to Bernal bilayer and rhombohedral trilayer graphene stacks. These results are presented and discussed in Sec.~\ref{Section-SummaryOfMainResults}. We use $\vex{k}\cdot\vex{p}$ Hamiltonians for these systems \cite{MacDonaldBandstructureBernal,MacDonaldBandstructureRhombo}, which we provide in Appendix~\ref{Appendix-HamiltoniansForStackedGrapheneSystems}. The band structure further gives the density of states and Fermi surface geometries depicted in Figs.~\ref{IntroductionSuperfigure-GappedBernalAndRhomboPhaseDiagram},\ref{ResultsSummary-GaplessBernalAndRhomboPhaseDiagram}. 

The nontrivial band geometry of these systems gives scattering manifolds that depend qualitatively on not only the Fermi level, but also the specific Bloch state in question, as depicted in Fig.~\ref{IntroductionSuperfigure-GappedBernalAndRhomboPhaseDiagram}. Since the bands are not isotropic and the phonon scattering cannot be considered ``quasi-elastic" \cite{SankarGrapheneKT1}, we need to find the full solution of Eq.~(\ref{RelaxationLengthSelfConsistency}). Solving Eq.~(\ref{RelaxationLengthSelfConsistency}) for the $\{l_k\}$ repeatedly for many values of $n$ and $T$, we calculate the resistivity data given in Figs.~\ref{ResultsSummary-LogRhoLowT},\ref{ResultsSummary-Slices}. Data showing how scattering lengths vary throughout the band structure are given in Fig.~\ref{KineticTheory-BernalScatteringLengths}.

The equipartition regime scaling coefficients, $c_{\vex{k}}$, are given for all the models of interest in Fig.~\ref{KineticTheory-EPLengthConstantPlots}. In the case of Dirac cone graphene, we see that there is a divergence of $c_{\vex{k}}$ at the Dirac point, arising from a vanishing set of scattering states. However, in all the other models under consideration, band curvature effects near charge neutrality more than compensate for the vanishing scattering manifolds and suppress $c_{\vex{k}}$.




\section{Nonlinear $T$-dependence of resistivity}
\label{Section-NonlinearHighTResistivity}

The ``common knowledge" of high-$T$ phonon scattering is that the resistivity scales linearly with $T$ above the BG crossover regime \cite{AshcroftAndMermin, Ziman, SankarGrapheneKT1, SankarGrapheneKT5}. While it is true that each individual relaxation length has the high-$T$ scaling of Eq.~(\ref{EquipartitionLengths}), the $T$-dependence of the resistivity itself can be quite nonlinear. Indeed, our calculations for BBG and RTG predict a nonlinear $T$-dependence of the resistivity, especially in the vicinity of the gap. [See Figs.~\ref{ResultsSummary-Slices},\ref{ResultsSummary-LogRhoHighT}.] 

Our calculations predict that the phonon scattering will crossover from the BG regime to the EP regime at an effective BG crossover temperature that varies from as high as $40\ K$ at high doping to as low as $<10\ K$ near charge neutrality. However, in the EP regime, we start to see sharp reductions in slope of the resistivity at temperatures as low as $40\ K$ [See Fig.~\ref{ResultsSummary-Slices}]. For dopings closer to charge neutrality, we see the resistivity peak and drop precipitously at $ T \approx 300\ K$. This behavior has already been observed in twisted bilayer graphene, \cite{Polshyn2019}, at temperatures and resistivity values qualitatively consistent with our results here.

We note that the non-linear $T$-dependence resembles the same sort of resistivity profiles that have been characterized as ``resistivity saturation" \cite{Poniatowski_2021, Sarkar_2018, Hwang_2019, Kivelson_1995} and are sometimes associated with a breakdown of kinetic theory at the Mott-Ioffe-Regel limit \cite{Gurvitch_1981, Millis_1999, Kivelson_1995, Hussey__2004}. However, we stress that our results are fully in the Boltzmann framework. The possibility that the apparent resistivity saturation type effect could arise purely from the electron-phonon coupling effects was pointed out in the literature before \cite{Millis_1999}, but the physics of this apparent saturation in the current work is qualitatively different, arising not from non-Boltzmann strong coupling physics, but from subtle band structure effects as discussed in our paper.

In the rest of this section, we provide some intuition for the non-linear $T$-dependence of the resistivity. As discussed above, the high-$T$ relaxation lengths are given in terms of the $n,T$-independent constants $\{c_{\vex{k}}\}.$ We can gain an understanding of the non-linearity of $\rho(T)$ by considering the function
\begin{align}
\label{GFunctionDefinition}
    \delta^{ij}g(\e)
    &\equiv 
    \frac{1}{L^2}
    \sum_{\vex{k}}
    \frac{
    v^i_{\vex{k}}v^j_{\vex{k}}
    }
    {|v_{\vex{k}}|}
    c_{\vex{k}}
    \delta[\e-\e_{\vex{k}}].
\end{align}
In the EP regime, we can write the high-$T$ conductivity in terms of $g(\e)$:
\begin{align}
\label{ConductivityFromG}
    \sigma = \frac{1}{(k_B T)^2} \int d\e g(\e) f(\e)[1-f(\e)] 
\end{align}
We see that even in the equipartition regime, we only have linear scaling of the resistivity if the integral over $\e$ in Eq.~(\ref{ConductivityFromG}) scales linearly with $T$, which will be true as long as $g(\e)$ has a good linear approximation in a window of width $k_B T$ about $\mu(n,T).$ The functions $g(\e)$ are plotted in Fig.~\ref{KineticTheory-GFunctionPlots}. We see that gapless Dirac cone graphene has a perfectly flat $g(\e),$ (though our figure shows finite-size effects near the Dirac point), giving the familiar, perfectly linear resistivity in the equipartition regime. On the other hand, we see that gapless Dirac cone graphene is the exception - all of the other systems studied exhibit band curvature that manifests nonlinearity in $g(\e).$ The gapless bilayer and trilayer systems exhibit $g(\e)$ that can be roughly approximated as linear over small $\e$-windows when sufficiently doped. However, we expect a qualitative change when $T \approx |\mu(n,T)|,$ and the integral in Eq.~(\ref{ConductivityFromG}) crosses the zero-energy point, where we expect the scaling of the integral in Eq.~(\ref{ConductivityFromG}) to crossover from linear-in-$T$ to quadratic-in-$T$. This would result in a crossover to a roughly $T$-independent resistivity when $T \approx \mu(n,T),$ which is indeed what we see in Fig.~(\ref{ResultsSummary-GaplessBernalAndRhomboPhaseDiagram}). All three gapped systems exhibit more curvature in $g(\e),$ even when far from charge neutrality, but may still be linearly approximated in a small $T$-window. However, sharp qualitative changes in $g(\e)$ occur at a band edge, so we expect sharp qualitative changes in the resistivity scaling when $T \approx |\mu(n,T)|-E_{gap}/2$ and again when $T \approx |\mu(n,T)|+E_{gap}/2$. For a system without a band gap, we would only expect a single kink. Figures~\ref{ResultsSummary-Slices} and \ref{AdditionalData-ParticleDopedResistivitySlices} - \ref{AdditionalData-NoFieldRhomboSlices} demonstrate this intuition. Two distinct kinks are visible in many resistivity curves in Figs.~\ref{ResultsSummary-Slices} and \ref{AdditionalData-ParticleDopedResistivitySlices}, which plot the data for gapped systems, while curves in Figs.~\ref{AdditionalData-NoFieldBernalSlices} and \ref{AdditionalData-NoFieldRhomboSlices} tend to have a single kink. 

We emphasize that the nonlinear $T$-dependence of the resistivity is in general due to the curvature of the bands and not necessarily related to interband excitations \cite{Polshyn2019}. For instance, in the hole-doped systems in Fig.~\ref{ResultsSummary-Slices}, with the potential difference at $\Delta = 0.07\ eV$, the gap is approximately $0.1\ eV$ wide. However, nonlinear $T$-dependence is seen at temperatures as low as $40\ K$, which is far to cold to excite appreciable states in the conduction band.

\section{Discussion and conclusions}
\label{Section-DiscussionAndConclusions}

We have calculated the electrical DC resistivity of Bernal bilayer and rhombohedral trilayer graphene systems, due to scattering off of acoustic phonons. We extend previous study by using a detailed $\vex{k}\cdot\vex{p}$ band structure and focusing our attention on the roles of geometric features of the band structure of these systems, including those affected by a displacement field.

We develop a thoroughly nontrivial transport theory for carrier resistivity due to electron-acoustic phonon interaction in experimentally relevant RTG and BBG multilayer graphene systems. The theory, while using the standard graphene acoustic phonons and the conventional electron-phonon deformation potential coupling, includes the full effects of RTG and BBG band structures (even including an applied electric displacement field) non-perturbatively by employing a full $\vex{k}\cdot\vex{p}$ description. The qualitative importance of the van Hove singularities and the anisotropies in the graphene band structures are exactly incorporated in the theory by iteratively solving the integral Boltzmann transport equation. This leads to several qualitatively new features in the resistivity (e.g. inapplicability of the simple Bloch-Gr\"{u}neisen criteria for linear versus non-linear resistivity in temperature, apparent resistivity saturation behavior at higher temperatures, and other features as discussed in this paper), which have not been discussed in the transport literature of electronic materials before in any context. We provide concrete predictions for the doping and temperature dependence of resistivity in RTG and BBG multilayers, finding that simple considerations for a Bloch-Gr\"{u}neisen temperature separating the linear-in-$T$ high-temperature resistivity from the non-linear low-temperature resistivity does not apply because of the band geometry introducing strong modifications of the resistivity behavior.

Our results are important in two contexts. First, the experiments in BBG and RTG have shown that the exotic superconductivity and the various interaction-driven correlated states are closely related to the nontrivial geometric features of the band structures, including Fermi-surface reshapes and Van Hove singularities. This spotlights the enhanced effects of band geometry on scattering processes in complex 2D systems. As 2D layered heterostructures are currently ascendant in condensed matter physics, it is important to study the relationship between band geometry and transport directly and to modify intuitions gained in three dimensions. Second, it is crucial in the investigation of the origin of the superconductivity in moir\'{e}less layered graphene systems to understand the relative importance of various scattering mechanisms. Our work provides a clear and concrete picture of how the resistivity should behave in a phonon-dominated system. If strong deviations from these results are seen in experiment, that could serve as evidence that the scattering mechanisms other than phonons play dominant roles in transport. This would point to directions for non-phonon pairing in the observed superconductivity. 

The doping and temperature dependence of the resistivity of these systems behave similarly and with many interesting features. We find that the BG crossover in the qualitative $T$-dependence of the scattering rates varies as a function of doping from as low as $5K$ to as high as $60K$. However, we note that this crossover temperature depends strongly on the geometric features of the band structure, and is sharply reduced by the emergence of the annular Fermi surface, which is related to the observed SC. Further, we find that band curvature effects also give rise to a non-linear $T$-dependence of the resistivity at temperatures in the intermediate range of $60K-300K$. While our results show an interesting sensitivity to changes in Fermi surface geometry, they are remarkably smooth at the Van Hove singularity.

Our results are qualitatively compatible with what is currently known in experiment \cite{AndreaYoungBernal, AndreaYoungRhombo, AndreaYoungRhombo2}. We have not yet seen evidence of the high-$T$ nonlinear equipartition resistivity in BBG or RTG, but very similar effects have been observed in twisted bilayer \cite{Polshyn2019} and trilayer \cite{Siriviboon_2021} graphene systems. While the BG crossover to linear scattering has not yet been observed in these systems at low temperatures, our results show that current experiment cannot rule out the possibility that these systems are dominated by phonon scattering. In particular, no linear-in-$T$ region has been observed below $20 K$ in RTG and the zero temperature resistivity varies from $20-70 \Omega$ (c.f. Fig.S6 in \cite{AndreaYoungRhombo}). Our predictions are compatible with these experimental results. However, our results do make it clear that extensive experimental resistivity data over wider ranges of doping and temperature (from $0$ to $300K$) should be sufficient to tell if there are strong deviations from the phonon-dominated picture. Comparison of our results with future additional experimental resistivity data could be a crucial step in discovering the origin of SC in these systems. Further, the low Fermi velocities and high density of states at the Van Hove singularity should enhance the effects of electron-electron interactions. Since our calculations do not predict sharp features to emerge at the Van Hove singularities in a purely phonon picture, observations of such features in the resistivity could serve as evidence for strong-coupling physics that could underlie the systems' superconductivity.




\acknowledgments

We thank Christopher David White, Jiabin Yu, Jay D. Sau, and Matthew S. Foster for helpful discussion. This work is supported by the Laboratory for Physical Sciences (S.M.D, Y.-Z.C., and S.D.S). It is also partially funded by JQI-NSF-PFC (Y.-Z.C.). F.W. is supported by startup funding of Wuhan University.

\appendix




\begin{widetext}

\section{Hamiltonians for stacked graphene systems}
\label{Appendix-HamiltoniansForStackedGrapheneSystems}

To calculate the band structure for the Bernal bilayer graphene stack, we use in Eq.~(\ref{SingleParticleHamiltonian}) the Hamiltonian introduced in \cite{MacDonaldBandstructureBernal}, and used also in \cite{AndreaYoungBernal, YZBernal, YZLongPaper}:

\begin{align}
    \label{BBLGHamiltonian}
    H^e = 
    \begin{bmatrix}
    -\Delta & v_0 \bar{k} & -v_4 \bar{k} & -v_3 k \\
    v_0 k & \Delta_2 - \Delta & t_1 & -v_4 \bar{k} & \\
    -v_4 k & t_1 & \Delta_2 + \Delta & v_0 \bar{k}\\
    -v_3 \bar{k} & -v_4 k & v_0 k & \Delta
    \end{bmatrix},
\end{align}
where we use the dimensionless, valley-dependent, (anti)holomorphic momenta $k \equiv a_0(\xi k_x + ik_y)$ and $\bar{k} = a_0(\xi k_x - ik_y)$ for valley $\xi \in \{\pm\}$, where $a_0$ is the lattice constant for graphene ($a_0 = 0.246\ nm$). The parameters take the following values (all quantities in $eV$): $\Delta_2 = 0.015, t_1 = 0.361, v_0 = 2.261, v_3 = 0.245, v_4 = 0.12$. The interlayer potential is $\Delta$, and in our calculations this is either set to $0.07\ eV$ or $0\ eV$. The basis for this matrix is $\{1A, 1B, 2A, 2B\},$ where $A,B$ correspond to sublattice and $1,2$ correspond to layer.

For the rhombohedral trilayer stack, we use the Hamiltonian introduced in \cite{MacDonaldBandstructureRhombo}, and used also in \cite{AndreaYoungRhombo, AndreaYoungRhombo2, YZRhombo, YZLongPaper}:
\begin{align}
    \label{RTLGHamiltonian}
    H^e = 
    \begin{bmatrix}
    \Delta_2 + \Delta + \delta & \gamma_2/2 &  v_0 \bar{k} & v_4 \bar{k} & v_3 k & 0 \\
    \gamma_2/2 &  \Delta_2 - \Delta + \delta & 0 &  v_3 \bar{k}& v_4 k & v_0 k \\
    v_0 k & 0 & \Delta_2 + \Delta & \gamma_1 & v_4 \bar{k} & 0 \\
    v_4 k & v_3 k & 
    \gamma_1 & -2\Delta_2 & v_0 \bar{k} & v_4 \bar{k}\\
    v_3 \bar{k} & v_4 \bar{k} &  v_4 k & v_0 k & -2\Delta_2 & \gamma_1\\
    0 & v_0 \bar{k} & 0 & v_4k & \gamma_1 & \Delta_2 - \Delta
    \end{bmatrix},
\end{align}

where we use the same notation ($k,\bar{k}$) as in Eq.~(\ref{BBLGHamiltonian}) and the following parameters (all quantities in $eV$): $\Delta_2 = -0.0023, \delta = -0.0105, v_j = \gamma_j\sqrt{3}/2, \gamma_0 = 3.1, \gamma_1 = 0.38, \gamma_3 = -0.29, \gamma_4 = -0.141$. Again, the interlayer potential is $\Delta$, and in our calculations this is either set to $0.07\ eV$ or $0\ eV$. The basis for the RTG Hamiltonian is $\{1A,3B,1B,2A,2B,3A\}.$
\end{widetext}




\section{Numerical implementation of resistivity calculation}
\label{Appendix-NumericalImplementationOfResistivityCalculation}

In our numerical calculations for $\{l_{\vex{k}}\}$, we usually retain approximately $10^6$ Bloch states, and must solve a rather large linear system [Eq.~(\ref{RelaxationLengthSelfConsistency})] for each pair of values $(n,T)$. In our main results [Figs.\ref{ResultsSummary-Slices},\ref{ResultsSummary-LogRhoLowT}], we do this on a 50-by-60 grid in $n-T$-space. This is necessary to understand the low-$T$ physics, but the EP regime can be studied much more efficiently since the $\{c_{\vex{k}}\}$ defined in Eq.~(\ref{EquipartitionLengths}) are independent of both $n,T$. Once we solve directly for the $c_{\vex{k}}$, calculating the EP approximation to the resistivity is as simple as computing $\mu(n,T)$ via Eq.~(\ref{SelfConsistentChemicalPotential}) and then using Eq.~(\ref{EquipartitionLengths}) in Eq.~(\ref{ResistivityDefinition}). This is how we compute the EP resistivity in Figs.\ref{ResultsSummary-Slices} and \ref{ResultsSummary-LogRhoHighT}.

We discuss the numerical solution of Eq.~(\ref{RelaxationLengthSelfConsistency}) in the main text. In order to discuss the existence and uniqueness of solutions to Eq.~(\ref{RelaxationLengthSelfConsistency}), as well as the convergence of iterative methods, we will re-cast this in the traditional notation of a linear operator problem. Letting $\vex{k}$ in the Brillouin zone act as a vector index, we define the vector $\hat{b}$ and the matrices $\hat{A},\hat{D}$, indexed by $\vex{k} \in BZ$.
\begin{align}
    \hat{A}_{\vex{k},\vex{k'}}
    &= 
    \mathcal{W}_{\vex{k}\rightarrow\vex{k'}}
    [1-f^0(\e_{\vex{k'}})]
    \cos\theta_{\vex{v}}
    \\
    \hat{D}_{\vex{k},\vex{k''}} 
    &= 
    \delta_{\vex{k},\vex{k''}}
    \sum_{\vex{k'}}
    \mathcal{W}_{\vex{k}\rightarrow\vex{k'}}
    [1-f^0(\e_{\vex{k'}})]
    \\
    \hat{b}_{\vex{k}}
    &=
    |\vex{v}_{\vex{k}}|L^2 [1-f^0(\e_{\vex{k}})]
\end{align}

With this notation, Eq.~(\ref{RelaxationLengthSelfConsistency}) takes the form 
\begin{align}
\label{MatrixScattering}
    (\hat{D}-\hat{A})\hat{l} = \hat{b}.
\end{align}
The solution for the relaxation lengths is then a matrix inversion problem. A unique solution exists if $\det[\hat{D}-\hat{A}] \neq 0$, which is always true in this case due to the diagonal dominance of $\hat{D}-\hat{A}$. Since our problem is large and we compute the matrix elements only as needed in the computation, Eq.~(\ref{MatrixScattering}) is most effectively solved via an iterative method. We set
\begin{align}
    \hat{l}^{i+1} \leftarrow \hat{D}^{-1}(\hat{A}\hat{l}^{i}+\hat{b})
\end{align}
repeatedly until convergence. This is simply a case of Gauss-Seidel iteration, which is guaranteed to converge to the unique solution. (This guarantee is again provided by diagonal dominance.)

Explicitly, in the $(i+1)^{\text{th}}$ iteration ($i \geq 0$), we define $\{l^{(i+1)}_{\vex{k}}\}$ in terms of $\{l^{(i)}_{\vex{k}}\}$ via
\begin{align}
\label{ExplicitIteration}
    l^{(i+1)}_{\vex{k}} = 
    \frac{
    |\vex{v}_{\vex{k}}| L^2 -
    \hbar v_p
    \sum_{\vex{k'}}
    |\vex{q}|
    \mathcal{C}_{\vex{k},\vex{k'}}
    \mathcal{F}_{\vex{k},\vex{k'}}^{\mu,T} 
    \cos\theta_{\vex{v}}
    l^{(i)}_{\vex{k'}}
    }
    {
    \hbar v_p
    \sum_{\vex{k'}}
    |\vex{q}|
    \mathcal{C}_{\vex{k},\vex{k'}}
    \mathcal{F}_{\vex{k},\vex{k'}}^{\mu,T} 
    }.
\end{align}
In order to optimize for quick convergence, we initialize the procedure using the explicit formula for an isotropic system with quasi-elastic scattering:
\begin{align}
\label{Initialization}
    l^{(0)}_{\vex{k}}
    =
    \left[
    \frac{\hbar v_p}{|\vex{v}_{\vex{k}}| L^2}
    \sum_{\vex{k'}}
    |\vex{q}|
    \mathcal{C}_{\vex{k},\vex{k'}}
    \mathcal{F}_{\vex{k},\vex{k'}}^{\mu,T} 
    \left(1-\cos\theta_{\vex{v}}\right)
    \right]^{-1}.
\end{align}

In practice, we find very quick convergence and only use two Gauss-Seidel iterations. We emphasize that our iterative algorithm is a numerical approach to solving the full BTE, as given in Eqs.~(\ref{RelaxationLengthSelfConsistency},\ref{RelaxationLengthSelfConsistency2}), which is different from yet equivalent to another commonly-employed technique of ``iterating the collision integral".

Additionally, to numerically solve Eq.~(\ref{RelaxationLengthSelfConsistency2}) on a discrete momentum grid, we must broaden the delta functions defining the scattering manifold [see Eq.~(\ref{EnergyConservation})]. In practice, we do this by broadening the delta function to a finite-width step function of a certain small ``tolerance". We then check that our results are independent of the tolerance variable. We note that our results are very insensitive to reasonable variation of the tolerance. We also emphasize that this procedure reproduces the known analytical results for a single Dirac cone with great accuracy.




\section{Relaxation time approximation in non-isotropic systems}
\label{Appendix-RelaxationTimeApproximationInNonIsotropicSystems}

In the case of elastic scattering and an isotropic band structure, it is well-known that the solution to the relaxation time approximation to the Boltzmann equation is also a solution to the full (linearized) Boltzmann equation \cite{AshcroftAndMermin}. In our case, we assume neither isotropy nor (quasi-)elasticity, which are both present in earlier treatments \cite{SankarGrapheneKT1, SankarGrapheneKT2, SankarGrapheneKT3, SankarGrapheneKT4, SankarGrapheneKT5}. In this appendix, we discuss the extent to which the relaxation time approach holds for our systems. 

The canonical ``relaxation time approximation" to the Boltzmann equation is the replacement of the collision integral for the scattering out of state $\vex{k}$ with the expression
\begin{align}
\label{RelaxationTimeApproximation}
\mathcal{I}_c[F_{\vex{k}}] 
\rightarrow 
\mathcal{I}^{\text{RT}}_c[F_{\vex{k}}]
\equiv 
\frac{-1}{\tau_{\vex{k}}}
\left[f(\e_{\vex{k}})-F_{\vex{k}}\right],
\end{align}
where $F_{\vex{k}}$ is the full non-equilibrium distribution function on the set of Bloch states and $f(\e_{\vex{k}})$ is the Fermi distribution function. This introduces the relaxation times as timescales for the occupation of state $\vex{k}$ to reach equilibrium.

In the absence of temperature gradients or external magnetic fields, the non-equilibrium distribution function may be written to linear order in $E$ in terms of the relaxation times as
\begin{align}
\label{FullDistributionFunction}
    F_{\vex{k}} \approx f(\e_{\vex{k}}) + \frac{1}{T}f(\e_{\vex{k}})[1-f(\e_{\vex{k}})]
    (e\vex{E}\cdot\vex{v}_{\vex{k}})
    \tau_{\vex{k}} \equiv F^1_{\vex{k}}
\end{align}
The distribution function in Eq.~(\ref{FullDistributionFunction}) is a solution of the Boltzmann equation under the approximation Eq.~(\ref{RelaxationTimeApproximation}) and calculating the current from the distribution function in Eq.~(\ref{FullDistributionFunction}) gives Eq.~(\ref{ResistivityDefinition}) in the main text. 

The relaxation time approximation is generally uncontrolled, and the true collision integral in the Boltzmann equation is 
\begin{align}
\label{FullCollisionIntegral}
    \mathcal{I}_c[F_{\vex{k}}] 
    &=
    -\sum_{\vex{k'}}
    \mathcal{W}_{\vex{k}\rightarrow\vex{k'}}
    F_{\vex{k}}[1-F_{\vex{k'}}]
    -
    \mathcal{W}_{\vex{k'}\rightarrow\vex{k}}
    F_{\vex{k'}}[1-F_{\vex{k}}].
\end{align}
However, if there exist $\{\tau_{\vex{k}}\}$ such that for $F^1_{\vex{k}}$ given by Eq.~(\ref{FullDistributionFunction}), we have $\mathcal{I}_c[F^1_{\vex{k}}] = \mathcal{I}^{RT}_c[F^1_{\vex{k}}]$ to first order in $\vex{E}$, then Eq.~(\ref{FullDistributionFunction}) is in fact a solution to the full linearized Boltzmann equation.

Evaluating Eq.~(\ref{FullCollisionIntegral}) on the distribution function $F^1_{\vex{k}}$ and using the principle of detailed balance, one may see that 
\begin{align}
\label{}
    \mathcal{I}_c[F^1_{\vex{k}}] 
    &=
    \frac{-e\vex{E}}{T}\cdot
    \sum_{\vex{k'}}
    \mathcal{W}_{\vex{k}\rightarrow\vex{k'}}
    \left[
    \begin{aligned}
    f(\e_{\vex{k}})[1-f(\e_{\vex{k'}})]
    \\
    \times
    \big(
    \tau_{\vex{k}}\vex{v}_{\vex{k}} - \tau_{\vex{k'}}\vex{v}_{\vex{k'}}
    \big)
    \end{aligned}
    \right].
\end{align}
Comparing with 
\begin{align}
\label{}
    \mathcal{I}^{RT}[F^1_{\vex{k}}] 
    &=
    -e\vex{E}\cdot\vex{v}_{\vex{k}}
    \frac{1}{T}f(\e_{\vex{k}})[1-f(\e_{\vex{k}})],
\end{align}
we find that Eq.~(\ref{RelaxationLengthSelfConsistency}) is necessary and sufficient for Eq.~(\ref{FullDistributionFunction}) to be a solution to the linearized Boltzmann equation. 



\section{Additional data}
\label{Appendix-AdditonalData}

In this final appendix, we compile additional data for the temperature and doping dependencies of the resistivity for BBG and RTG. We provide the zero displacement field ($\Delta = 0$) counterparts to Fig.~\ref{ResultsSummary-Slices}, as well as particle-doped data complementing Fig.~\ref{ResultsSummary-Slices}.

\begin{figure*}[t!]
\includegraphics*[angle=0,width=.9\textwidth]{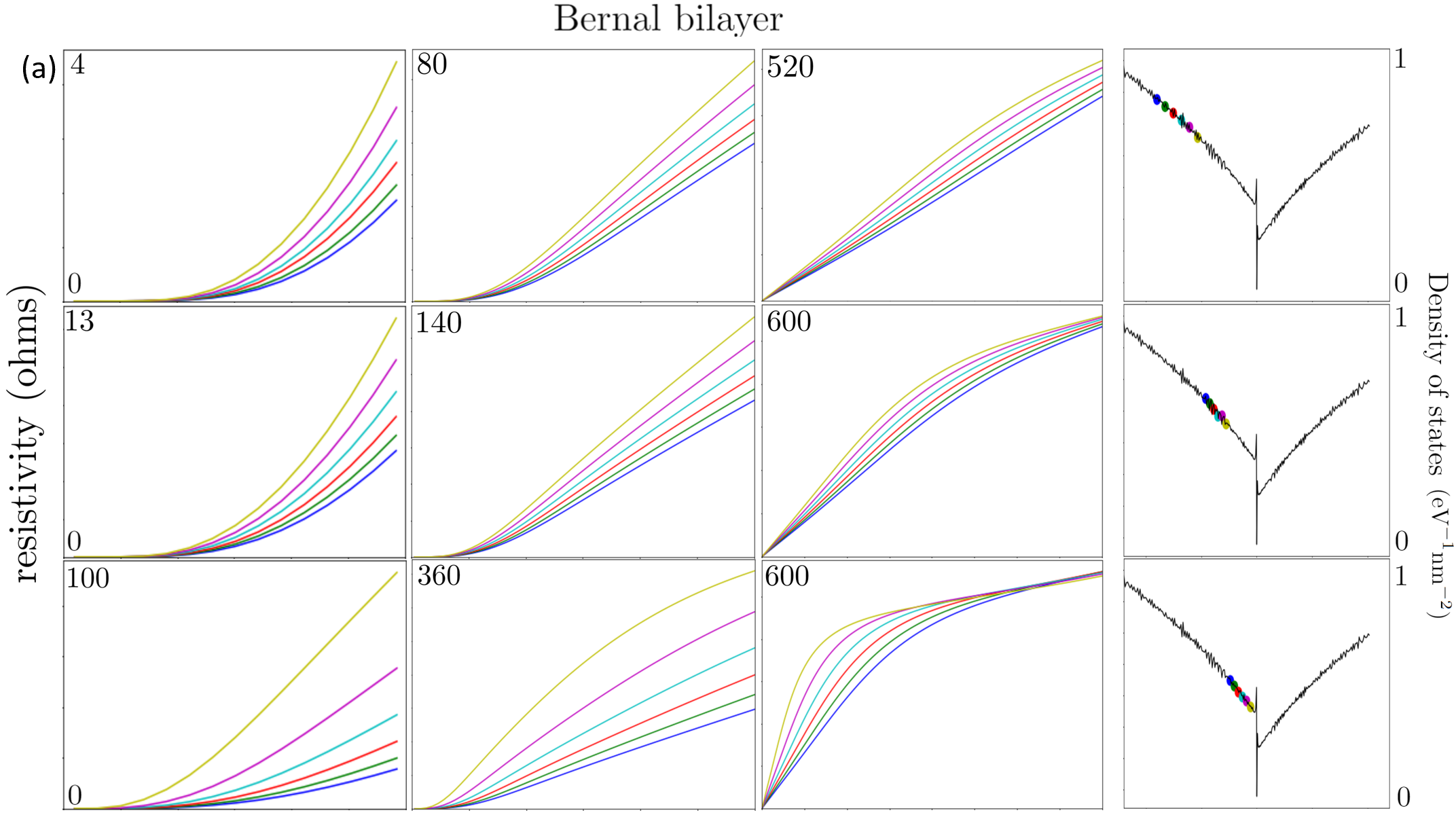}
\includegraphics*[angle=0,width=.9\textwidth]{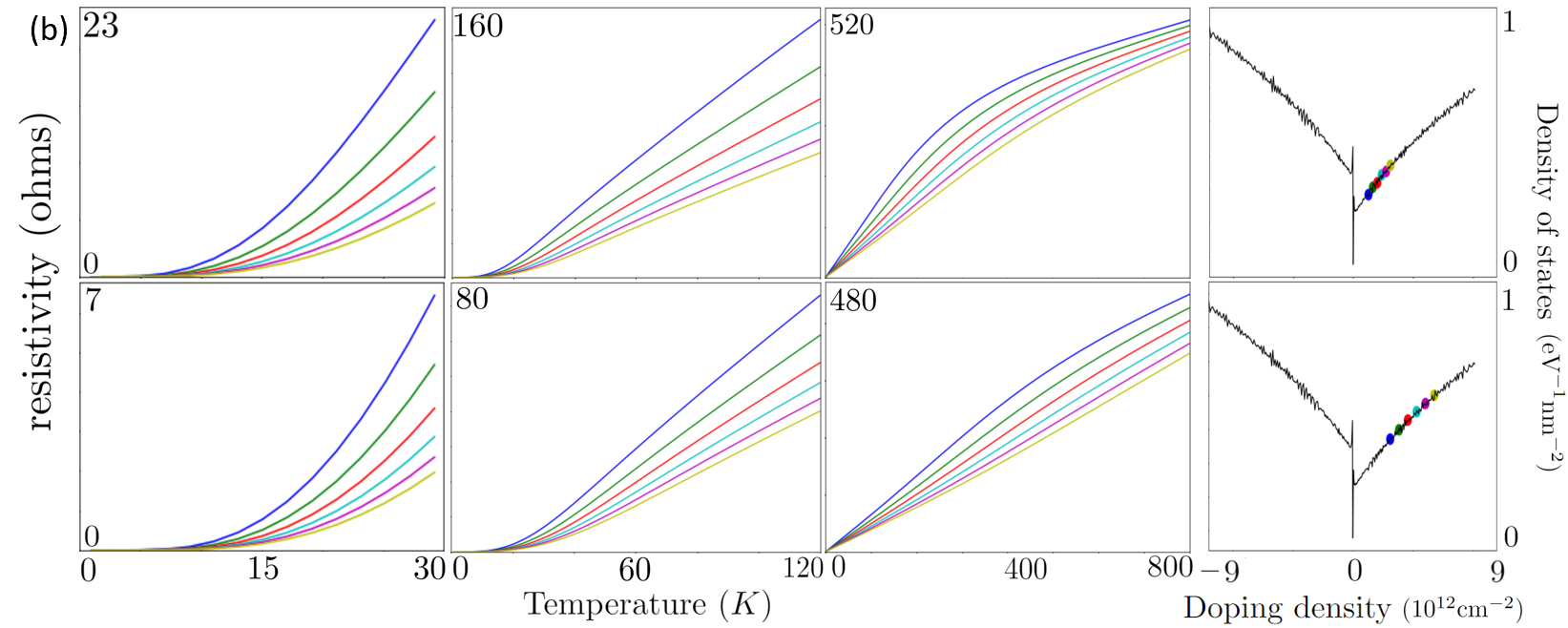}
\caption{
We plot resistivity data over temperature for hole-doped and electron-doped Bernal bilayer systems in the absence of a displacement field ($\Delta = 0$). This should be compared with the Bernal bilayer data in Figs.~\ref{ResultsSummary-Slices} and \ref{AdditionalData-ParticleDopedResistivitySlices}. As with Fig.~\ref{ResultsSummary-Slices}, the leftmost two columns give the results of our full numerical calculation for the resistivity of the two systems at various doping levels up to $30K$ and $120K$, respectively. The third column gives the high-$T$ results in the EP regime. The far-right column denotes the doping values corresponding to the resistivity curves.
} 
\label{AdditionalData-NoFieldBernalSlices}
\end{figure*}


\begin{figure*}[t!]
\includegraphics*[angle=0,width=.9\textwidth]{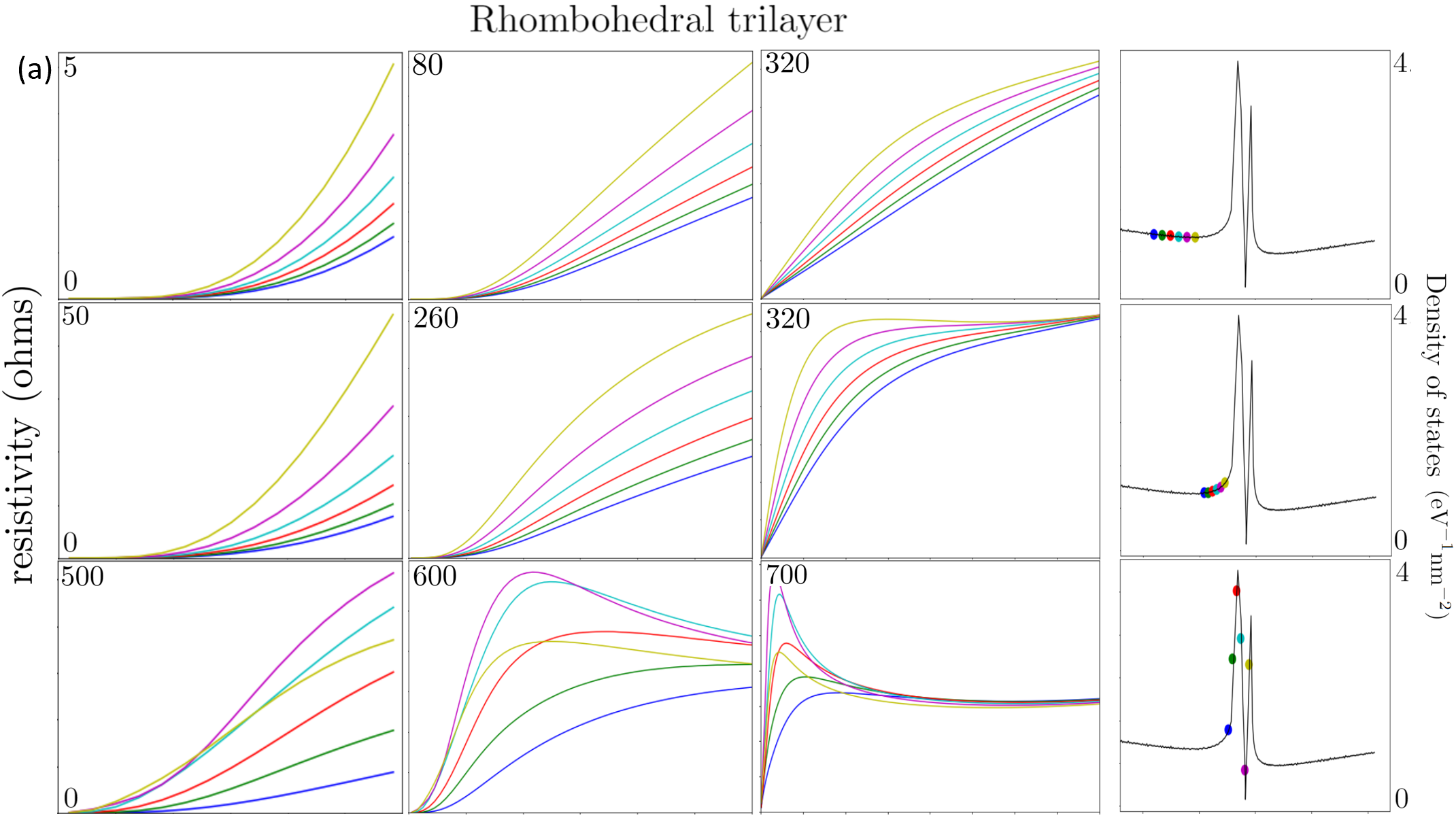}
\includegraphics*[angle=0,width=.91\textwidth]{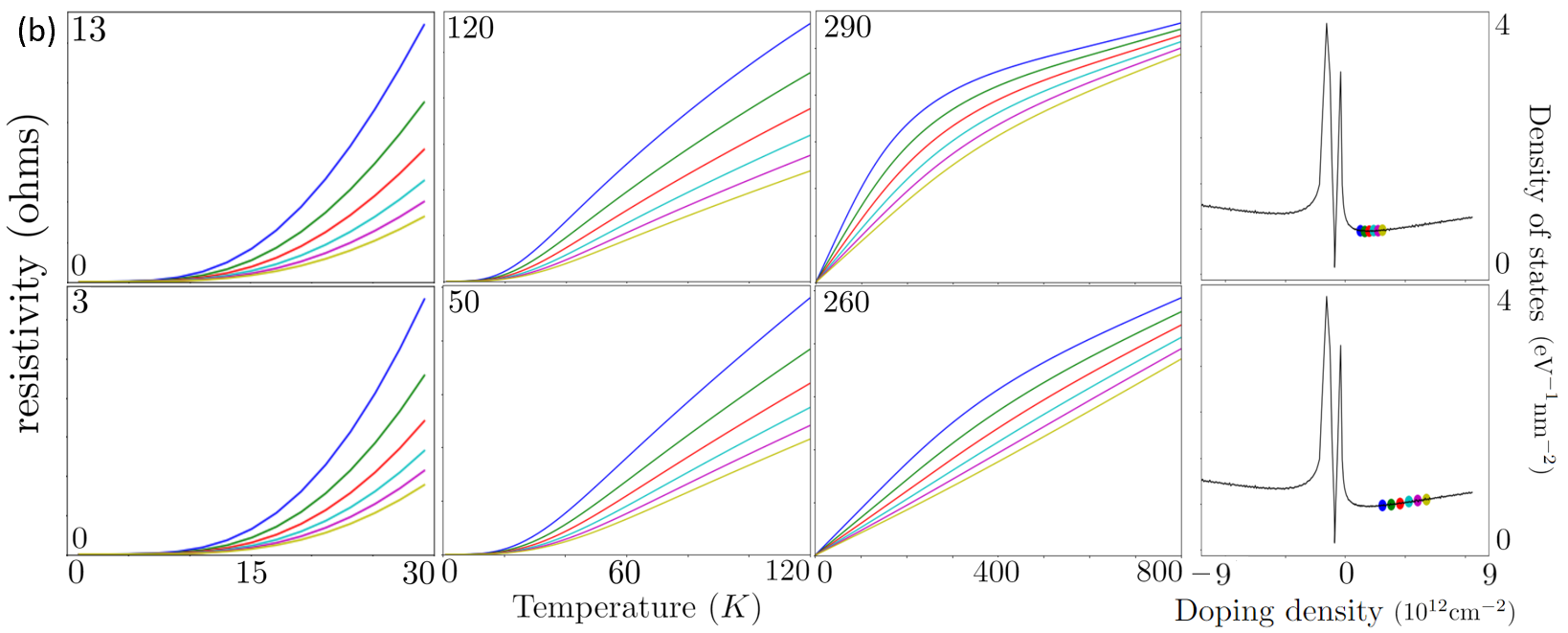}
\caption{
We plot resistivity data over temperature for hole-doped and electron-doped rhombohedral trilayer systems in the absence of a displacement field ($\Delta = 0$). This should be compared with the rhombohedral trilayer data in Figs.~\ref{ResultsSummary-Slices} and \ref{AdditionalData-ParticleDopedResistivitySlices}. As with Fig.~\ref{ResultsSummary-Slices}, the leftmost two columns give the results of our full numerical calculation for the resistivity of the two systems at various doping levels up to $30K$ and $120K$, respectively. The third column gives the high-$T$ results in the EP regime. The far-right column denotes the doping values corresponding to the resistivity curves.
} 
\label{AdditionalData-NoFieldRhomboSlices}
\end{figure*}

\begin{figure*}[t!]
\includegraphics*[angle=0,width=.9\textwidth]{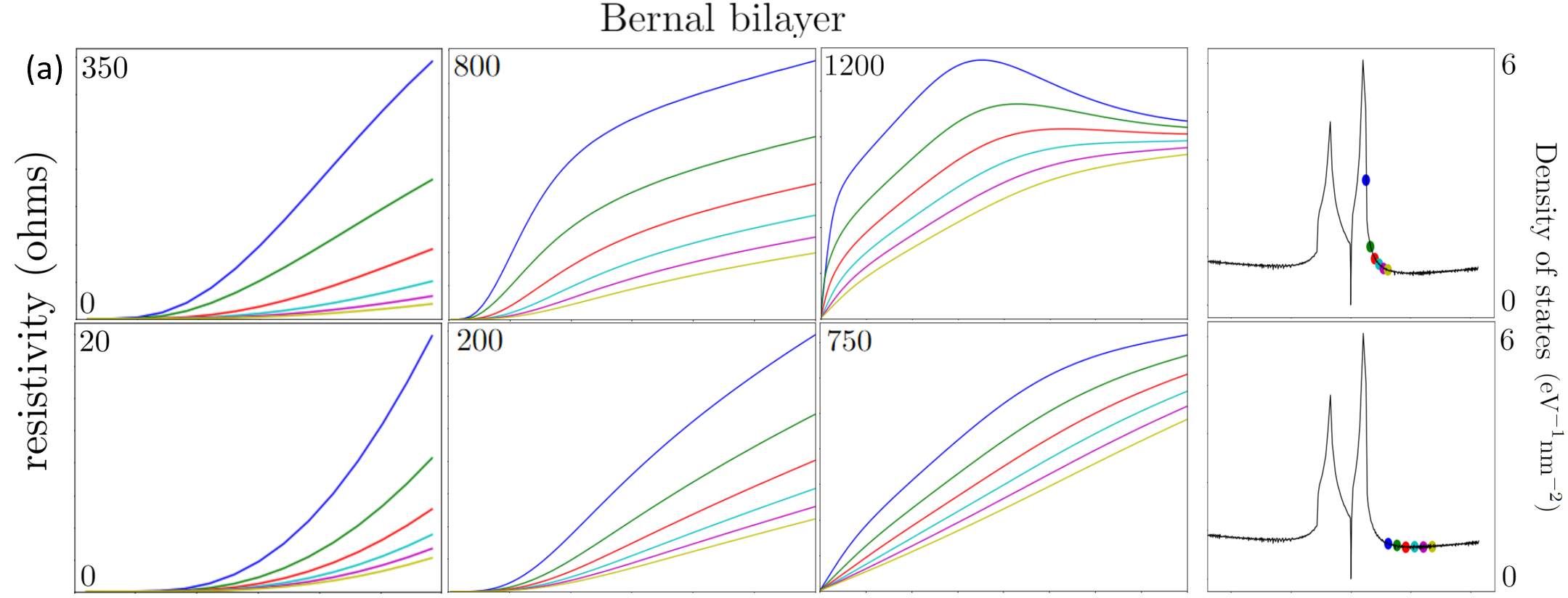}
\includegraphics*[angle=0,width=.9\textwidth]{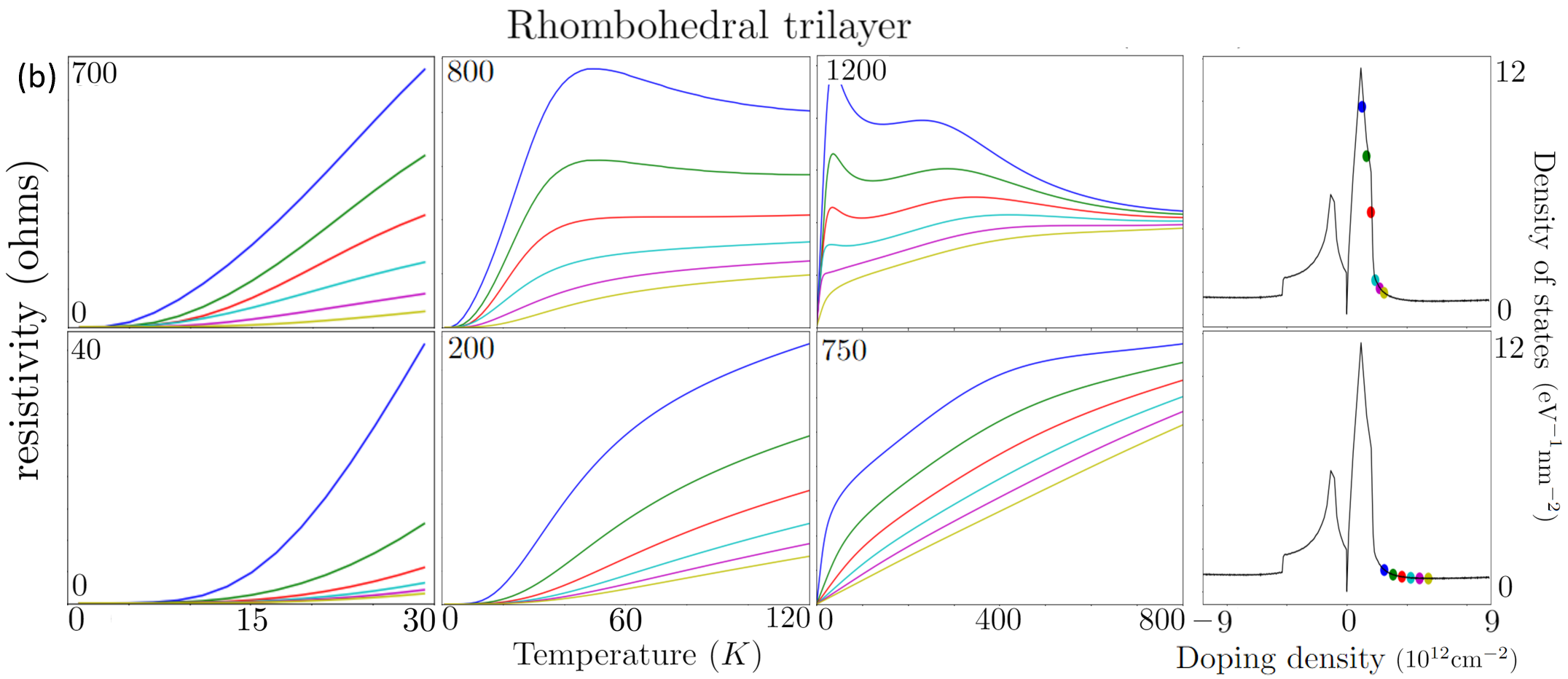}
\caption{
We plot resistivity data over temperature for electron-doped samples, complementing the hole-doped data presented in Fig.~\ref{ResultsSummary-Slices}. We emphasize that the results are analogous to the hole-doped side. As with Fig.~\ref{ResultsSummary-Slices}, the leftmost two columns give the results of our full numerical calculation for the resistivity of the two systems at various doping levels up to $30K$ and $120K$, respectively. The third column gives the high-T results in the EP regime. The far-right column denotes the doping values corresponding to the resistivity curves.
} 
\label{AdditionalData-ParticleDopedResistivitySlices}
\end{figure*}

\FloatBarrier


\bibliography{BIB}

\begin{thebibliography}{77}
\expandafter\ifx\csname natexlab\endcsname\relax\def\natexlab#1{#1}\fi
\expandafter\ifx\csname bibnamefont\endcsname\relax
  \def\bibnamefont#1{#1}\fi
\expandafter\ifx\csname bibfnamefont\endcsname\relax
  \def\bibfnamefont#1{#1}\fi
\expandafter\ifx\csname citenamefont\endcsname\relax
  \def\citenamefont#1{#1}\fi
\expandafter\ifx\csname url\endcsname\relax
  \def\url#1{\texttt{#1}}\fi
\expandafter\ifx\csname urlprefix\endcsname\relax\def\urlprefix{URL }\fi
\providecommand{\bibinfo}[2]{#2}
\providecommand{\eprint}[2][]{\url{#2}}

\bibitem[{\citenamefont{Geim and Grigorieva}(2013)}]{Geim_2013}
\bibinfo{author}{\bibfnamefont{A.~K.} \bibnamefont{Geim}} \bibnamefont{and}
  \bibinfo{author}{\bibfnamefont{I.~V.} \bibnamefont{Grigorieva}},
  \bibinfo{journal}{Nature} \textbf{\bibinfo{volume}{499}},
  \bibinfo{pages}{419} (\bibinfo{year}{2013}),
  \urlprefix\url{https://doi.org/10.1038%2Fnature12385}.

\bibitem[{\citenamefont{Novoselov et~al.}(2006)\citenamefont{Novoselov, McCann,
  Morozov, Fal'ko, Katsnelson, Zeitler, Jiang, Schedin, and
  Geim}}]{Novoselov_2006}
\bibinfo{author}{\bibfnamefont{K.~S.} \bibnamefont{Novoselov}},
  \bibinfo{author}{\bibfnamefont{E.}~\bibnamefont{McCann}},
  \bibinfo{author}{\bibfnamefont{S.~V.} \bibnamefont{Morozov}},
  \bibinfo{author}{\bibfnamefont{V.~I.} \bibnamefont{Fal'ko}},
  \bibinfo{author}{\bibfnamefont{M.~I.} \bibnamefont{Katsnelson}},
  \bibinfo{author}{\bibfnamefont{U.}~\bibnamefont{Zeitler}},
  \bibinfo{author}{\bibfnamefont{D.}~\bibnamefont{Jiang}},
  \bibinfo{author}{\bibfnamefont{F.}~\bibnamefont{Schedin}}, \bibnamefont{and}
  \bibinfo{author}{\bibfnamefont{A.~K.} \bibnamefont{Geim}},
  \bibinfo{journal}{Nature Physics} \textbf{\bibinfo{volume}{2}},
  \bibinfo{pages}{177} (\bibinfo{year}{2006}),
  \urlprefix\url{https://doi.org/10.1038%2Fnphys245}.

\bibitem[{\citenamefont{Bistritzer and MacDonald}(2011)}]{Bistritzer_2011}
\bibinfo{author}{\bibfnamefont{R.}~\bibnamefont{Bistritzer}} \bibnamefont{and}
  \bibinfo{author}{\bibfnamefont{A.~H.} \bibnamefont{MacDonald}},
  \bibinfo{journal}{Proceedings of the National Academy of Sciences}
  \textbf{\bibinfo{volume}{108}}, \bibinfo{pages}{12233}
  (\bibinfo{year}{2011}),
  \urlprefix\url{https://doi.org/10.1073%2Fpnas.1108174108}.

\bibitem[{\citenamefont{Morell et~al.}(2010)\citenamefont{Morell, Correa,
  Vargas, Pacheco, and Barticevic}}]{Morell_2010}
\bibinfo{author}{\bibfnamefont{E.~S.} \bibnamefont{Morell}},
  \bibinfo{author}{\bibfnamefont{J.~D.} \bibnamefont{Correa}},
  \bibinfo{author}{\bibfnamefont{P.}~\bibnamefont{Vargas}},
  \bibinfo{author}{\bibfnamefont{M.}~\bibnamefont{Pacheco}}, \bibnamefont{and}
  \bibinfo{author}{\bibfnamefont{Z.}~\bibnamefont{Barticevic}},
  \bibinfo{journal}{Physical Review B} \textbf{\bibinfo{volume}{82}}
  (\bibinfo{year}{2010}),
  \urlprefix\url{https://doi.org/10.1103%2Fphysrevb.82.121407}.

\bibitem[{\citenamefont{Li et~al.}(2019)\citenamefont{Li, Wu, and
  MacDonald}}]{Li_2019}
\bibinfo{author}{\bibfnamefont{X.}~\bibnamefont{Li}},
  \bibinfo{author}{\bibfnamefont{F.}~\bibnamefont{Wu}}, \bibnamefont{and}
  \bibinfo{author}{\bibfnamefont{A.~H.} \bibnamefont{MacDonald}},
  \emph{\bibinfo{title}{Electronic structure of single-twist trilayer
  graphene}} (\bibinfo{year}{2019}),
  \urlprefix\url{https://arxiv.org/abs/1907.12338}.

\bibitem[{\citenamefont{Kim et~al.}(2017)\citenamefont{Kim, DaSilva, Huang,
  Fallahazad, Larentis, Taniguchi, Watanabe, LeRoy, MacDonald, and
  Tutuc}}]{Kim_2017}
\bibinfo{author}{\bibfnamefont{K.}~\bibnamefont{Kim}},
  \bibinfo{author}{\bibfnamefont{A.}~\bibnamefont{DaSilva}},
  \bibinfo{author}{\bibfnamefont{S.}~\bibnamefont{Huang}},
  \bibinfo{author}{\bibfnamefont{B.}~\bibnamefont{Fallahazad}},
  \bibinfo{author}{\bibfnamefont{S.}~\bibnamefont{Larentis}},
  \bibinfo{author}{\bibfnamefont{T.}~\bibnamefont{Taniguchi}},
  \bibinfo{author}{\bibfnamefont{K.}~\bibnamefont{Watanabe}},
  \bibinfo{author}{\bibfnamefont{B.~J.} \bibnamefont{LeRoy}},
  \bibinfo{author}{\bibfnamefont{A.~H.} \bibnamefont{MacDonald}},
  \bibnamefont{and} \bibinfo{author}{\bibfnamefont{E.}~\bibnamefont{Tutuc}},
  \bibinfo{journal}{Proceedings of the National Academy of Sciences}
  \textbf{\bibinfo{volume}{114}}, \bibinfo{pages}{3364} (\bibinfo{year}{2017}),
  \urlprefix\url{https://doi.org/10.1073%2Fpnas.1620140114}.

\bibitem[{\citenamefont{Cao et~al.}(2018{\natexlab{a}})\citenamefont{Cao,
  Fatemi, Demir, Fang, Tomarken, Luo, Sanchez-Yamagishi, Watanabe, Taniguchi,
  Kaxiras et~al.}}]{Cao_2018a}
\bibinfo{author}{\bibfnamefont{Y.}~\bibnamefont{Cao}},
  \bibinfo{author}{\bibfnamefont{V.}~\bibnamefont{Fatemi}},
  \bibinfo{author}{\bibfnamefont{A.}~\bibnamefont{Demir}},
  \bibinfo{author}{\bibfnamefont{S.}~\bibnamefont{Fang}},
  \bibinfo{author}{\bibfnamefont{S.~L.} \bibnamefont{Tomarken}},
  \bibinfo{author}{\bibfnamefont{J.~Y.} \bibnamefont{Luo}},
  \bibinfo{author}{\bibfnamefont{J.~D.} \bibnamefont{Sanchez-Yamagishi}},
  \bibinfo{author}{\bibfnamefont{K.}~\bibnamefont{Watanabe}},
  \bibinfo{author}{\bibfnamefont{T.}~\bibnamefont{Taniguchi}},
  \bibinfo{author}{\bibfnamefont{E.}~\bibnamefont{Kaxiras}},
  \bibnamefont{et~al.}, \bibinfo{journal}{Nature}
  \textbf{\bibinfo{volume}{556}}, \bibinfo{pages}{80}
  (\bibinfo{year}{2018}{\natexlab{a}}),
  \urlprefix\url{https://doi.org/10.1038%2Fnature26154}.

\bibitem[{\citenamefont{Cao et~al.}(2018{\natexlab{b}})\citenamefont{Cao,
  Fatemi, Fang, Watanabe, Taniguchi, Kaxiras, and Jarillo-Herrero}}]{Cao_2018b}
\bibinfo{author}{\bibfnamefont{Y.}~\bibnamefont{Cao}},
  \bibinfo{author}{\bibfnamefont{V.}~\bibnamefont{Fatemi}},
  \bibinfo{author}{\bibfnamefont{S.}~\bibnamefont{Fang}},
  \bibinfo{author}{\bibfnamefont{K.}~\bibnamefont{Watanabe}},
  \bibinfo{author}{\bibfnamefont{T.}~\bibnamefont{Taniguchi}},
  \bibinfo{author}{\bibfnamefont{E.}~\bibnamefont{Kaxiras}}, \bibnamefont{and}
  \bibinfo{author}{\bibfnamefont{P.}~\bibnamefont{Jarillo-Herrero}},
  \bibinfo{journal}{Nature} \textbf{\bibinfo{volume}{556}}, \bibinfo{pages}{43}
  (\bibinfo{year}{2018}{\natexlab{b}}),
  \urlprefix\url{https://doi.org/10.1038%2Fnature26160}.

\bibitem[{\citenamefont{Cao et~al.}(2020{\natexlab{a}})\citenamefont{Cao,
  Chowdhury, Rodan-Legrain, Rubies-Bigorda, Watanabe, Taniguchi, Senthil, and
  Jarillo-Herrero}}]{Cao2020PRL}
\bibinfo{author}{\bibfnamefont{Y.}~\bibnamefont{Cao}},
  \bibinfo{author}{\bibfnamefont{D.}~\bibnamefont{Chowdhury}},
  \bibinfo{author}{\bibfnamefont{D.}~\bibnamefont{Rodan-Legrain}},
  \bibinfo{author}{\bibfnamefont{O.}~\bibnamefont{Rubies-Bigorda}},
  \bibinfo{author}{\bibfnamefont{K.}~\bibnamefont{Watanabe}},
  \bibinfo{author}{\bibfnamefont{T.}~\bibnamefont{Taniguchi}},
  \bibinfo{author}{\bibfnamefont{T.}~\bibnamefont{Senthil}}, \bibnamefont{and}
  \bibinfo{author}{\bibfnamefont{P.}~\bibnamefont{Jarillo-Herrero}},
  \bibinfo{journal}{Phys. Rev. Lett.} \textbf{\bibinfo{volume}{124}},
  \bibinfo{pages}{076801} (\bibinfo{year}{2020}{\natexlab{a}}),
  \urlprefix\url{https://link.aps.org/doi/10.1103/PhysRevLett.124.076801}.

\bibitem[{\citenamefont{Cao et~al.}(2021)\citenamefont{Cao, Park, Watanabe,
  Taniguchi, and Jarillo-Herrero}}]{Cao2021}
\bibinfo{author}{\bibfnamefont{Y.}~\bibnamefont{Cao}},
  \bibinfo{author}{\bibfnamefont{J.~M.} \bibnamefont{Park}},
  \bibinfo{author}{\bibfnamefont{K.}~\bibnamefont{Watanabe}},
  \bibinfo{author}{\bibfnamefont{T.}~\bibnamefont{Taniguchi}},
  \bibnamefont{and}
  \bibinfo{author}{\bibfnamefont{P.}~\bibnamefont{Jarillo-Herrero}},
  \bibinfo{journal}{Nature} \textbf{\bibinfo{volume}{595}},
  \bibinfo{pages}{526} (\bibinfo{year}{2021}).

\bibitem[{\citenamefont{Yankowitz et~al.}(2019)\citenamefont{Yankowitz, Chen,
  Polshyn, Zhang, Watanabe, Taniguchi, Graf, Young, and Dean}}]{Yankowitz_2019}
\bibinfo{author}{\bibfnamefont{M.}~\bibnamefont{Yankowitz}},
  \bibinfo{author}{\bibfnamefont{S.}~\bibnamefont{Chen}},
  \bibinfo{author}{\bibfnamefont{H.}~\bibnamefont{Polshyn}},
  \bibinfo{author}{\bibfnamefont{Y.}~\bibnamefont{Zhang}},
  \bibinfo{author}{\bibfnamefont{K.}~\bibnamefont{Watanabe}},
  \bibinfo{author}{\bibfnamefont{T.}~\bibnamefont{Taniguchi}},
  \bibinfo{author}{\bibfnamefont{D.}~\bibnamefont{Graf}},
  \bibinfo{author}{\bibfnamefont{A.~F.} \bibnamefont{Young}}, \bibnamefont{and}
  \bibinfo{author}{\bibfnamefont{C.~R.} \bibnamefont{Dean}},
  \bibinfo{journal}{Science} \textbf{\bibinfo{volume}{363}},
  \bibinfo{pages}{1059} (\bibinfo{year}{2019}),
  \urlprefix\url{https://doi.org/10.1126%2Fscience.aav1910}.

\bibitem[{\citenamefont{Kerelsky et~al.}(2019)\citenamefont{Kerelsky, McGilly,
  Kennes, Xian, Yankowitz, Chen, Watanabe, Taniguchi, Hone, Dean
  et~al.}}]{Kerelsky_2019}
\bibinfo{author}{\bibfnamefont{A.}~\bibnamefont{Kerelsky}},
  \bibinfo{author}{\bibfnamefont{L.~J.} \bibnamefont{McGilly}},
  \bibinfo{author}{\bibfnamefont{D.~M.} \bibnamefont{Kennes}},
  \bibinfo{author}{\bibfnamefont{L.}~\bibnamefont{Xian}},
  \bibinfo{author}{\bibfnamefont{M.}~\bibnamefont{Yankowitz}},
  \bibinfo{author}{\bibfnamefont{S.}~\bibnamefont{Chen}},
  \bibinfo{author}{\bibfnamefont{K.}~\bibnamefont{Watanabe}},
  \bibinfo{author}{\bibfnamefont{T.}~\bibnamefont{Taniguchi}},
  \bibinfo{author}{\bibfnamefont{J.}~\bibnamefont{Hone}},
  \bibinfo{author}{\bibfnamefont{C.}~\bibnamefont{Dean}}, \bibnamefont{et~al.},
  \bibinfo{journal}{Nature} \textbf{\bibinfo{volume}{572}}, \bibinfo{pages}{95}
  (\bibinfo{year}{2019}),
  \urlprefix\url{https://doi.org/10.1038%2Fs41586-019-1431-9}.

\bibitem[{\citenamefont{Lu et~al.}(2019)\citenamefont{Lu, Stepanov, Yang, Xie,
  Aamir, Das, Urgell, Watanabe, Taniguchi, Zhang et~al.}}]{Lu_2019}
\bibinfo{author}{\bibfnamefont{X.}~\bibnamefont{Lu}},
  \bibinfo{author}{\bibfnamefont{P.}~\bibnamefont{Stepanov}},
  \bibinfo{author}{\bibfnamefont{W.}~\bibnamefont{Yang}},
  \bibinfo{author}{\bibfnamefont{M.}~\bibnamefont{Xie}},
  \bibinfo{author}{\bibfnamefont{M.~A.} \bibnamefont{Aamir}},
  \bibinfo{author}{\bibfnamefont{I.}~\bibnamefont{Das}},
  \bibinfo{author}{\bibfnamefont{C.}~\bibnamefont{Urgell}},
  \bibinfo{author}{\bibfnamefont{K.}~\bibnamefont{Watanabe}},
  \bibinfo{author}{\bibfnamefont{T.}~\bibnamefont{Taniguchi}},
  \bibinfo{author}{\bibfnamefont{G.}~\bibnamefont{Zhang}},
  \bibnamefont{et~al.}, \bibinfo{journal}{Nature}
  \textbf{\bibinfo{volume}{574}}, \bibinfo{pages}{653} (\bibinfo{year}{2019}),
  \urlprefix\url{https://doi.org/10.1038%2Fs41586-019-1695-0}.

\bibitem[{\citenamefont{Stepanov et~al.}(2020)\citenamefont{Stepanov, Das, Lu,
  Fahimniya, Watanabe, Taniguchi, Koppens, Lischner, Levitov, and
  Efetov}}]{Stepanov2020untying}
\bibinfo{author}{\bibfnamefont{P.}~\bibnamefont{Stepanov}},
  \bibinfo{author}{\bibfnamefont{I.}~\bibnamefont{Das}},
  \bibinfo{author}{\bibfnamefont{X.}~\bibnamefont{Lu}},
  \bibinfo{author}{\bibfnamefont{A.}~\bibnamefont{Fahimniya}},
  \bibinfo{author}{\bibfnamefont{K.}~\bibnamefont{Watanabe}},
  \bibinfo{author}{\bibfnamefont{T.}~\bibnamefont{Taniguchi}},
  \bibinfo{author}{\bibfnamefont{F.~H.} \bibnamefont{Koppens}},
  \bibinfo{author}{\bibfnamefont{J.}~\bibnamefont{Lischner}},
  \bibinfo{author}{\bibfnamefont{L.}~\bibnamefont{Levitov}}, \bibnamefont{and}
  \bibinfo{author}{\bibfnamefont{D.~K.} \bibnamefont{Efetov}},
  \bibinfo{journal}{Nature} \textbf{\bibinfo{volume}{583}},
  \bibinfo{pages}{375} (\bibinfo{year}{2020}).

\bibitem[{\citenamefont{Sharpe et~al.}(2019)\citenamefont{Sharpe, Fox, Barnard,
  Finney, Watanabe, Taniguchi, Kastner, and Goldhaber-Gordon}}]{Sharpe_2019}
\bibinfo{author}{\bibfnamefont{A.~L.} \bibnamefont{Sharpe}},
  \bibinfo{author}{\bibfnamefont{E.~J.} \bibnamefont{Fox}},
  \bibinfo{author}{\bibfnamefont{A.~W.} \bibnamefont{Barnard}},
  \bibinfo{author}{\bibfnamefont{J.}~\bibnamefont{Finney}},
  \bibinfo{author}{\bibfnamefont{K.}~\bibnamefont{Watanabe}},
  \bibinfo{author}{\bibfnamefont{T.}~\bibnamefont{Taniguchi}},
  \bibinfo{author}{\bibfnamefont{M.~A.} \bibnamefont{Kastner}},
  \bibnamefont{and}
  \bibinfo{author}{\bibfnamefont{D.}~\bibnamefont{Goldhaber-Gordon}},
  \bibinfo{journal}{Science} \textbf{\bibinfo{volume}{365}},
  \bibinfo{pages}{605} (\bibinfo{year}{2019}),
  \urlprefix\url{https://doi.org/10.1126%2Fscience.aaw3780}.

\bibitem[{\citenamefont{Chen et~al.}(2020)\citenamefont{Chen, Sharpe, Fox,
  Zhang, Wang, Jiang, Lyu, Li, Watanabe, Taniguchi et~al.}}]{Chen_2020}
\bibinfo{author}{\bibfnamefont{G.}~\bibnamefont{Chen}},
  \bibinfo{author}{\bibfnamefont{A.~L.} \bibnamefont{Sharpe}},
  \bibinfo{author}{\bibfnamefont{E.~J.} \bibnamefont{Fox}},
  \bibinfo{author}{\bibfnamefont{Y.-H.} \bibnamefont{Zhang}},
  \bibinfo{author}{\bibfnamefont{S.}~\bibnamefont{Wang}},
  \bibinfo{author}{\bibfnamefont{L.}~\bibnamefont{Jiang}},
  \bibinfo{author}{\bibfnamefont{B.}~\bibnamefont{Lyu}},
  \bibinfo{author}{\bibfnamefont{H.}~\bibnamefont{Li}},
  \bibinfo{author}{\bibfnamefont{K.}~\bibnamefont{Watanabe}},
  \bibinfo{author}{\bibfnamefont{T.}~\bibnamefont{Taniguchi}},
  \bibnamefont{et~al.}, \bibinfo{journal}{Nature}
  \textbf{\bibinfo{volume}{579}}, \bibinfo{pages}{56} (\bibinfo{year}{2020}),
  \urlprefix\url{https://doi.org/10.1038%2Fs41586-020-2049-7}.

\bibitem[{\citenamefont{Rozen et~al.}(2021)\citenamefont{Rozen, Park, Zondiner,
  Cao, Rodan-Legrain, Taniguchi, Watanabe, Oreg, Stern, Berg
  et~al.}}]{Rozen2021entropic}
\bibinfo{author}{\bibfnamefont{A.}~\bibnamefont{Rozen}},
  \bibinfo{author}{\bibfnamefont{J.~M.} \bibnamefont{Park}},
  \bibinfo{author}{\bibfnamefont{U.}~\bibnamefont{Zondiner}},
  \bibinfo{author}{\bibfnamefont{Y.}~\bibnamefont{Cao}},
  \bibinfo{author}{\bibfnamefont{D.}~\bibnamefont{Rodan-Legrain}},
  \bibinfo{author}{\bibfnamefont{T.}~\bibnamefont{Taniguchi}},
  \bibinfo{author}{\bibfnamefont{K.}~\bibnamefont{Watanabe}},
  \bibinfo{author}{\bibfnamefont{Y.}~\bibnamefont{Oreg}},
  \bibinfo{author}{\bibfnamefont{A.}~\bibnamefont{Stern}},
  \bibinfo{author}{\bibfnamefont{E.}~\bibnamefont{Berg}}, \bibnamefont{et~al.},
  \bibinfo{journal}{Nature} \textbf{\bibinfo{volume}{592}},
  \bibinfo{pages}{214} (\bibinfo{year}{2021}).

\bibitem[{\citenamefont{Zhou et~al.}(2022)\citenamefont{Zhou, Holleis, Saito,
  Cohen, Huynh, Patterson, Yang, Taniguchi, Watanabe, and
  Young}}]{AndreaYoungBernal}
\bibinfo{author}{\bibfnamefont{H.}~\bibnamefont{Zhou}},
  \bibinfo{author}{\bibfnamefont{L.}~\bibnamefont{Holleis}},
  \bibinfo{author}{\bibfnamefont{Y.}~\bibnamefont{Saito}},
  \bibinfo{author}{\bibfnamefont{L.}~\bibnamefont{Cohen}},
  \bibinfo{author}{\bibfnamefont{W.}~\bibnamefont{Huynh}},
  \bibinfo{author}{\bibfnamefont{C.~L.} \bibnamefont{Patterson}},
  \bibinfo{author}{\bibfnamefont{F.}~\bibnamefont{Yang}},
  \bibinfo{author}{\bibfnamefont{T.}~\bibnamefont{Taniguchi}},
  \bibinfo{author}{\bibfnamefont{K.}~\bibnamefont{Watanabe}}, \bibnamefont{and}
  \bibinfo{author}{\bibfnamefont{A.~F.} \bibnamefont{Young}},
  \bibinfo{journal}{Science} \textbf{\bibinfo{volume}{375}},
  \bibinfo{pages}{774} (\bibinfo{year}{2022}),
  \urlprefix\url{https://doi.org/10.1126%2Fscience.abm8386}.

\bibitem[{\citenamefont{Zhou et~al.}(2021{\natexlab{a}})\citenamefont{Zhou,
  Xie, Taniguchi, Watanabe, and Young}}]{AndreaYoungRhombo}
\bibinfo{author}{\bibfnamefont{H.}~\bibnamefont{Zhou}},
  \bibinfo{author}{\bibfnamefont{T.}~\bibnamefont{Xie}},
  \bibinfo{author}{\bibfnamefont{T.}~\bibnamefont{Taniguchi}},
  \bibinfo{author}{\bibfnamefont{K.}~\bibnamefont{Watanabe}}, \bibnamefont{and}
  \bibinfo{author}{\bibfnamefont{A.~F.} \bibnamefont{Young}},
  \bibinfo{journal}{Nature} \textbf{\bibinfo{volume}{598}},
  \bibinfo{pages}{434} (\bibinfo{year}{2021}{\natexlab{a}}),
  \urlprefix\url{https://doi.org/10.1038%2Fs41586-021-03926-0}.

\bibitem[{\citenamefont{Zhou et~al.}(2021{\natexlab{b}})\citenamefont{Zhou,
  Xie, Ghazaryan, Holder, Ehrets, Spanton, Taniguchi, Watanabe, Berg, Serbyn
  et~al.}}]{AndreaYoungRhombo2}
\bibinfo{author}{\bibfnamefont{H.}~\bibnamefont{Zhou}},
  \bibinfo{author}{\bibfnamefont{T.}~\bibnamefont{Xie}},
  \bibinfo{author}{\bibfnamefont{A.}~\bibnamefont{Ghazaryan}},
  \bibinfo{author}{\bibfnamefont{T.}~\bibnamefont{Holder}},
  \bibinfo{author}{\bibfnamefont{J.~R.} \bibnamefont{Ehrets}},
  \bibinfo{author}{\bibfnamefont{E.~M.} \bibnamefont{Spanton}},
  \bibinfo{author}{\bibfnamefont{T.}~\bibnamefont{Taniguchi}},
  \bibinfo{author}{\bibfnamefont{K.}~\bibnamefont{Watanabe}},
  \bibinfo{author}{\bibfnamefont{E.}~\bibnamefont{Berg}},
  \bibinfo{author}{\bibfnamefont{M.}~\bibnamefont{Serbyn}},
  \bibnamefont{et~al.}, \bibinfo{journal}{Nature}
  \textbf{\bibinfo{volume}{598}}, \bibinfo{pages}{429}
  (\bibinfo{year}{2021}{\natexlab{b}}),
  \urlprefix\url{https://doi.org/10.1038%2Fs41586-021-03938-w}.

\bibitem[{\citenamefont{Serlin et~al.}(2020)\citenamefont{Serlin, Tschirhart,
  Polshyn, Zhang, Zhu, Watanabe, Taniguchi, Balents, and Young}}]{Serlin_2020}
\bibinfo{author}{\bibfnamefont{M.}~\bibnamefont{Serlin}},
  \bibinfo{author}{\bibfnamefont{C.~L.} \bibnamefont{Tschirhart}},
  \bibinfo{author}{\bibfnamefont{H.}~\bibnamefont{Polshyn}},
  \bibinfo{author}{\bibfnamefont{Y.}~\bibnamefont{Zhang}},
  \bibinfo{author}{\bibfnamefont{J.}~\bibnamefont{Zhu}},
  \bibinfo{author}{\bibfnamefont{K.}~\bibnamefont{Watanabe}},
  \bibinfo{author}{\bibfnamefont{T.}~\bibnamefont{Taniguchi}},
  \bibinfo{author}{\bibfnamefont{L.}~\bibnamefont{Balents}}, \bibnamefont{and}
  \bibinfo{author}{\bibfnamefont{A.~F.} \bibnamefont{Young}},
  \bibinfo{journal}{Science} \textbf{\bibinfo{volume}{367}},
  \bibinfo{pages}{900} (\bibinfo{year}{2020}),
  \urlprefix\url{https://doi.org/10.1126%2Fscience.aay5533}.

\bibitem[{\citenamefont{Wu et~al.}(2018)\citenamefont{Wu, MacDonald, and
  Martin}}]{Wu_2018}
\bibinfo{author}{\bibfnamefont{F.}~\bibnamefont{Wu}},
  \bibinfo{author}{\bibfnamefont{A.}~\bibnamefont{MacDonald}},
  \bibnamefont{and} \bibinfo{author}{\bibfnamefont{I.}~\bibnamefont{Martin}},
  \bibinfo{journal}{Physical Review Letters} \textbf{\bibinfo{volume}{121}}
  (\bibinfo{year}{2018}),
  \urlprefix\url{https://doi.org/10.1103%2Fphysrevlett.121.257001}.

\bibitem[{\citenamefont{Wu et~al.}(2019{\natexlab{a}})\citenamefont{Wu, Lovorn,
  Tutuc, Martin, and MacDonald}}]{Wu_2019_TIPRL}
\bibinfo{author}{\bibfnamefont{F.}~\bibnamefont{Wu}},
  \bibinfo{author}{\bibfnamefont{T.}~\bibnamefont{Lovorn}},
  \bibinfo{author}{\bibfnamefont{E.}~\bibnamefont{Tutuc}},
  \bibinfo{author}{\bibfnamefont{I.}~\bibnamefont{Martin}}, \bibnamefont{and}
  \bibinfo{author}{\bibfnamefont{A.~H.} \bibnamefont{MacDonald}},
  \bibinfo{journal}{Phys. Rev. Lett.} \textbf{\bibinfo{volume}{122}},
  \bibinfo{pages}{086402} (\bibinfo{year}{2019}{\natexlab{a}}),
  \urlprefix\url{https://link.aps.org/doi/10.1103/PhysRevLett.122.086402}.

\bibitem[{\citenamefont{Tschirhart et~al.}(2022)\citenamefont{Tschirhart,
  Redekop, Li, Li, Jiang, Arp, Sheekey, Taniguchi, Watanabe, Mak
  et~al.}}]{KinFaiMak_TopologyTMD}
\bibinfo{author}{\bibfnamefont{C.~L.} \bibnamefont{Tschirhart}},
  \bibinfo{author}{\bibfnamefont{E.}~\bibnamefont{Redekop}},
  \bibinfo{author}{\bibfnamefont{L.}~\bibnamefont{Li}},
  \bibinfo{author}{\bibfnamefont{T.}~\bibnamefont{Li}},
  \bibinfo{author}{\bibfnamefont{S.}~\bibnamefont{Jiang}},
  \bibinfo{author}{\bibfnamefont{T.}~\bibnamefont{Arp}},
  \bibinfo{author}{\bibfnamefont{O.}~\bibnamefont{Sheekey}},
  \bibinfo{author}{\bibfnamefont{T.}~\bibnamefont{Taniguchi}},
  \bibinfo{author}{\bibfnamefont{K.}~\bibnamefont{Watanabe}},
  \bibinfo{author}{\bibfnamefont{K.~F.} \bibnamefont{Mak}},
  \bibnamefont{et~al.}, \emph{\bibinfo{title}{Intrinsic spin hall torque in a
  moire chern magnet}} (\bibinfo{year}{2022}),
  \urlprefix\url{https://arxiv.org/abs/2205.02823}.

\bibitem[{\citenamefont{Polshyn et~al.}(2020)\citenamefont{Polshyn, Zhu, Kumar,
  Zhang, Yang, Tschirhart, Serlin, Watanabe, Taniguchi, MacDonald
  et~al.}}]{Polshyn_2020}
\bibinfo{author}{\bibfnamefont{H.}~\bibnamefont{Polshyn}},
  \bibinfo{author}{\bibfnamefont{J.}~\bibnamefont{Zhu}},
  \bibinfo{author}{\bibfnamefont{M.~A.} \bibnamefont{Kumar}},
  \bibinfo{author}{\bibfnamefont{Y.}~\bibnamefont{Zhang}},
  \bibinfo{author}{\bibfnamefont{F.}~\bibnamefont{Yang}},
  \bibinfo{author}{\bibfnamefont{C.~L.} \bibnamefont{Tschirhart}},
  \bibinfo{author}{\bibfnamefont{M.}~\bibnamefont{Serlin}},
  \bibinfo{author}{\bibfnamefont{K.}~\bibnamefont{Watanabe}},
  \bibinfo{author}{\bibfnamefont{T.}~\bibnamefont{Taniguchi}},
  \bibinfo{author}{\bibfnamefont{A.~H.} \bibnamefont{MacDonald}},
  \bibnamefont{et~al.}, \bibinfo{journal}{Nature}
  \textbf{\bibinfo{volume}{588}}, \bibinfo{pages}{66} (\bibinfo{year}{2020}),
  \urlprefix\url{https://doi.org/10.1038%2Fs41586-020-2963-8}.

\bibitem[{\citenamefont{Jaoui et~al.}(2021)\citenamefont{Jaoui, Das,
  Di~Battista, Díez-Mérida, Lu, Watanabe, Taniguchi, Ishizuka, Levitov, and
  Efetov}}]{TBLGStrangeMetalExperiment1}
\bibinfo{author}{\bibfnamefont{A.}~\bibnamefont{Jaoui}},
  \bibinfo{author}{\bibfnamefont{I.}~\bibnamefont{Das}},
  \bibinfo{author}{\bibfnamefont{G.}~\bibnamefont{Di~Battista}},
  \bibinfo{author}{\bibfnamefont{J.}~\bibnamefont{Díez-Mérida}},
  \bibinfo{author}{\bibfnamefont{X.}~\bibnamefont{Lu}},
  \bibinfo{author}{\bibfnamefont{K.}~\bibnamefont{Watanabe}},
  \bibinfo{author}{\bibfnamefont{T.}~\bibnamefont{Taniguchi}},
  \bibinfo{author}{\bibfnamefont{H.}~\bibnamefont{Ishizuka}},
  \bibinfo{author}{\bibfnamefont{L.}~\bibnamefont{Levitov}}, \bibnamefont{and}
  \bibinfo{author}{\bibfnamefont{D.~K.} \bibnamefont{Efetov}},
  \emph{\bibinfo{title}{Quantum critical behavior in magic-angle twisted
  bilayer graphene}} (\bibinfo{year}{2021}),
  \urlprefix\url{https://arxiv.org/abs/2108.07753}.

\bibitem[{\citenamefont{Polshyn
  et~al.}(2019{\natexlab{a}})\citenamefont{Polshyn, Yankowitz, Chen, Zhang,
  Watanabe, Taniguchi, Dean, and Young}}]{TBLGStrangeMetalExperiment2}
\bibinfo{author}{\bibfnamefont{H.}~\bibnamefont{Polshyn}},
  \bibinfo{author}{\bibfnamefont{M.}~\bibnamefont{Yankowitz}},
  \bibinfo{author}{\bibfnamefont{S.}~\bibnamefont{Chen}},
  \bibinfo{author}{\bibfnamefont{Y.}~\bibnamefont{Zhang}},
  \bibinfo{author}{\bibfnamefont{K.}~\bibnamefont{Watanabe}},
  \bibinfo{author}{\bibfnamefont{T.}~\bibnamefont{Taniguchi}},
  \bibinfo{author}{\bibfnamefont{C.~R.} \bibnamefont{Dean}}, \bibnamefont{and}
  \bibinfo{author}{\bibfnamefont{A.~F.} \bibnamefont{Young}},
  \bibinfo{journal}{Nature} \textbf{\bibinfo{volume}{15}}
  (\bibinfo{year}{2019}{\natexlab{a}}),
  \urlprefix\url{https://doi.org/10.1038/s41567-019-0596-3}.

\bibitem[{\citenamefont{Cao et~al.}(2020{\natexlab{b}})\citenamefont{Cao,
  Chowdhury, Rodan-Legrain, Rubies-Bigorda, Watanabe, Taniguchi, Senthil, and
  Jarillo-Herrero}}]{TBLGStrangeMetalExperiment3}
\bibinfo{author}{\bibfnamefont{Y.}~\bibnamefont{Cao}},
  \bibinfo{author}{\bibfnamefont{D.}~\bibnamefont{Chowdhury}},
  \bibinfo{author}{\bibfnamefont{D.}~\bibnamefont{Rodan-Legrain}},
  \bibinfo{author}{\bibfnamefont{O.}~\bibnamefont{Rubies-Bigorda}},
  \bibinfo{author}{\bibfnamefont{K.}~\bibnamefont{Watanabe}},
  \bibinfo{author}{\bibfnamefont{T.}~\bibnamefont{Taniguchi}},
  \bibinfo{author}{\bibfnamefont{T.}~\bibnamefont{Senthil}}, \bibnamefont{and}
  \bibinfo{author}{\bibfnamefont{P.}~\bibnamefont{Jarillo-Herrero}},
  \bibinfo{journal}{Phys. Rev. Lett.} \textbf{\bibinfo{volume}{124}},
  \bibinfo{pages}{076801} (\bibinfo{year}{2020}{\natexlab{b}}),
  \urlprefix\url{https://link.aps.org/doi/10.1103/PhysRevLett.124.076801}.

\bibitem[{\citenamefont{Sarma and Wu}(2022)}]{SankarFengchengStrangeMetal}
\bibinfo{author}{\bibfnamefont{S.~D.} \bibnamefont{Sarma}} \bibnamefont{and}
  \bibinfo{author}{\bibfnamefont{F.}~\bibnamefont{Wu}},
  \emph{\bibinfo{title}{Strange metallicity of moiré twisted bilayer
  graphene}} (\bibinfo{year}{2022}),
  \urlprefix\url{https://arxiv.org/abs/2201.10270}.

\bibitem[{\citenamefont{Zhang et~al.}(2022)\citenamefont{Zhang, Polski,
  Thomson, Lantagne-Hurtubise, Lewandowski, Zhou, Watanabe, Taniguchi, Alicea,
  and Nadj-Perge}}]{CalTechBernal}
\bibinfo{author}{\bibfnamefont{Y.}~\bibnamefont{Zhang}},
  \bibinfo{author}{\bibfnamefont{R.}~\bibnamefont{Polski}},
  \bibinfo{author}{\bibfnamefont{A.}~\bibnamefont{Thomson}},
  \bibinfo{author}{\bibfnamefont{Ã.}~\bibnamefont{Lantagne-Hurtubise}},
  \bibinfo{author}{\bibfnamefont{C.}~\bibnamefont{Lewandowski}},
  \bibinfo{author}{\bibfnamefont{H.}~\bibnamefont{Zhou}},
  \bibinfo{author}{\bibfnamefont{K.}~\bibnamefont{Watanabe}},
  \bibinfo{author}{\bibfnamefont{T.}~\bibnamefont{Taniguchi}},
  \bibinfo{author}{\bibfnamefont{J.}~\bibnamefont{Alicea}}, \bibnamefont{and}
  \bibinfo{author}{\bibfnamefont{S.}~\bibnamefont{Nadj-Perge}},
  \emph{\bibinfo{title}{Spin-orbit enhanced superconductivity in bernal bilayer
  graphene}} (\bibinfo{year}{2022}),
  \urlprefix\url{https://arxiv.org/abs/2205.05087}.

\bibitem[{\citenamefont{Polski et~al.}(2022)\citenamefont{Polski, Zhang, Peng,
  Arora, Choi, Kim, Watanabe, Taniguchi, Refael, von Oppen
  et~al.}}]{CalTechSymmetryBreakingTBLG}
\bibinfo{author}{\bibfnamefont{R.}~\bibnamefont{Polski}},
  \bibinfo{author}{\bibfnamefont{Y.}~\bibnamefont{Zhang}},
  \bibinfo{author}{\bibfnamefont{Y.}~\bibnamefont{Peng}},
  \bibinfo{author}{\bibfnamefont{H.~S.} \bibnamefont{Arora}},
  \bibinfo{author}{\bibfnamefont{Y.}~\bibnamefont{Choi}},
  \bibinfo{author}{\bibfnamefont{H.}~\bibnamefont{Kim}},
  \bibinfo{author}{\bibfnamefont{K.}~\bibnamefont{Watanabe}},
  \bibinfo{author}{\bibfnamefont{T.}~\bibnamefont{Taniguchi}},
  \bibinfo{author}{\bibfnamefont{G.}~\bibnamefont{Refael}},
  \bibinfo{author}{\bibfnamefont{F.}~\bibnamefont{von Oppen}},
  \bibnamefont{et~al.}, \emph{\bibinfo{title}{Hierarchy of symmetry breaking
  correlated phases in twisted bilayer graphene}} (\bibinfo{year}{2022}),
  \urlprefix\url{https://arxiv.org/abs/2205.05225}.

\bibitem[{\citenamefont{Arora et~al.}(2020)\citenamefont{Arora, Polski, Zhang,
  Thomson, Choi, Kim, Lin, Wilson, Xu, Chu et~al.}}]{CalTechTwisted}
\bibinfo{author}{\bibfnamefont{H.~S.} \bibnamefont{Arora}},
  \bibinfo{author}{\bibfnamefont{R.}~\bibnamefont{Polski}},
  \bibinfo{author}{\bibfnamefont{Y.}~\bibnamefont{Zhang}},
  \bibinfo{author}{\bibfnamefont{A.}~\bibnamefont{Thomson}},
  \bibinfo{author}{\bibfnamefont{Y.}~\bibnamefont{Choi}},
  \bibinfo{author}{\bibfnamefont{H.}~\bibnamefont{Kim}},
  \bibinfo{author}{\bibfnamefont{Z.}~\bibnamefont{Lin}},
  \bibinfo{author}{\bibfnamefont{I.~Z.} \bibnamefont{Wilson}},
  \bibinfo{author}{\bibfnamefont{X.}~\bibnamefont{Xu}},
  \bibinfo{author}{\bibfnamefont{J.-H.} \bibnamefont{Chu}},
  \bibnamefont{et~al.}, \bibinfo{journal}{Nature}
  \textbf{\bibinfo{volume}{583}}, \bibinfo{pages}{379} (\bibinfo{year}{2020}),
  \urlprefix\url{https://doi.org/10.1038%2Fs41586-020-2473-8}.

\bibitem[{\citenamefont{Xie and MacDonald}(2020)}]{Xie_2020}
\bibinfo{author}{\bibfnamefont{M.}~\bibnamefont{Xie}} \bibnamefont{and}
  \bibinfo{author}{\bibfnamefont{A.}~\bibnamefont{MacDonald}},
  \bibinfo{journal}{Physical Review Letters} \textbf{\bibinfo{volume}{124}}
  (\bibinfo{year}{2020}),
  \urlprefix\url{https://doi.org/10.1103%2Fphysrevlett.124.097601}.

\bibitem[{\citenamefont{Andrei and MacDonald}(2020)}]{MacDonaldTLBGReview}
\bibinfo{author}{\bibfnamefont{E.~Y.} \bibnamefont{Andrei}} \bibnamefont{and}
  \bibinfo{author}{\bibfnamefont{A.~H.} \bibnamefont{MacDonald}},
  \bibinfo{journal}{Nature Materials} \textbf{\bibinfo{volume}{19}},
  \bibinfo{pages}{1265} (\bibinfo{year}{2020}),
  \urlprefix\url{https://doi.org/10.1038%2Fs41563-020-00840-0}.

\bibitem[{\citenamefont{Li et~al.}(2021)\citenamefont{Li, Jiang, Li, Zhang,
  Kang, Zhu, Watanabe, Taniguchi, Chowdhury, Fu et~al.}}]{Li_2021}
\bibinfo{author}{\bibfnamefont{T.}~\bibnamefont{Li}},
  \bibinfo{author}{\bibfnamefont{S.}~\bibnamefont{Jiang}},
  \bibinfo{author}{\bibfnamefont{L.}~\bibnamefont{Li}},
  \bibinfo{author}{\bibfnamefont{Y.}~\bibnamefont{Zhang}},
  \bibinfo{author}{\bibfnamefont{K.}~\bibnamefont{Kang}},
  \bibinfo{author}{\bibfnamefont{J.}~\bibnamefont{Zhu}},
  \bibinfo{author}{\bibfnamefont{K.}~\bibnamefont{Watanabe}},
  \bibinfo{author}{\bibfnamefont{T.}~\bibnamefont{Taniguchi}},
  \bibinfo{author}{\bibfnamefont{D.}~\bibnamefont{Chowdhury}},
  \bibinfo{author}{\bibfnamefont{L.}~\bibnamefont{Fu}}, \bibnamefont{et~al.},
  \bibinfo{journal}{Nature} \textbf{\bibinfo{volume}{597}},
  \bibinfo{pages}{350} (\bibinfo{year}{2021}),
  \urlprefix\url{https://doi.org/10.1038%2Fs41586-021-03853-0}.

\bibitem[{\citenamefont{Ghiotto et~al.}(2021)\citenamefont{Ghiotto, Shih,
  Pereira, Rhodes, Kim, Zang, Millis, Watanabe, Taniguchi, Hone
  et~al.}}]{Ghiotto_2021}
\bibinfo{author}{\bibfnamefont{A.}~\bibnamefont{Ghiotto}},
  \bibinfo{author}{\bibfnamefont{E.-M.} \bibnamefont{Shih}},
  \bibinfo{author}{\bibfnamefont{G.~S. S.~G.} \bibnamefont{Pereira}},
  \bibinfo{author}{\bibfnamefont{D.~A.} \bibnamefont{Rhodes}},
  \bibinfo{author}{\bibfnamefont{B.}~\bibnamefont{Kim}},
  \bibinfo{author}{\bibfnamefont{J.}~\bibnamefont{Zang}},
  \bibinfo{author}{\bibfnamefont{A.~J.} \bibnamefont{Millis}},
  \bibinfo{author}{\bibfnamefont{K.}~\bibnamefont{Watanabe}},
  \bibinfo{author}{\bibfnamefont{T.}~\bibnamefont{Taniguchi}},
  \bibinfo{author}{\bibfnamefont{J.~C.} \bibnamefont{Hone}},
  \bibnamefont{et~al.}, \bibinfo{journal}{Nature}
  \textbf{\bibinfo{volume}{597}}, \bibinfo{pages}{345} (\bibinfo{year}{2021}),
  \urlprefix\url{https://doi.org/10.1038%2Fs41586-021-03815-6}.

\bibitem[{\citenamefont{Pan et~al.}(2020)\citenamefont{Pan, Wu, and
  Sarma}}]{Pan_2020}
\bibinfo{author}{\bibfnamefont{H.}~\bibnamefont{Pan}},
  \bibinfo{author}{\bibfnamefont{F.}~\bibnamefont{Wu}}, \bibnamefont{and}
  \bibinfo{author}{\bibfnamefont{S.~D.} \bibnamefont{Sarma}},
  \bibinfo{journal}{Physical Review B} \textbf{\bibinfo{volume}{102}}
  (\bibinfo{year}{2020}),
  \urlprefix\url{https://doi.org/10.1103%2Fphysrevb.102.201104}.

\bibitem[{\citenamefont{Pan and Sarma}(2021)}]{Pan_2021}
\bibinfo{author}{\bibfnamefont{H.}~\bibnamefont{Pan}} \bibnamefont{and}
  \bibinfo{author}{\bibfnamefont{S.~D.} \bibnamefont{Sarma}},
  \bibinfo{journal}{Physical Review Letters} \textbf{\bibinfo{volume}{127}}
  (\bibinfo{year}{2021}),
  \urlprefix\url{https://doi.org/10.1103%2Fphysrevlett.127.096802}.

\bibitem[{\citenamefont{Morales-Dur{\'{a}}n
  et~al.}(2021)\citenamefont{Morales-Dur{\'{a}}n, MacDonald, and
  Potasz}}]{Morales_Dur_n_2021}
\bibinfo{author}{\bibfnamefont{N.}~\bibnamefont{Morales-Dur{\'{a}}n}},
  \bibinfo{author}{\bibfnamefont{A.~H.} \bibnamefont{MacDonald}},
  \bibnamefont{and} \bibinfo{author}{\bibfnamefont{P.}~\bibnamefont{Potasz}},
  \bibinfo{journal}{Physical Review B} \textbf{\bibinfo{volume}{103}}
  (\bibinfo{year}{2021}),
  \urlprefix\url{https://doi.org/10.1103%2Fphysrevb.103.l241110}.

\bibitem[{\citenamefont{Ahn and Sarma}(2022)}]{Ahn_2022}
\bibinfo{author}{\bibfnamefont{S.}~\bibnamefont{Ahn}} \bibnamefont{and}
  \bibinfo{author}{\bibfnamefont{S.~D.} \bibnamefont{Sarma}},
  \bibinfo{journal}{Physical Review B} \textbf{\bibinfo{volume}{105}}
  (\bibinfo{year}{2022}),
  \urlprefix\url{https://doi.org/10.1103%2Fphysrevb.105.115114}.

\bibitem[{\citenamefont{Kerelsky et~al.}(2021)\citenamefont{Kerelsky,
  Rubio-Verd{\'u}, Xian, Kennes, Halbertal, Finney, Song, Turkel, Wang,
  Watanabe et~al.}}]{Kerelsky2021moireless}
\bibinfo{author}{\bibfnamefont{A.}~\bibnamefont{Kerelsky}},
  \bibinfo{author}{\bibfnamefont{C.}~\bibnamefont{Rubio-Verd{\'u}}},
  \bibinfo{author}{\bibfnamefont{L.}~\bibnamefont{Xian}},
  \bibinfo{author}{\bibfnamefont{D.~M.} \bibnamefont{Kennes}},
  \bibinfo{author}{\bibfnamefont{D.}~\bibnamefont{Halbertal}},
  \bibinfo{author}{\bibfnamefont{N.}~\bibnamefont{Finney}},
  \bibinfo{author}{\bibfnamefont{L.}~\bibnamefont{Song}},
  \bibinfo{author}{\bibfnamefont{S.}~\bibnamefont{Turkel}},
  \bibinfo{author}{\bibfnamefont{L.}~\bibnamefont{Wang}},
  \bibinfo{author}{\bibfnamefont{K.}~\bibnamefont{Watanabe}},
  \bibnamefont{et~al.}, \bibinfo{journal}{Proceedings of the National Academy
  of Sciences} \textbf{\bibinfo{volume}{118}} (\bibinfo{year}{2021}).

\bibitem[{\citenamefont{Khalaf et~al.}(2019)\citenamefont{Khalaf, Kruchkov,
  Tarnopolsky, and Vishwanath}}]{Khalaf2019}
\bibinfo{author}{\bibfnamefont{E.}~\bibnamefont{Khalaf}},
  \bibinfo{author}{\bibfnamefont{A.~J.} \bibnamefont{Kruchkov}},
  \bibinfo{author}{\bibfnamefont{G.}~\bibnamefont{Tarnopolsky}},
  \bibnamefont{and}
  \bibinfo{author}{\bibfnamefont{A.}~\bibnamefont{Vishwanath}},
  \bibinfo{journal}{Phys. Rev. B} \textbf{\bibinfo{volume}{100}},
  \bibinfo{pages}{085109} (\bibinfo{year}{2019}),
  \urlprefix\url{https://link.aps.org/doi/10.1103/PhysRevB.100.085109}.

\bibitem[{\citenamefont{Chou et~al.}(2022{\natexlab{a}})\citenamefont{Chou, Wu,
  Sau, and Sarma}}]{YZBernal}
\bibinfo{author}{\bibfnamefont{Y.-Z.} \bibnamefont{Chou}},
  \bibinfo{author}{\bibfnamefont{F.}~\bibnamefont{Wu}},
  \bibinfo{author}{\bibfnamefont{J.~D.} \bibnamefont{Sau}}, \bibnamefont{and}
  \bibinfo{author}{\bibfnamefont{S.~D.} \bibnamefont{Sarma}},
  \bibinfo{journal}{Physical Review B} \textbf{\bibinfo{volume}{105}}
  (\bibinfo{year}{2022}{\natexlab{a}}),
  \urlprefix\url{https://doi.org/10.1103%2Fphysrevb.105.l100503}.

\bibitem[{\citenamefont{Chou et~al.}(2021)\citenamefont{Chou, Wu, Sau, and
  Sarma}}]{YZRhombo}
\bibinfo{author}{\bibfnamefont{Y.-Z.} \bibnamefont{Chou}},
  \bibinfo{author}{\bibfnamefont{F.}~\bibnamefont{Wu}},
  \bibinfo{author}{\bibfnamefont{J.~D.} \bibnamefont{Sau}}, \bibnamefont{and}
  \bibinfo{author}{\bibfnamefont{S.~D.} \bibnamefont{Sarma}},
  \bibinfo{journal}{Physical Review Letters} \textbf{\bibinfo{volume}{127}}
  (\bibinfo{year}{2021}),
  \urlprefix\url{https://doi.org/10.1103%2Fphysrevlett.127.187001}.

\bibitem[{\citenamefont{Chou et~al.}(2022{\natexlab{b}})\citenamefont{Chou, Wu,
  Sau, and Sarma}}]{YZLongPaper}
\bibinfo{author}{\bibfnamefont{Y.-Z.} \bibnamefont{Chou}},
  \bibinfo{author}{\bibfnamefont{F.}~\bibnamefont{Wu}},
  \bibinfo{author}{\bibfnamefont{J.~D.} \bibnamefont{Sau}}, \bibnamefont{and}
  \bibinfo{author}{\bibfnamefont{S.~D.} \bibnamefont{Sarma}},
  \emph{\bibinfo{title}{Acoustic-phonon-mediated superconductivity in
  moiréless graphene multilayers}} (\bibinfo{year}{2022}{\natexlab{b}}),
  \urlprefix\url{https://arxiv.org/abs/2204.09811}.

\bibitem[{\citenamefont{You and Vishwanath}(2021)}]{EEAshvin}
\bibinfo{author}{\bibfnamefont{Y.-Z.} \bibnamefont{You}} \bibnamefont{and}
  \bibinfo{author}{\bibfnamefont{A.}~\bibnamefont{Vishwanath}},
  \emph{\bibinfo{title}{Kohn-luttinger superconductivity and inter-valley
  coherence in rhombohedral trilayer graphene}} (\bibinfo{year}{2021}),
  \urlprefix\url{https://arxiv.org/abs/2109.04669}.

\bibitem[{\citenamefont{Ghazaryan et~al.}(2021)\citenamefont{Ghazaryan, Holder,
  Serbyn, and Berg}}]{EEBerg}
\bibinfo{author}{\bibfnamefont{A.}~\bibnamefont{Ghazaryan}},
  \bibinfo{author}{\bibfnamefont{T.}~\bibnamefont{Holder}},
  \bibinfo{author}{\bibfnamefont{M.}~\bibnamefont{Serbyn}}, \bibnamefont{and}
  \bibinfo{author}{\bibfnamefont{E.}~\bibnamefont{Berg}},
  \bibinfo{journal}{Phys. Rev. Lett.} \textbf{\bibinfo{volume}{127}},
  \bibinfo{pages}{247001} (\bibinfo{year}{2021}),
  \urlprefix\url{https://link.aps.org/doi/10.1103/PhysRevLett.127.247001}.

\bibitem[{\citenamefont{Szab{\'{o}} and Roy}(2022)}]{EEBitan}
\bibinfo{author}{\bibfnamefont{A.~L.} \bibnamefont{Szab{\'{o}}}}
  \bibnamefont{and} \bibinfo{author}{\bibfnamefont{B.}~\bibnamefont{Roy}},
  \bibinfo{journal}{Physical Review B} \textbf{\bibinfo{volume}{105}}
  (\bibinfo{year}{2022}),
  \urlprefix\url{https://doi.org/10.1103%2Fphysrevb.105.l081407}.

\bibitem[{\citenamefont{Cea et~al.}(2022)\citenamefont{Cea, Pantale\'on, Phong,
  and Guinea}}]{EEGuinea}
\bibinfo{author}{\bibfnamefont{T.}~\bibnamefont{Cea}},
  \bibinfo{author}{\bibfnamefont{P.~A.} \bibnamefont{Pantale\'on}},
  \bibinfo{author}{\bibfnamefont{V.~o.~T.} \bibnamefont{Phong}},
  \bibnamefont{and} \bibinfo{author}{\bibfnamefont{F.}~\bibnamefont{Guinea}},
  \bibinfo{journal}{Phys. Rev. B} \textbf{\bibinfo{volume}{105}},
  \bibinfo{pages}{075432} (\bibinfo{year}{2022}),
  \urlprefix\url{https://link.aps.org/doi/10.1103/PhysRevB.105.075432}.

\bibitem[{\citenamefont{Dong and Levitov}(2021)}]{EELeonid}
\bibinfo{author}{\bibfnamefont{Z.}~\bibnamefont{Dong}} \bibnamefont{and}
  \bibinfo{author}{\bibfnamefont{L.}~\bibnamefont{Levitov}},
  \emph{\bibinfo{title}{Superconductivity in the vicinity of an
  isospin-polarized state in a cubic dirac band}} (\bibinfo{year}{2021}),
  \urlprefix\url{https://arxiv.org/abs/2109.01133}.

\bibitem[{\citenamefont{Chatterjee et~al.}(2021)\citenamefont{Chatterjee, Wang,
  Berg, and Zaletel}}]{EEZalatel}
\bibinfo{author}{\bibfnamefont{S.}~\bibnamefont{Chatterjee}},
  \bibinfo{author}{\bibfnamefont{T.}~\bibnamefont{Wang}},
  \bibinfo{author}{\bibfnamefont{E.}~\bibnamefont{Berg}}, \bibnamefont{and}
  \bibinfo{author}{\bibfnamefont{M.~P.} \bibnamefont{Zaletel}},
  \emph{\bibinfo{title}{Inter-valley coherent order and isospin fluctuation
  mediated superconductivity in rhombohedral trilayer graphene}}
  (\bibinfo{year}{2021}), \urlprefix\url{https://arxiv.org/abs/2109.00002}.

\bibitem[{\citenamefont{Szab\'o and Roy}(2022)}]{Szabo2021BBG}
\bibinfo{author}{\bibfnamefont{A.~L.} \bibnamefont{Szab\'o}} \bibnamefont{and}
  \bibinfo{author}{\bibfnamefont{B.}~\bibnamefont{Roy}},
  \bibinfo{journal}{Phys. Rev. B} \textbf{\bibinfo{volume}{105}},
  \bibinfo{pages}{L201107} (\bibinfo{year}{2022}),
  \urlprefix\url{https://link.aps.org/doi/10.1103/PhysRevB.105.L201107}.

\bibitem[{\citenamefont{Qin et~al.}(2022)\citenamefont{Qin, Huang, Wolf, Wei,
  Blinov, and MacDonald}}]{Qin2022}
\bibinfo{author}{\bibfnamefont{W.}~\bibnamefont{Qin}},
  \bibinfo{author}{\bibfnamefont{C.}~\bibnamefont{Huang}},
  \bibinfo{author}{\bibfnamefont{T.}~\bibnamefont{Wolf}},
  \bibinfo{author}{\bibfnamefont{N.}~\bibnamefont{Wei}},
  \bibinfo{author}{\bibfnamefont{I.}~\bibnamefont{Blinov}}, \bibnamefont{and}
  \bibinfo{author}{\bibfnamefont{A.~H.} \bibnamefont{MacDonald}},
  \bibinfo{journal}{arXiv preprint arXiv:2203.09083}  (\bibinfo{year}{2022}).

\bibitem[{\citenamefont{Dai et~al.}(2022)\citenamefont{Dai, Ma, Zhang, and
  Ma}}]{Dai2022}
\bibinfo{author}{\bibfnamefont{H.}~\bibnamefont{Dai}},
  \bibinfo{author}{\bibfnamefont{R.}~\bibnamefont{Ma}},
  \bibinfo{author}{\bibfnamefont{X.}~\bibnamefont{Zhang}}, \bibnamefont{and}
  \bibinfo{author}{\bibfnamefont{T.}~\bibnamefont{Ma}}, \bibinfo{journal}{arXiv
  preprint arXiv:2204.06222}  (\bibinfo{year}{2022}).

\bibitem[{\citenamefont{Hwang and Sarma}(2008)}]{SankarGrapheneKT1}
\bibinfo{author}{\bibfnamefont{E.~H.} \bibnamefont{Hwang}} \bibnamefont{and}
  \bibinfo{author}{\bibfnamefont{S.~D.} \bibnamefont{Sarma}},
  \bibinfo{journal}{Physical Review B} \textbf{\bibinfo{volume}{77}}
  (\bibinfo{year}{2008}),
  \urlprefix\url{https://doi.org/10.1103%2Fphysrevb.77.115449}.

\bibitem[{\citenamefont{Min et~al.}(2011)\citenamefont{Min, Hwang, and
  Sarma}}]{SankarGrapheneKT2}
\bibinfo{author}{\bibfnamefont{H.}~\bibnamefont{Min}},
  \bibinfo{author}{\bibfnamefont{E.~H.} \bibnamefont{Hwang}}, \bibnamefont{and}
  \bibinfo{author}{\bibfnamefont{S.~D.} \bibnamefont{Sarma}},
  \bibinfo{journal}{Physical Review B} \textbf{\bibinfo{volume}{83}}
  (\bibinfo{year}{2011}),
  \urlprefix\url{https://doi.org/10.1103%2Fphysrevb.83.161404}.

\bibitem[{\citenamefont{Wu et~al.}(2019{\natexlab{b}})\citenamefont{Wu, Hwang,
  and Sarma}}]{SankarGrapheneKT3}
\bibinfo{author}{\bibfnamefont{F.}~\bibnamefont{Wu}},
  \bibinfo{author}{\bibfnamefont{E.}~\bibnamefont{Hwang}}, \bibnamefont{and}
  \bibinfo{author}{\bibfnamefont{S.~D.} \bibnamefont{Sarma}},
  \bibinfo{journal}{Physical Review B} \textbf{\bibinfo{volume}{99}}
  (\bibinfo{year}{2019}{\natexlab{b}}),
  \urlprefix\url{https://doi.org/10.1103%2Fphysrevb.99.165112}.

\bibitem[{\citenamefont{Li et~al.}(2020)\citenamefont{Li, Wu, and
  Sarma}}]{SankarGrapheneKT4}
\bibinfo{author}{\bibfnamefont{X.}~\bibnamefont{Li}},
  \bibinfo{author}{\bibfnamefont{F.}~\bibnamefont{Wu}}, \bibnamefont{and}
  \bibinfo{author}{\bibfnamefont{S.~D.} \bibnamefont{Sarma}},
  \bibinfo{journal}{Physical Review B} \textbf{\bibinfo{volume}{101}}
  (\bibinfo{year}{2020}),
  \urlprefix\url{https://doi.org/10.1103%2Fphysrevb.101.245436}.

\bibitem[{\citenamefont{Hwang and
  Sarma}(2019{\natexlab{a}})}]{SankarGrapheneKT5}
\bibinfo{author}{\bibfnamefont{E.~H.} \bibnamefont{Hwang}} \bibnamefont{and}
  \bibinfo{author}{\bibfnamefont{S.~D.} \bibnamefont{Sarma}},
  \bibinfo{journal}{Physical Review B} \textbf{\bibinfo{volume}{99}}
  (\bibinfo{year}{2019}{\natexlab{a}}),
  \urlprefix\url{https://doi.org/10.1103%2Fphysrevb.99.085105}.

\bibitem[{\citenamefont{Ziman}(1960)}]{Ziman}
\bibinfo{author}{\bibfnamefont{J.~M.} \bibnamefont{Ziman}},
  \emph{\bibinfo{title}{Electrons and Phonons}} (\bibinfo{publisher}{Oxford
  University Press}, \bibinfo{year}{1960}), ISBN \bibinfo{isbn}{9780198507796}.

\bibitem[{\citenamefont{Ashcroft and Mermin}(1976)}]{AshcroftAndMermin}
\bibinfo{author}{\bibfnamefont{N.~W.} \bibnamefont{Ashcroft}} \bibnamefont{and}
  \bibinfo{author}{\bibfnamefont{N.~D.} \bibnamefont{Mermin}},
  \emph{\bibinfo{title}{Solid State Physics}} (\bibinfo{publisher}{Harcourt
  College Publishers}, \bibinfo{year}{1976}), ISBN
  \bibinfo{isbn}{9780030839931}.

\bibitem[{\citenamefont{Efetov and Kim}(2010)}]{Efetov_2010}
\bibinfo{author}{\bibfnamefont{D.~K.} \bibnamefont{Efetov}} \bibnamefont{and}
  \bibinfo{author}{\bibfnamefont{P.}~\bibnamefont{Kim}},
  \bibinfo{journal}{Physical Review Letters} \textbf{\bibinfo{volume}{105}}
  (\bibinfo{year}{2010}),
  \urlprefix\url{https://doi.org/10.1103%2Fphysrevlett.105.256805}.

\bibitem[{\citenamefont{Jung and
  MacDonald}(2014)}]{MacDonaldBandstructureBernal}
\bibinfo{author}{\bibfnamefont{J.}~\bibnamefont{Jung}} \bibnamefont{and}
  \bibinfo{author}{\bibfnamefont{A.~H.} \bibnamefont{MacDonald}},
  \bibinfo{journal}{Phys. Rev. B} \textbf{\bibinfo{volume}{89}},
  \bibinfo{pages}{035405} (\bibinfo{year}{2014}),
  \urlprefix\url{https://link.aps.org/doi/10.1103/PhysRevB.89.035405}.

\bibitem[{\citenamefont{Zhang et~al.}(2010)\citenamefont{Zhang, Sahu, Min, and
  MacDonald}}]{MacDonaldBandstructureRhombo}
\bibinfo{author}{\bibfnamefont{F.}~\bibnamefont{Zhang}},
  \bibinfo{author}{\bibfnamefont{B.}~\bibnamefont{Sahu}},
  \bibinfo{author}{\bibfnamefont{H.}~\bibnamefont{Min}}, \bibnamefont{and}
  \bibinfo{author}{\bibfnamefont{A.~H.} \bibnamefont{MacDonald}},
  \bibinfo{journal}{Physical Review B} \textbf{\bibinfo{volume}{82}}
  (\bibinfo{year}{2010}),
  \urlprefix\url{https://doi.org/10.1103%2Fphysrevb.82.035409}.

\bibitem[{\citenamefont{Polshyn
  et~al.}(2019{\natexlab{b}})\citenamefont{Polshyn, Yankowitz, Chen, Zhang,
  Watanabe, Taniguchi, Dean, and Young}}]{Polshyn2019}
\bibinfo{author}{\bibfnamefont{H.}~\bibnamefont{Polshyn}},
  \bibinfo{author}{\bibfnamefont{M.}~\bibnamefont{Yankowitz}},
  \bibinfo{author}{\bibfnamefont{S.}~\bibnamefont{Chen}},
  \bibinfo{author}{\bibfnamefont{Y.}~\bibnamefont{Zhang}},
  \bibinfo{author}{\bibfnamefont{K.}~\bibnamefont{Watanabe}},
  \bibinfo{author}{\bibfnamefont{T.}~\bibnamefont{Taniguchi}},
  \bibinfo{author}{\bibfnamefont{C.~R.} \bibnamefont{Dean}}, \bibnamefont{and}
  \bibinfo{author}{\bibfnamefont{A.~F.} \bibnamefont{Young}},
  \bibinfo{journal}{Nature Physics} \textbf{\bibinfo{volume}{15}},
  \bibinfo{pages}{1011} (\bibinfo{year}{2019}{\natexlab{b}}).

\bibitem[{\citenamefont{Siriviboon et~al.}(2021)\citenamefont{Siriviboon, Lin,
  Liu, Scammell, Liu, Rhodes, Watanabe, Taniguchi, Hone, Scheurer
  et~al.}}]{Siriviboon_2021}
\bibinfo{author}{\bibfnamefont{P.}~\bibnamefont{Siriviboon}},
  \bibinfo{author}{\bibfnamefont{J.-X.} \bibnamefont{Lin}},
  \bibinfo{author}{\bibfnamefont{X.}~\bibnamefont{Liu}},
  \bibinfo{author}{\bibfnamefont{H.~D.} \bibnamefont{Scammell}},
  \bibinfo{author}{\bibfnamefont{S.}~\bibnamefont{Liu}},
  \bibinfo{author}{\bibfnamefont{D.}~\bibnamefont{Rhodes}},
  \bibinfo{author}{\bibfnamefont{K.}~\bibnamefont{Watanabe}},
  \bibinfo{author}{\bibfnamefont{T.}~\bibnamefont{Taniguchi}},
  \bibinfo{author}{\bibfnamefont{J.}~\bibnamefont{Hone}},
  \bibinfo{author}{\bibfnamefont{M.~S.} \bibnamefont{Scheurer}},
  \bibnamefont{et~al.}, \emph{\bibinfo{title}{A new flavor of correlation and
  superconductivity in small twist-angle trilayer graphene}}
  (\bibinfo{year}{2021}), \urlprefix\url{https://arxiv.org/abs/2112.07127}.

\bibitem[{\citenamefont{Sohier et~al.}(2014)\citenamefont{Sohier, Calandra,
  Park, Bonini, Marzari, and Mauri}}]{Sohier2014}
\bibinfo{author}{\bibfnamefont{T.}~\bibnamefont{Sohier}},
  \bibinfo{author}{\bibfnamefont{M.}~\bibnamefont{Calandra}},
  \bibinfo{author}{\bibfnamefont{C.-H.} \bibnamefont{Park}},
  \bibinfo{author}{\bibfnamefont{N.}~\bibnamefont{Bonini}},
  \bibinfo{author}{\bibfnamefont{N.}~\bibnamefont{Marzari}}, \bibnamefont{and}
  \bibinfo{author}{\bibfnamefont{F.}~\bibnamefont{Mauri}},
  \bibinfo{journal}{Phys. Rev. B} \textbf{\bibinfo{volume}{90}},
  \bibinfo{pages}{125414} (\bibinfo{year}{2014}),
  \urlprefix\url{https://link.aps.org/doi/10.1103/PhysRevB.90.125414}.

\bibitem[{\citenamefont{Xie and Foster}(2016)}]{Xie2016}
\bibinfo{author}{\bibfnamefont{H.-Y.} \bibnamefont{Xie}} \bibnamefont{and}
  \bibinfo{author}{\bibfnamefont{M.~S.} \bibnamefont{Foster}},
  \bibinfo{journal}{Phys. Rev. B} \textbf{\bibinfo{volume}{93}},
  \bibinfo{pages}{195103} (\bibinfo{year}{2016}),
  \urlprefix\url{https://link.aps.org/doi/10.1103/PhysRevB.93.195103}.

\bibitem[{\citenamefont{Ghahari et~al.}(2016)\citenamefont{Ghahari, Xie,
  Taniguchi, Watanabe, Foster, and Kim}}]{Ghahari2016}
\bibinfo{author}{\bibfnamefont{F.}~\bibnamefont{Ghahari}},
  \bibinfo{author}{\bibfnamefont{H.-Y.} \bibnamefont{Xie}},
  \bibinfo{author}{\bibfnamefont{T.}~\bibnamefont{Taniguchi}},
  \bibinfo{author}{\bibfnamefont{K.}~\bibnamefont{Watanabe}},
  \bibinfo{author}{\bibfnamefont{M.~S.} \bibnamefont{Foster}},
  \bibnamefont{and} \bibinfo{author}{\bibfnamefont{P.}~\bibnamefont{Kim}},
  \bibinfo{journal}{Phys. Rev. Lett.} \textbf{\bibinfo{volume}{116}},
  \bibinfo{pages}{136802} (\bibinfo{year}{2016}),
  \urlprefix\url{https://link.aps.org/doi/10.1103/PhysRevLett.116.136802}.

\bibitem[{\citenamefont{Coleman}(2015)}]{Coleman2015introduction}
\bibinfo{author}{\bibfnamefont{P.}~\bibnamefont{Coleman}},
  \emph{\bibinfo{title}{Introduction to Many-Body Physics}}
  (\bibinfo{publisher}{Cambridge University Press}, \bibinfo{year}{2015}), ISBN
  \bibinfo{isbn}{9780521864886}.

\bibitem[{\citenamefont{Poniatowski et~al.}(2021)\citenamefont{Poniatowski,
  Sarkar, Sarma, and Greene}}]{Poniatowski_2021}
\bibinfo{author}{\bibfnamefont{N.~R.} \bibnamefont{Poniatowski}},
  \bibinfo{author}{\bibfnamefont{T.}~\bibnamefont{Sarkar}},
  \bibinfo{author}{\bibfnamefont{S.~D.} \bibnamefont{Sarma}}, \bibnamefont{and}
  \bibinfo{author}{\bibfnamefont{R.~L.} \bibnamefont{Greene}},
  \bibinfo{journal}{Physical Review B} \textbf{\bibinfo{volume}{103}}
  (\bibinfo{year}{2021}),
  \urlprefix\url{https://doi.org/10.1103%2Fphysrevb.103.l020501}.

\bibitem[{\citenamefont{Sarkar et~al.}(2018)\citenamefont{Sarkar, Greene, and
  Sarma}}]{Sarkar_2018}
\bibinfo{author}{\bibfnamefont{T.}~\bibnamefont{Sarkar}},
  \bibinfo{author}{\bibfnamefont{R.~L.} \bibnamefont{Greene}},
  \bibnamefont{and} \bibinfo{author}{\bibfnamefont{S.~D.} \bibnamefont{Sarma}},
  \bibinfo{journal}{Physical Review B} \textbf{\bibinfo{volume}{98}}
  (\bibinfo{year}{2018}),
  \urlprefix\url{https://doi.org/10.1103%2Fphysrevb.98.224503}.

\bibitem[{\citenamefont{Hwang and Sarma}(2019{\natexlab{b}})}]{Hwang_2019}
\bibinfo{author}{\bibfnamefont{E.~H.} \bibnamefont{Hwang}} \bibnamefont{and}
  \bibinfo{author}{\bibfnamefont{S.~D.} \bibnamefont{Sarma}},
  \bibinfo{journal}{Physical Review B} \textbf{\bibinfo{volume}{99}}
  (\bibinfo{year}{2019}{\natexlab{b}}),
  \urlprefix\url{https://doi.org/10.1103%2Fphysrevb.99.085105}.

\bibitem[{\citenamefont{Emery and Kivelson}(1995)}]{Kivelson_1995}
\bibinfo{author}{\bibfnamefont{V.~J.} \bibnamefont{Emery}} \bibnamefont{and}
  \bibinfo{author}{\bibfnamefont{S.~A.} \bibnamefont{Kivelson}},
  \bibinfo{journal}{Phys. Rev. Lett.} \textbf{\bibinfo{volume}{74}},
  \bibinfo{pages}{3253} (\bibinfo{year}{1995}),
  \urlprefix\url{https://link.aps.org/doi/10.1103/PhysRevLett.74.3253}.

\bibitem[{\citenamefont{Gurvitch}(1981)}]{Gurvitch_1981}
\bibinfo{author}{\bibfnamefont{M.}~\bibnamefont{Gurvitch}},
  \bibinfo{journal}{Phys. Rev. B} \textbf{\bibinfo{volume}{24}},
  \bibinfo{pages}{7404} (\bibinfo{year}{1981}),
  \urlprefix\url{https://link.aps.org/doi/10.1103/PhysRevB.24.7404}.

\bibitem[{\citenamefont{Millis et~al.}(1999)\citenamefont{Millis, Hu, and
  Sarma}}]{Millis_1999}
\bibinfo{author}{\bibfnamefont{A.~J.} \bibnamefont{Millis}},
  \bibinfo{author}{\bibfnamefont{J.}~\bibnamefont{Hu}}, \bibnamefont{and}
  \bibinfo{author}{\bibfnamefont{S.~D.} \bibnamefont{Sarma}},
  \bibinfo{journal}{Physical Review Letters} \textbf{\bibinfo{volume}{82}},
  \bibinfo{pages}{2354} (\bibinfo{year}{1999}),
  \urlprefix\url{https://doi.org/10.1103%2Fphysrevlett.82.2354}.

\bibitem[{\citenamefont{Hussey et~al.}(2004)\citenamefont{Hussey, Takenaka, and
  Takagi}}]{Hussey__2004}
\bibinfo{author}{\bibfnamefont{N.~E.} \bibnamefont{Hussey}},
  \bibinfo{author}{\bibfnamefont{K.}~\bibnamefont{Takenaka}}, \bibnamefont{and}
  \bibinfo{author}{\bibfnamefont{H.}~\bibnamefont{Takagi}},
  \bibinfo{journal}{Philosophical Magazine} \textbf{\bibinfo{volume}{84}},
  \bibinfo{pages}{2847} (\bibinfo{year}{2004}),
  \urlprefix\url{https://doi.org/10.1080%2F14786430410001716944}.

\end{thebibliography}

\end{document}